\documentclass[
reprint,
superscriptaddress,
altaffilletter,
nofootinbib,
amsmath,
amssymb,
aps,
prd]{revtex4-1}

\usepackage{times}
\usepackage{mathtools}
\usepackage{graphicx}
\usepackage{dcolumn}
\newcolumntype{d}[1]{D{.}{.}{#1}}
\usepackage{bm}
\usepackage{soul}
\usepackage{verbatim} 
\usepackage{cprotect} 
\usepackage{url}
\usepackage{subfigure} 
\usepackage{tikz}
\usepackage{afterpage}
\usepackage{nicefrac}
\usepackage{tabularx}
\newcolumntype{L}{>{\raggedright\arraybackslash}X}
\usepackage{multirow}
\usepackage[english]{babel}
\usepackage{natbib} 
\usepackage{glossaries} 
\usepackage{hyperref}
\usepackage{aas_macros} 
\usepackage{comment}
\usepackage{float}
\usepackage{placeins}
\usepackage{ulem}
\graphicspath{{figures/}}

\usepackage{xcolor} 

\newcommand{\orcid}[1]{\href{https://orcid.org/#1}{
    \includegraphics[width=10pt]{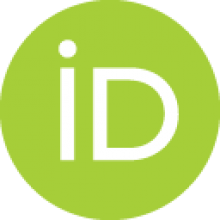}}}
\begin{document}

\title{Distinct neutrino signatures and onset condition of quark deconfinement in accretion-induced collapse of white dwarfs}
\author{Juno C. L. Chan\:\orcid{0000-0002-3377-4737}}
\email{chun.lung.chan@nbi.ku.dk}
\affiliation{Center of Gravity, Niels Bohr Institute, Blegdamsvej 17, 2100 Copenhagen, Denmark}

\author{Harry Ho-Yin Ng\:\orcid{0000-0003-3453-7394}}
\affiliation{Institut f\"ur Theoretische Physik, Goethe Universit\"at,
  Max-von-Laue-Str. 1, 60438 Frankfurt am Main, Germany}

\author{Patrick Chi-Kit Cheong\:\orcid{0000-0003-1449-3363}}
\affiliation{Center for Nonlinear Studies, Los Alamos National Laboratory, Los Alamos, NM 87545, USA}
\affiliation{Department of Physics \& Astronomy, University of New Hampshire, 9 Library Way, Durham NH 03824, USA}
\affiliation{Department of Physics, University of California, Berkeley, Berkeley, CA 94720, USA}

\date{\today}

\begin{abstract}
We present the first general relativistic,
neutrino-radiation hydrodynamics simulations of
accretion-induced collapse (AIC) extending to seconds
after core bounce, using realistic hadron–quark hybrid equations of state (EOSs).
A first-order QCD phase transition (PT) triggers a
second dynamical collapse and the formation of a quasistable protohybrid star (PHS)
with a deconfined quark core and a distinctive second neutrino burst.
We find that the thermally suppressed onset of the mixed phase
allows low-mass protoneutron stars to enter the hadron-quark mixed phase during long-term evolution,
even for hybrid EOSs with high onset densities.
In contrast to core-collapse supernovae (CCSNe), AIC models exhibit a tightly constrained onset mass
with minimal EOS dependence,
owing to the absence of a massive envelope and
thus the reduced postbounce accretion.
This enhances the sensitivity of neutrino observables in AIC to hybrid EOS properties.
We establish empirical relations between PT onset density and neutrino signatures,
revealing a distinct behavior in AIC not seen in CCSNe.
Our results suggest that a single Galactic AIC neutrino detection could place strong constraints on QCD PT thresholds,
hybrid EOS characteristics,
and the existence of PHSs.
PT in AIC may also produce 
gravitational waves, gamma-ray bursts,
and $r$-process elements,
motivating multidimensional simulations with rotation,
magnetic fields, and improved microphysics for realistic multimessenger predictions.
\end{abstract}

\maketitle

\section{Introduction}

White dwarfs (WDs) are compact stellar remnants supported by electron 
degeneracy pressure. 
The outcome of an accreting WD is determined 
by its chemical composition, initial mass, and accretion rate.
Carbon-oxygen WDs with masses of $\lesssim 1.2 \,M_{\odot}$ and high accretion rates
are prone to carbon deflagration, leading to a type Ia supernova that
leaves no remnant~\cite{2007MNRAS.380..933Y,2009ApJ...692..324S,2009ApJ...705..693S,2013ApJ...776...97M}.
In contrast, oxygen-neon WDs with masses of $\gtrsim 1.2 \,M_{\odot}$ and
slower accretion rates can continue to gain mass until they reach
the Chandrasekhar limit beyond $1.44\,M_{\odot}$~\cite{1931ApJ....74...81C}.
This triggers a gravitational collapse due to electron capture, followed by a core bounce,
resulting in the formation of a
protoneutron star (PNS) remnant--a phenomenon known as accretion-induced
collapse (AIC) and the resulting explosion, known as an electron-capture supernova (SN)
~\cite{1991ApJ...367L..19N,Yoon:2005cz}. 

AIC can occur via two main channels: mass transfer from 
a companion star to a WD in a binary system~\cite{nomoto1991conditions,Tauris:2013zna,Wang:2018wje}, 
or binary WD merger~\cite{Schwab:2016lep,Wu:2018pbt,Ruiter:2019tam,Liu:2020qny,Wang:2020pzc,Schwab:2020law}.
The resulting super-Chandrasekhar WDs are often 
rapidly rotating~\cite{Piersanti:2002sb,Uenishi:2003sk,Saio:2004gz} and highly magnetized
~\cite{2015SSRv..191..111F,2020IAUS..357...60K,2020AdSpR..66.1025F,2015ApJ...806L...1Z,Pakmor:2024efc} in both channels.
The collapse of such progenitors is expected
to produce multimessenger signals, including gravitational waves and neutrinos~\cite{Fryer:2001zw,Dimmelmeier:2008iq,Abdikamalov:2009aq,LongoMicchi:2023khv}, 
and to explain the origin of millisecond pulsars~\cite{Tauris:2013zna}.
The ejecta mass driven by neutrino heating can potentially power
rapid neutron capture (r-process) nucleosynthesis~\cite{Fryer:1998jb,Dessart:2006gd,Batziou:2024ory,Cheong:2024hrd,Yip:2024akb}.
Furthermore, these collapses can potentially power gamma-ray bursts
~\cite{Yi:1997qb,Metzger:2007cd,Perley:2008ay,Cheong:2024hrd},
fast radio bursts~\cite{Waxman:2017zme,Margalit:2019hke} and
fast blue optical transients~\cite{Lyutikov:2018pkv,Lyutikov:2022xri}.

The galactic rate of AIC events is estimated to be
$(0.3$–$0.9) \times 10^{-3}~\rm{yr}^{-1}$ from mass transfer onto a WD~\cite{Wang:2018wje},
and $(1.4$–$8.9) \times 10^{-3}~\rm{yr}^{-1}$ from binary WD mergers~\cite{Liu:2020qny,Wang:2020pzc}.
Although the occurrence rates remains uncertain, there is observational evidence of AIC~\cite{McBrien:2019wfv,Moriya:2019ees,Gillanders:2020fhm}.
Moreover, several systems have been proposed as potential AIC progenitors or remnants,
based on their total mass, magnetic field strength, or binary configuration~\cite{Kato:2003qz,2019Natur.569..684G,2020A&A...644L...8O,Caiazzo:2021xkk,Luo:2024qwa,Zhao:2024gmz,Mereghetti:2024mxn}.

The progenitor mass of AIC depends on the formation channel.
AIC resulting from mass transfer onto a single WD typically occurs near the Chandrasekhar limit~\cite{Wang:2018wje},
whereas AIC from binary WD mergers can reach up to $\sim 2.5~M_{\odot}$~\cite{Liu:2020qny,Wang:2020pzc}.
This is substantially smaller than the progenitor mass of SN systems with a massive envelope, such as core-collapse supernovae (CCSNe),
which range from $8$–$100~M_{\odot}$~\cite{Woosley:2002zz}.
Besides, the low-mass nature leads to a considerable astrophysical 
interest owing to its characteristically low-mass progenitors.
It is expected to explain the formation of low-mass neutron stars~\cite{Leung:2019ctw},
and low-mass millisecond pulsars or x-ray binaries~\cite{Ablimit:2014vka}.
The ejecta mass, which depends on the magnetic field strength
and rotation period~\cite{Cheong:2024hrd,Kuroda:2025iyj},
is bounded by the difference between the initial WD and the final PNS masses,
typically ranging up to $\sim10^{-1~}M_{\odot}$~\cite{LongoMicchi:2023khv,Cheong:2024hrd,Kuroda:2025iyj}.
This is generally smaller than that of CCSNe
associated with the collapse of massive stars,
due to the low-mass progenitors of AIC.
Moreover, the absence of a massive envelope allows the ejecta to maintain their initial high velocities.

The absence of a massive envelope and the low-mass nature of the pregenitor of an AIC system
lead to a short accretion phase after core bounce.
Thereafter, neutrino emission from the surface carries away thermal energy,
and the resulting loss of thermal pressure drives the contraction of the PNS~\cite{Burrows:1986me,Janka:2012wk}.
This process, known as Kelvin-Helmholtz cooling, proceeds over a timescale of several seconds.
Consequently, the density and temperature at the core of the PNS can increase continuously,
though at a slower rate than during the accretion phase.
Several-second-scale simulations of the PNS cooling phase
have shown that the central temperature can rise to $40-60~\rm{MeV}$ within the first ten seconds,
before eventually decreasing~\cite{Burrows:1986me,1989ASIB..216..731W,Pons:1998mm,Roberts:2012zza}.
This long-term evolution naturally raises the question of whether the
conditions within AIC systems become suitable for triggering 
a quantum chromodynamics (QCD) phase transition (PT).

A conservative estimate is to utilize the phase diagram of a hadron-quark hybrid equation of state (EOS)
to examine whether the onset of the hadron-quark mixed phase can be reached by a 
typical PNS near nuclear saturation density $\rho_{\rm{sat}}$
and temperatures of $\sim 10-20~\rm{MeV}$.
For some hybrid EOSs with onset conditions 
near $\rho_{\rm{sat}}$~\cite{Fischer:2010wp,Bastian:2020unt,Blacker:2023afl},
it is reasonable to expect that QCD PT can take place in the low-mass AIC system.
A key remaining question is whether the thermodynamic conditions 
in AIC systems can reach the PT onset threshold for certain hybrid EOSs,
particularly in cases where the onset density of PT 
is suppressed at higher temperatures~\cite{Bauswein:2019skm,Miao:2020yjk,Bastian:2020unt,Blacker:2020nlq,Blacker:2023afl,Zhu:2025vmz}.
  
In similar compact stellar systems such as CCSNe, hybrid EOSs featuring a first-order QCD PT
have been shown to induce a second collapse, halted by the formation of a stiffer deconfined quark core,
and to generate a secondary hydrodynamical shock accompanied by a neutrino burst~\cite{Sagert:2008ka,Fischer:2017lag,Zha:2020gjw,Zha:2021fbi,Fischer:2021tvv,Kuroda:2021eiv,Jakobus:2022ucs,Janka:2022krt,Bauswein:2022vtq,Largani:2023oyk}.
Moreover, such a collapse can produce a strong burst of gravitational waves~\cite{Zha:2020gjw},
and the explosion driven by the QCD PT 
has been proposed as a potential r-process site~\cite{Fischer:2020xjl},
with its properties closely linked to the characteristics of the hybrid EOS~\cite{Largani:2023oyk,Huang:2024xff}.
These findings highlight the astrophysical and observational significance of studying a first order QCD PT in the AIC system.
The properties of the PT, such as the onset conditions and the location of the critical point, are a central topic in nuclear astrophysics for understanding the behavior of dense matter. 
These aspects have been extensively investigated through theoretical studies of dense matter EOS~\cite{Bluhm:2024uhj,Clarke:2024ugt,Shah:2024img}, gravitational-wave astronomy~\cite{Chatziioannou:2019yko,Pang:2020ilf,Fujimoto:2022xhv}, and 
observation~\cite{Li:2024sft,Tang:2025xib}.
Since the primary difference between AIC and CCSNe is the absence of a massive envelope,
it is essential to investigate how the dynamics of a QCD PT may unfold in the AIC scenario,
where the lack of an extended envelope could potentially alter the resulting evolution and observables.

In this work, we present the first seconds-long general relativistic
neutrino-radiation simulations of the AIC systems that employ realistic hadron-quark
hybrid EOSs to explore the influence first-order QCD
PT, the existence of deconfined quark matter and the properties of hybrid stars.

Unlike the isolated neutron star setups adopted in Ref.~\cite{Lin:2005zda,Mallick:2020bdc}, 
our super-Chandrasekhar mass WD model follows the collapse 
and formation of a PNS embedded in an accreting envelope. 
This leads to a strongly nonequilibrium environment with evolving density, 
temperature, and electron fraction regulated by neutrino transport, 
which is absent in equilibrium neutron star configurations. 
As a result, the PT in our work is determined self-consistently by 
the collapse dynamics rather than imposed on a preexisting neutron star.

This paper is structured as follows. 
In Sec.~\ref{sec:setup}, we present our numerical setup. 
Our main results are presented in Sec.~\ref{sec:results}, 
followed by a discussion of future research directions in Sec.~\ref{sec:discussion}.

\section{Numerical methods and setups}\label{sec:setup}

We use a set of hadron-quark hybrid EOSs, 
known as relativistic density functional (RDF) EOSs, 
where both quark and hadron matter are described 
within the RDF formalism~\cite{compose,Bastian:2020unt}.
In the RDF formalism, both the hadronic and quark models are described by a set of coupling parameters that determine the underlying interaction potentials~\cite{Bastian:2020unt}.
A first-order PT connects the hadronic phase to the 
stiffer deconfined quark phase, forming mixed nuclear-quark matter
constructed via Maxwell reconstruction under the assumption of
mechanical and chemical equilibrium. 
For each set of coupling parameters, the latent heat is set 
by the pressure–density jump between 
hadronic and quark branches, 
and controlled by quark couplings tuned 
to fix the onset density and density jump~\cite{Bastian:2020unt}.
The RDF EOSs incorporate varying onset densities 
and latent heats, with deconfined quark matter modeled 
using a string-flip approach. 
These EOSs are consistent with astrophysical constraints,
including precise high-mass neutron star
measurements~\cite{Antoniadis:2013pzd} and recent NICER
results~\cite{Fonseca:2021wxt}.
These EOSs have been recently utilized in astrophysical simulations
~\cite{Blacker:2020nlq, Fischer:2021tvv,Kuroda:2021eiv,Jakobus:2022ucs,Largani:2023oyk}.
Detailed properties are listed in Table~\ref{model} 
and can also be found in Ref.~\cite{Bastian:2020unt,Largani:2023oyk}.

Since a given set of coupling parameters 
does not produce systematic changes in key observables, 
such as the onset density and density jump (latent heat) of 
the hadron–quark PT, 
we select four hybrid EOSs with notably 
different onset densities to simplify the comparison. 
The corresponding latent heats do not vary in an equally systematic manner, 
so we restrict our systematic analysis to the onset conditions.

We focus on the double degenerate binary merger channel for AIC~\cite{Schwab:2016lep,Wu:2018pbt,Ruiter:2019tam,Liu:2020qny,Wang:2020pzc,Schwab:2020law}, in which the merger of two WDs—including double ONeMg, ONeMg-CO, or double CO configurations—produces a super-Chandrasekhar remnant that undergoes AIC rather than thermonuclear explosion.
In particular, for double CO WD mergers, nonstandard evolutionary pathways involving off centre carbon ignition can transform the merger remnant into an ONe-core configuration that may subsequently collapse to a neutron star~\cite{Wang:2020pzc,Schwab:2020law}.

All progenitors are generated using \texttt{RNS}~\cite{Stergioulas:1994ea}.
They are all spherically symmetric, nonrotating, and in beta equilibrium.
The central energy density $\epsilon_c / c^{2}$ is $10^{9}~\rm{g~cm^{-3}}$.
The gravitational mass and baryonic mass are $1.53~M_{\odot}$ 
and $1.54~M_{\odot}$, respectively, with a radius of $2800~\rm{km}$. 
The use of massive WD initial models is commonly adopted in AIC studies
~\cite{Dessart:2006gd,Dessart:2007kh,LongoMicchi:2023khv,Cheong:2024hrd,Batziou:2024ory}.
Although rotation is known to significantly affect the dynamics of AIC
~\cite{LongoMicchi:2023khv,Cheong:2024hrd,Batziou:2024ory}, 
we neglect rotational degrees of freedom in this study to isolate the effects of 
the hadron-quark EOS on collapse dynamics and neutrino signals. The impact of rotation will be explored in future work.
Since \texttt{RNS} does not solve for temperature, all progenitors generated are cold and require additional construction before the evolution.

\begin{table}[]
        \centering
        \begin{tabular}{ccccc}
          \hline
          EOS  & $\rho_{\rm{onset}}$ & $M_{\rm{max}}$ & $M_{\rm{onset}}$ \\
          & [$10^{14} $~g~${\rm cm}^{-3}$]& [$M_{\odot} $] & [$M_{\odot} $]   \\
         \hline \hline
	 \texttt{RDF-1.9} & 4.6 & 2.16 & 0.81  \\
	 \texttt{RDF-1.2} & 7.3 & 2.15 & 1.37  \\
	 \texttt{RDF-1.5} &  8.2 & 2.03 & 1.46  \\
         \texttt{RDF-1.1} &  8.8  & 2.13 & 1.55 \\   
        \hline
        \end{tabular}
	\caption{The hybrid EOSs employed in our simulations.
			$\rho_{\rm{onset}}$ and $M_{\rm{onset}}$ denote the onset
			density and onset gravitational mass of the PT starting to reach the 
			mixed phase, where these three quantities are defined under
			 $T=0~\rm{MeV}$ and beta equilibrium~\cite{Largani:2023oyk}).  The
			maximum gravitational mass of the nonrotating model is denoted as
			$M_{\rm{max}}$ for each EOS.
			}
        \label{model}
\end{table}

\begin{table*}[t!]
\centering
\setlength{\tabcolsep}{2pt}
\begin{tabular}{c@{\hspace{0.02cm}}c@{\hspace{0.02cm}}cc@{\hspace{0.02cm}}c@{\hspace{0.02cm}}ccc|cccccccccccccccc}
\hline
\hline
EOS & $t_{\rm{mixed}}$ & $\rho_{\rm{mixed,c}}$ & $T_{\rm{mixed,c}}$ & $M^{\rm b}_{\rm{mixed}}$ & $\mathcal{L}_{\rm{mixed,tot}}$ & $t_{\rm{Q}}$  & $M^{\rm b}_{t_{\rm{Q}}}$ & $M^{\rm b}_{\rm{Q}}$ & $M^{\rm b}_{\rm{MM}}$ & $M^{\rm b}_{\rm{PHS}}$ & $r_{\rm{Q}}$ &  $r_{\rm{MM}}$ & $r_{\rm{PHS}}$  & $\mathcal{L}_{\nu_e}$ & $\mathcal{L}_{\bar\nu_e}$ &  $\mathcal{L}_{\nu_x}$& $\langle{\epsilon_{\nu_e} \rangle }$ &$ \langle {\epsilon_{\bar\nu_e}\rangle }$ & $ \langle {\epsilon_{\nu_x}}\rangle$ \\ \hline 
& $[\rm{s}]$ & [$10^{14}$ & [$\rm{MeV}$] & [$M_{\odot} $] & [$10^{51}$ &  $[\rm{s}]$ & [$M_{\odot} $] & [$M_{\odot} $] & [$M_{\odot} $] & [$M_{\odot} $] & [$\rm{km}$] & [$\rm{km}$] & [$\rm{km}$] &   & [$10^{51}$ & &  & [$\rm{MeV}$] & \\ 
           & & ${\rm g~cm}^{-3}$] &  &  & $\rm{erg~s^{-1}}$] &  &  &  & & & &  &  &   & $\rm{erg~s^{-1}}$] &  &  &  &  \\ \hline
	 \texttt{RDF-1.9} &0.58 & 4.1 & 17.8 & 1.53364 & 41.9 & 1.11 & 1.53443 & 1.082  & 0.15 & 1.524 & 5.93 & 6.66 & 12.7  & 148 & 837 & 501 & 28.8 & 59.6 & 104.7 \\ \hline 
	 \texttt{RDF-1.2} &1.56 & 5.3 & 23.4 & 1.53459 & 22.6 & 2.91 & 1.53397 &  0.92 & 0.23 & 1.531 & 5.42  & 6.34 & 10.7 & 194 & 412 & 187 & 44 & 50.5 & 91.8 \\ \hline 
	 \texttt{RDF-1.5} &3.06 & 5.94 & 26.6 & 1.53468 & 14.4 & 4.27 & 1.53346 & 0.875 & 0.19 & 1.533 & 4.92 & 5.61 & 9.64 & 49.8 & 65.2 & 21.2 & 31.8 & 34.1 & 46.9 \\ \hline 
         \texttt{RDF-1.1} &3.32 & 6.03 & 27.5 & 1.53469 & 14.2 & 6.44 & 1.53397 & 0.288 & 0.61 & 1.533 & 4.14 & 5.29 & 9.82  & 6.59 & 16 & 8.03 & 20.8 & 24.3 & 24.2 \\  \hline 
\end{tabular}
\caption{\label{table_model} Various quantities extracted at the moment of first
		reaching the mixed nuclear-quark phase (columns in the left part) and 
		$300~\rm{ms}$ after the first entry into 
		the deconfined quark phase (columns in the right part) 
		in our AIC simulations, using different RDF EOS.
		$t_{\rm mixed}$, $\rho_{\rm{mixed,c}}$, $T_{\rm{mixed,c}}$ 
		, $M^{\rm b}_{\rm{mixed}}$ and $\mathcal{L}_{\rm{mixed,tot}}$ denote the 
		the time after the first core bounce at $t=t_{\rm b}$, 
		central density, central temperature, the enclosed baryon mass of the PNS
		and the total neutrino luminosity for the system first reaching the mixed 
		nuclear-quark phase.
		$t_{Q}$ and $M^{\rm b}_{t\rm{_Q}}$ are the time with respect to $t_{\rm b}$
		and the enclosed baryon mass of the PHS
		when the PHS first reaching the deconfined quark phase.
		Baryon mass (radius) of the quark core, the mixed-phase mantle
		(MM) and the new-born protohybrid star (PHS), 
		are expressed as $M^{\rm b}_{\rm Q}$
		($r_{\rm{Q}}$), $M^{\rm b}_{\rm{MM}}$ ($r_{\rm{MM}}$) and
		$M^{\rm b}_{\rm PHS}$
		($r_{\rm{PHS}}$).
		The values of neutrino luminosity and averaged energy 
		are evaluated at $r=500~\rm{km}$
		The values of the second neutrino burst are represented as $\mathcal{L}_{\nu_l}$ 
		and $ \langle{\epsilon_{\nu_l}}\rangle$, respectively for electron
		neutrino $\nu_e$, electron antineutrino $\bar{\nu}_e$ and
		a collective species describing heavy
		lepton neutrinos and their antineutrinos as $\nu_x$.
		Note that the PNS or PHS are defined as the regions of 
		$\rho > 10^{11}~\rm{g~cm^{-3}}$.}
\end{table*}

Our simulations employ the general relativistic 
neutrino-radiation hydrodynamics code \texttt{Gmunu}
~\cite{Cheong:2020dxe,Cheong:2020kpv,Cheong:2022vev,Cheong:2023fgh,Ng:2023syk}, 
which solves Einstein’s field equations under 
the conformal flatness condition
~\cite{Cordero-Carrion:2008grk,Cheong:2020dxe,Ng:2023yyg}.

The initial progenitor is constructed with 
a fixed temperature of $0.01~\rm{MeV}$ 
corresponding to the minimum of the EOS table 
and an electron fraction of $Y_e = 0.5$ in \texttt{Gmunu}. 
To model the temperature profile of a 
hot white dwarf prior to collapse,
we superimpose a temperature profile using the relation
$T = T_c (\rho/\rho_c)^{0.35}$ 
with $T_c = 5 \times10^{9}~\rm{K}$ 
corresponding to $\sim 0.43~\rm{MeV}$ following the approach of Refs.~\cite{Dessart:2006gd,Dessart:2007kh,Cheong:2024hrd}.
The addition of this temperature profile together with the specific metric initialization reduces the gravitational mass to $M_g = 1.52~M_{\odot}$. 
This inconsistency in $M_g$ arises 
because \texttt{Gmunu} adopts the metric initialization method of Ref.~\cite{Cordero-Carrion:2008grk},
 which fixes the initial primitive variables after the thermal profile is imposed. 
 However the metric initialization should instead be performed by fixing the conformal factor and the conserved variables as done for the hot hypermassive neutron star postmerger simulation 
in Ref.~\cite{Ng:2023yyg}, which will be adopted in the future work.

We adopt the energy-integrated two-moment neutrino transport
scheme~\cite{2022MNRAS.512.1499R,Cheong:2024buu}, and the
implicit-explicit Runge-Kutta method
IMEX-SSP2(2,2,2)~\cite{pareschi2005implicit} for time integration.
Details of the implementation and comparisons with 
similar numerical schemes~\cite{Foucart:2016rxm,Andresen:2024mtt} 
are provided in the appendix of~\cite{Cheong:2024buu}.
For the neutrino microphysics, we employ the neutrino
library \texttt{Weakhub}~\cite{Ng:2023syk} and consider the conventional
set of weak interactions defined in~\cite{Ng:2023syk} but excluding
inelastic neutrino-electron scattering.  
Neutrino opacities are tabulated after averaging in energy space~\cite{Cheong:2024buu}, where the neutrino energy is logarithmically
discretized into 20 bins within the range $[0.1,400]~\rm{MeV}$. The
emissivities are recalculated on the fly using Kirchhoff’s law.

The collapse of WD is triggered by the electron capture, in which the 
electron capture rates on heavy nuclei are uncertain~\cite{Nagakura:2018qpg}
during the collapse of WD.
Our initial models are constructed with a central density of $\rho_c = 10^{9}~\rm{g\cdot cm^{-3}}$.
Also, energy-integrated neutrino transport schemes neglect energy-coupled interactions, 
such as neutrino-electron inelastic scattering, which are crucial during the collapse phase.
To bridge this modeling gap while maintaining computational efficiency, 
we apply an effective deleptonization scheme~\cite{Liebendoerfer:2005gm} to mimic the proton 
fraction $Y_p$ profile as a function of rest-mass density $\rho$ during the collapsing phase,
which is widely used in AIC simulations
~\cite{Abdikamalov:2009aq,LongoMicchi:2023khv,Cheong:2024hrd}.
We then switch to the two-moment neutrino transport when the core
bounces (defined as when the matter specific entropy $s \geq3~\rm{k_b/baryon}$ in the core region), 
and enable the coupling between neutrinos and
the fluid before core bounce.

The computational domain extends from $0$ to $3 \times 10^{3}~\rm{km}$ 
at the outer boundary, with a radial resolution of $N_{r} = 128$. 
We allow a $10$ AMR level and the finest grid spacing at the center of the star is $45.8~\rm{m}$. 
To demonstrate the robustness of our results, 
we present a comparison with a recent CCSN PT simulation~\cite{Zha:2020gjw} 
and include a resolution dependence test of our AIC setup in Appendix~\ref{sec:codetest} 
and~\ref{sec:resolution_dependence} respectively.

\begin{figure*}[]
\includegraphics[width=\textwidth, height=0.26\textwidth]{ 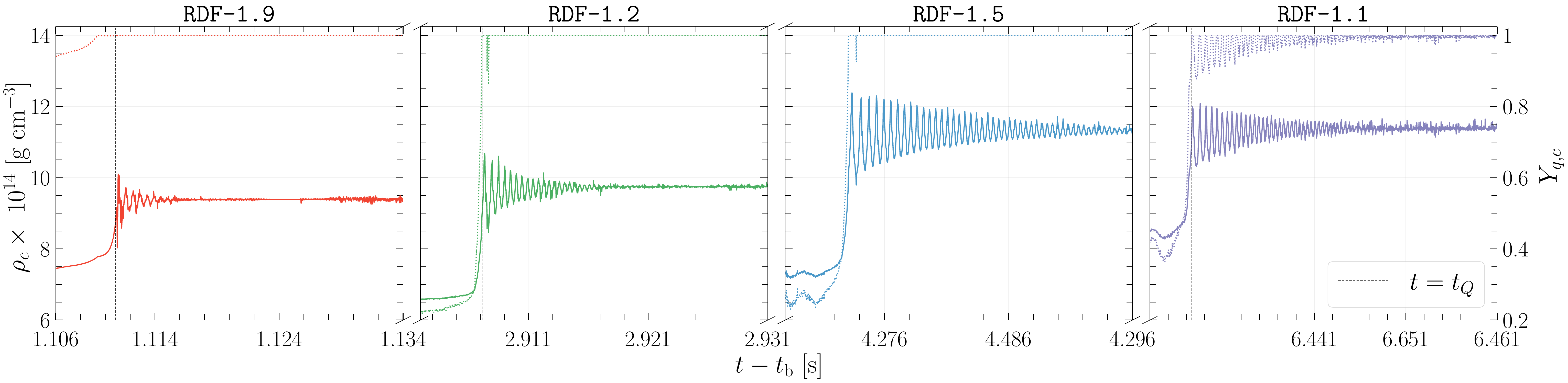}
\caption{\label{central_density_quark_fraction} 
		Time evolution of the central density $\rho_{c}$ (solid lines) and
		central quark fraction $Y_{q, c}$ (dotted lines) after
 		the first core bounce for all models.
		The second bounce for $t -t_{\rm{Q}}<23~\rm{ms}$ are shown.
		Dashed lines correspond to the time at $t=t_{\rm{Q}}$.
}
\end{figure*}

\section{Results}\label{sec:results}
The unstable initial WD undergoes a collapse and
an electron capture following the effective deleptonization
scheme, which reduces the electron degeneracy pressure 
in the WD.
Electron capture becomes significant at $\sim 5\times10^{9}~\rm{g~cm}^{-3}$
and triggers a gravitational collapse with a rapid increase in central density $\rho_c$.
When $\rho_c$ exceed nuclear saturation density ($\rho_c \gtrsim 10^{14}~\rm{g~cm}^{-3}$), hadronic matter stiffens due to the 
strong nuclear force and 
halt the collapse, lead to a core bounce 
and an outward shock at $t = t_{\rm{b}}$. 
This produces a $\nu_e$ burst with $\mathcal{L}_{\nu_e} \sim 10^{53}~\rm{ergs}^{-1}$ 
over $\sim 10~\rm{ms}$, which qualitatively consistent with
the result in 
Refs.~\cite{LongoMicchi:2023khv,Cheong:2024hrd,Batziou:2024ory} 
and detailed in Appendix~\ref{sec:comparison}.

The PNS gradually loses energy through 
weaker neutrino emissions across all species, 
maintaining $\mathcal{L}_{\nu_l} \sim 10^{51}~\rm{ergs^{-1}}$
for several seconds. 
This sustained emission depletes thermal energy and pressure, 
leading to gradual contraction via Kelvin-Helmholtz cooling, 
which slowly increases $\rho_c$ and $T_c$.

The contraction accelerates upon reaching the onset of a first-order
PT, forming mixed hadron-quark phase matter, i.e. 
quark fraction $Y_q > 0$, which softens the EOS. 
The left columns of Table~\ref{table_model} show the
time $t_{\rm{mixed}}$, central density $\rho_{\rm{mixed,c}}$, central
temperature $T_{\rm{mixed,c}}$, the enclosed baryonic mass of the PNS
$M^{\rm b}_{\rm{mixed}}$ and the total neutrino luminosity $\mathcal{L}_{\rm{mixed,tot}}$
when the system first reaches the mixed phase with $Y_q > 0$.

\begin{figure}[]
\includegraphics[width=\columnwidth, height=0.45\textwidth]{ 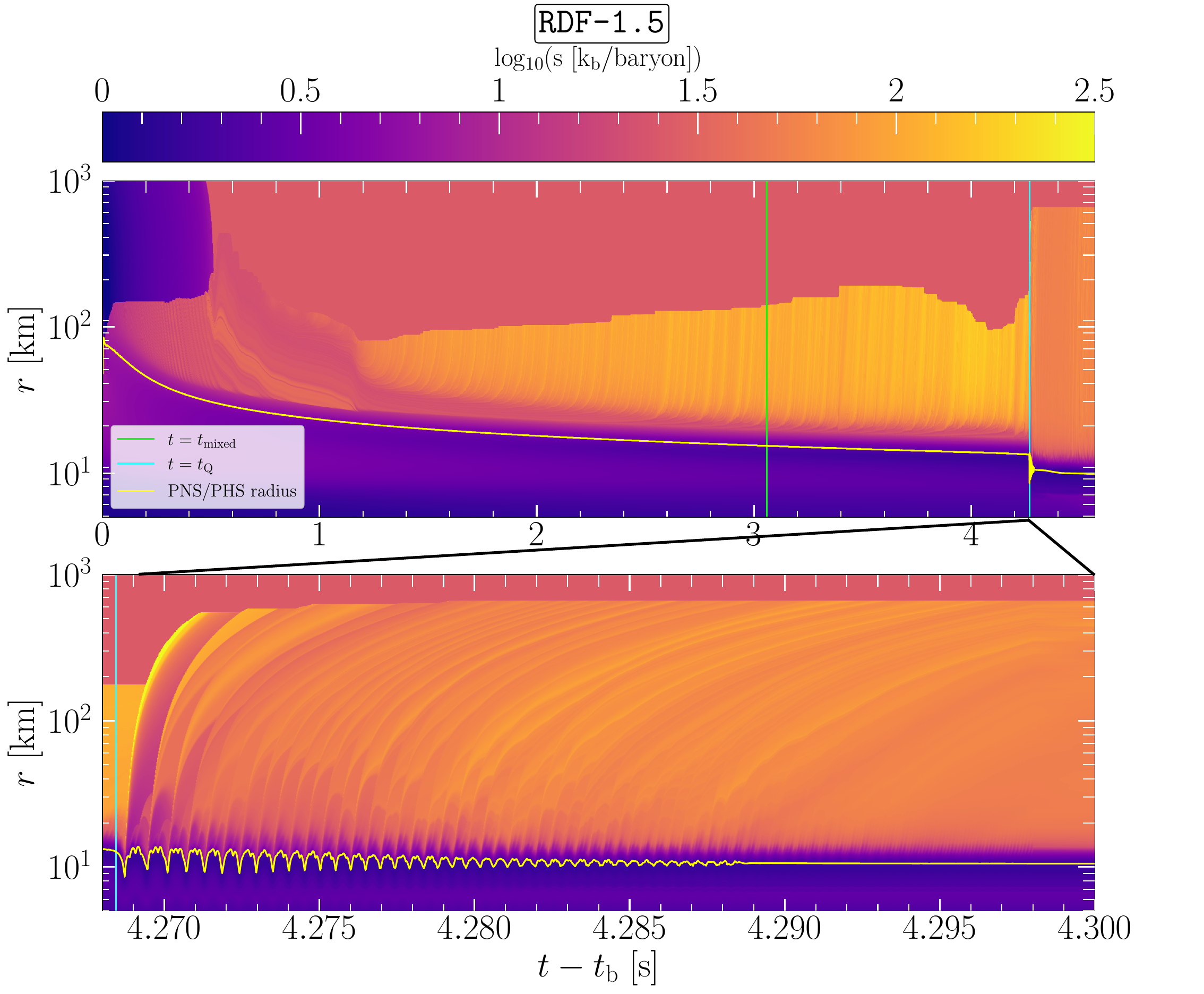}
\caption{\label{entropy_profile} Specific entropy $s$ profile of the model with
		\texttt{RDF-1.5} EOS. 
		\textit{Top panel:} from $t=t_{\rm b}$ to $300~\rm{ms}$ after $t_{\rm Q}$.
		\textit{Bottom panel}: from $t = t_{\rm{Q}}$ to
		$32~\rm{ms}$ after $t_{\rm{Q}}$
		The vertical lines indicate time moments of $t = t_{\rm mixed}$ (lime) and 
		$t = t_{\rm{Q}}$ (cyan). 
		The yellow solid represents the radius
		of PNS or PHS.
		}
\end{figure}

For all models,
$M^{\rm b}_{\rm{mixed}}$ differs by under $0.07\%$
across models, indicating that the accretion phase is effectively ended and a steady
PNS has formed at this point (see discussion in Fig.~\ref{entropy_profile}).
The small variation in $M^{\rm b}_{\rm{mixed}}$ indicates a degree of universality in the 
PT onset due to the absence of a massive envelope.

The negligible variation in onset mass distinguishes PT in AIC from that in CCSNe,
where the presence of a massive envelope leads to continuous mass accretion
and possibly different onset mass condition.
These considerations motivate an investigation of the potential difference in PT dynamics;
we perform a set of CCSN simulations using the lowest-mass progenitor model,
$s12$ from Ref.~\cite{Woosley:2002zz}, with the same set of hybrid EOSs.
We focus the discussion of PT in AIC in the main text, and thus
the numerical setup and detailed comparison
are provided in Appendix~\ref{sec:ccsn_pt}.

Due to high $T_{c}$ and low $Y_p$ at the
PNS core, the modified onset density $\rho_{\rm{mixed,c}}$ is lower than
$\rho_{\rm{onset}}$ in Table~\ref{model}, which is defined under $T =
0~\rm{MeV}$ and beta equilibrium~\citep{Largani:2023oyk}. For example,
$\rho_{\rm{mixed}}$ is $30\%$ ($10\%$) lower than $\rho_{\rm{onset}}$ in
model $\texttt{RDF-1.1}$ ($\texttt{RDF-1.9}$), corresponding to
$T_{\rm{mixed,c}}$ of $27.5~\rm{MeV}$ ($17.8~\rm{MeV}$) at $t = t_{\rm{mixed}}$.

As contraction accelerates at the onset of a first-order PT, 
the emergence of a mixed phase 
with nonzero quark fraction ($Y_q > 0$) further softens the EOS.
In our case, this PT-induced contraction proceeds concurrently 
with the Kelvin-Helmholtz cooling, where models with higher onset densities receive less assistance from cooling, 
as indicated by the substantial decrease in  $\mathcal{L}_{\rm{mixed,tot}}$.

Figure~\ref{central_density_quark_fraction}
illustrates the evolution of central density $\rho_c$ (solid) 
and quark fraction $Y_{q,c}$ (dotted) through three phases:
the mixed phase ($Y_{q,c} < 1$), 
the onset of deconfined quark matter ($Y_{q,c} = 1$) at $t_{\rm{Q}}$ 
(first entry into the deconfined quark phase relative to $t_{\rm b}$, 
see Table~\ref{table_model}), 
and the quasistable PHS phase.

In the first phase, increasing $Y_q$ accelerates collapse as the EOS softens. 
A greater fraction of mixed-phase matter leads to faster collapse, 
meaning higher $\rho_{\rm mixed, c}$ results in a longer time to reach $t_{\rm Q}$. 
As the core approaches the stiffer deconfined quark phase 
at $t \approx t_{\rm{Q}}$, a second dynamical collapse happens 
within $0.5~\rm{ms}$. Notably, the PHS collapse with enclosed
baryon mass at $t_{\rm{Q}}$ with $M^{\rm{b}}_{t_{\rm{Q}}}\sim1.53~M_{\odot}$ 
across all models (See Table~\ref{table_model}),
highlighting that PT-induced contraction in AIC
is not subject to mass accretion, 
enhancing its sensitivity to properties of the hybrid EOS,
in contrast to the behavior of mixed phase observed in CCSNe
(see Appendix~\ref{sec:ccsn_pt}).

For all RDF EOSs, the system’s small baryonic mass ($\sim 1.54~M_{\odot}$) 
and the stiffness of deconfined quark matter 
disfavor black hole formation. 
Instead, the stiff quark core halts the infall of mixed-phase material,
generating a second outward-propagating shock. 
This shock, at high density and temperature, 
converts infalling hadronic matter into the mixed phase 
and accelerates as 
it reaches the lower-density hadronic envelope near the PNS surface.

The second shock energizes the hadronic envelope, 
enhances $e^+$ captures, and produces 
a second neutrino burst 
(see detailed discussion on the second burst later).
This second burst from PT is consistent with
Appendixes~\ref{sec:codetest} and~\ref{sec:ccsn_pt} 
and with the previous studies of PT in CCSNe
~\citep{Sagert:2008ka,Zha:2020gjw,Kuroda:2021eiv,Fischer:2021tvv,Zha:2021fbi,Jakobus:2022ucs,Largani:2023oyk}.

A quasistable PHS forms in each case, 
comprising a quark core, a mixed-phase mantle, 
and a residual hadronic envelope with baryonic masses 
$M^{\rm b}_{\rm{Q}}$, $M^{\rm b}_{\rm{mixed}}$, and $M^{\rm b}_{\rm{PHS}}$, respectively, 
as listed in Table~\ref{table_model} (right columns). 
The overcompression of the quark core before the second bounce, 
combined with dynamical changes in quark and mixed-phase matter, 
induces strong perturbations in the newborn PHS, 
leading to rapid oscillations that damp out within $\sim 23~\rm{ms}$ 
in Fig.~\ref{central_density_quark_fraction}.

To examine the dynamics around the second bounce, 
we focus on the \texttt{RDF-1.5} EOS case. 
The upper panel of Fig.~\ref{entropy_profile} shows the specific entropy $s$
evolution from $t - t_{\rm b}$ to $300~\rm{ms}$ after $t_{\rm Q}$.
The first core bounce generates a shock that expands 
to $\sim 90~\rm{km}$ within $5~\rm{ms}$. 
By $t - t_{\rm{b}} \sim 10$–$100~\rm{ms}$, 
the shock loses energy, stalls at $\sim 100~\rm{km}$, 
and forms an accreting envelope with $s \sim 7$–$15~k_{b}/\rm{baryon}$. 

After core bounce, the PNS mass increases 
from $0.92M_{\odot}$ at $t = t_{\rm{b}}$ to $1.534~M_{\odot}$
by $t - t_{\rm{b}} \sim 600~\rm{ms}$ 
due to the accretion of infalling shocked matter 
and material from the outer envelope. 

As accretion diminishes, 
the system reaches a steady state with 
a baryonic mass of $1.535~M_{\odot}$ at $t - t_{\rm{b}} \approx 1.2~\rm{s}$. 
Despite entering the mixed phase, 
the PNS continues to contract (see yellow line in Fig.~\ref{entropy_profile}), 
reinforcing that neutrino cooling and mixed-phase matter soften the EOS. 
Additionally, the high-entropy matter 
($s \sim 25$–$50~k_b/\rm{baryon}$) observed 
at $30$–$100~\rm{km}$ after $t - t_{\rm b} \sim 1~\rm{s}$ 
arises from interactions between the accretion envelope 
and fallback of ejected matter from the first shock.

\begin{figure}[]
\includegraphics[width=0.9\columnwidth, height=0.6\textwidth]{ 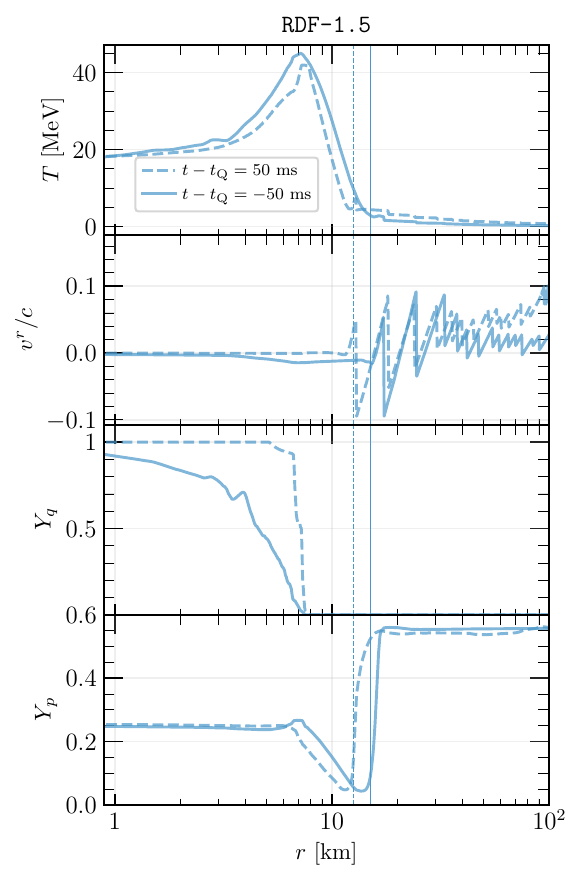}
\caption{\label{profile} Radial snapshots of temperature $T$, radial
		velocity $v^r/c$, proton fraction $Y_p$, and quark fraction $Y_q$ of
		\texttt{RDF-1.5} model at $50~\rm{ms}$ before (solid) and $50~\rm{ms}$
		after (dashed) the time moment of $t_{\rm{Q}}$. Vertical lines correspond
		to the density at $\rho = 10^{11}~\rm{g~cm^{-3}}$.
}
\end{figure}

Immediately after $t = t_{\rm{Q}}$, 
the newly formed PHS core undergoes a second bounce, 
generating a shock that propagates to $\sim 500~\rm{km}$ 
in $2.4~\rm{ms}$ at $\sim 0.5~c$ 
(see Fig.~\ref{entropy_profile}, lower panel). 
As it pushes against dense matter beyond the accretion envelope, 
heating it to $\sim 50~k_{b}/\rm{baryon}$, 
it loses energy and stalls at $\sim 500~\rm{km}$. 
Before stalling, PHS oscillations produce multiple weaker shocks at 
$\sim 0.25~c$ that continue pushing matter outward, 
expanding the envelope to $\sim 600~\rm{km}$ within $4~\rm{ms}$.
Since the accreting envelope in AIC systems is only 
$\sim 0.1\%$ of the total mass, 
the second shock loses less energy and propagates 
beyond the first shock stall ($\sim 100~\rm{km}$), 
reaching at least $500~\rm{km}$ (\texttt{RDF-1.1}) 
and up to $1500~\rm{km}$ (\texttt{RDF-1.9}). 
Following the second bounce, the PHS rapidly contracts, 
with the radius shrinking by $\sim 20\%$ at $50~\rm{ms}$ 
after $t_{\rm Q}$ and continuing to contract gradually. 
This shock, propagating through the low-proton-fraction region
(see the bottom panel of Fig.~\ref{profile}), 
generates a second neutrino burst.

Figure~\ref{profile} shows the radial hydrodynamical profiles at $50~\rm{ms}$ before and 
after $t_{\rm Q}$ for the \texttt{RDF-1.5} model.
The vertical lines indicate the radii of PNS (dashed) and 
PHS (solid), respectively.
Note that by $t-t_{\rm Q} = 50~\rm{ms}$, the PHS attains a quasisteady
state and is much more compact (see also Fig.~\ref{central_density_quark_fraction}).

The PHS exhibits a cooler core compared to 
the PNS at $t-t_{\rm Q} = -50~\rm{ms}$, 
as the thermal energy is converted to latent heat during the formation of
deconfined quark matter. In contrast, 
the shocked envelope outside the PHS (for $r > 20~\rm{km}$) is slightly hotter.

Unlike the PNS at $t-t_{\rm Q} = -50~\rm{ms}$, which features oscillating
matter in its envelope, 
the quasisteady PHS continues to eject materials with $v^r > 0$ as long as
$300~\rm{ms}$.
In these outer layers, the velocity can exceed $0.1~c$ at $r>100~\rm{km}$.
This outflow is driven by the heavily perturbed PHS, where neutrino interactions
become active again in the shocked regions,
These neutrinos deposit energy via
reabsorption in the outer layers, leading to the ejection of about $0.1\%$ of the PHS
mass by $t-t_{\rm Q} = 300~\rm{ms}$.

The fractions of deconfined quark matter, mixed-phase matter and hadronic
matter in the PHS are highly nonlinear and depend on the EOS properties that 
determine the collapse dynamics and the strength of the second bounce 
(the properties of PHS can be found in the right columns of Table~\ref{table_model}).
In all models, the mixed-phase mantle is 
sandwiched between the 
hadronic envelope and deconfined-quark core with a small size of at most $2~\rm{km}$ due to its softness among all cases. Even 
a special case of \texttt{RDF-1.1}, where 
the mass of the mixed-phase mantle exceeds that of the quark core, follows this pattern.

Particular attention should be paid to this special case of $\texttt{RDF-1.1}$. 
Its PNS collapse driven by the PT is weaker and slower due to its 
high $\rho_{\rm mixed, c}$ and low $\mathcal{L}_{\rm{mixed,tot}}$, leading to a longer timescale of cooling and 
thus slower velocity of collapsing matter before $t_{\rm Q}$.
The slower contraction is reflected in a small 
but notable decrease in both $Y_{q,c}$ and $\rho_c$ at $t - t_{\rm{Q}} \approx
-3.8~\rm{ms}$ in Fig.~\ref{central_density_quark_fraction}.
These result in a much lighter quark core surrounded by
a significantly more mixed-phase mantle.
The result of \texttt{RDF-1.1} case is nearly a critical case 
where the progenitor mass with $1.51~M_{\odot}$ (gravitational mass)
is close to the onset mass of the EOS with $1.57~M_{\odot}$ shown in
Table~\ref{model}, and it is also the reason for oscillating in $Y_{q,c}$ 
after the second bounce.

Two additional robust features appear in all models. 
First, a cavity with a low proton fraction ($Y_p \sim 0.1$) forms at about $t -
t_{\mathrm{b}} \sim 10~\mathrm{ms}$, coinciding with the $\nu_e$ burst
shortly after the core bounce, and this cavity persists until the PHS forms 
(see the $Y_p$ profile in Fig.~\ref{profile}). The second
shock, originating from the deconfined quark core beneath the
cavity, passes through this low $Y_p$  region, emitting primarily $\bar{\nu}_e$
via positron ($e^+$) capture. Consequently, the second burst is
predominantly composed of $\bar{\nu}_e$ for all EOSs, similar to the behavior reported in
CCSN simulations with a QCD PT~\citep{Fischer:2017lag,Largani:2023oyk}. In contrast, the first
neutrino burst, which results from the first bounce, is dominated by
$\nu_e$ emission due to the proton-rich environment of the newly
formed PNS.

Finally, the high temperature of the PHS leads to a greater integrated
emission of $\nu_x$ compared to $\nu_e$ during the second burst.
Furthermore, the second burst
generally produces neutrinos with higher average energy across all species
compared to the first burst.

\begin{figure}[t!]
\includegraphics[width=1\columnwidth, height=0.46\textwidth]{ 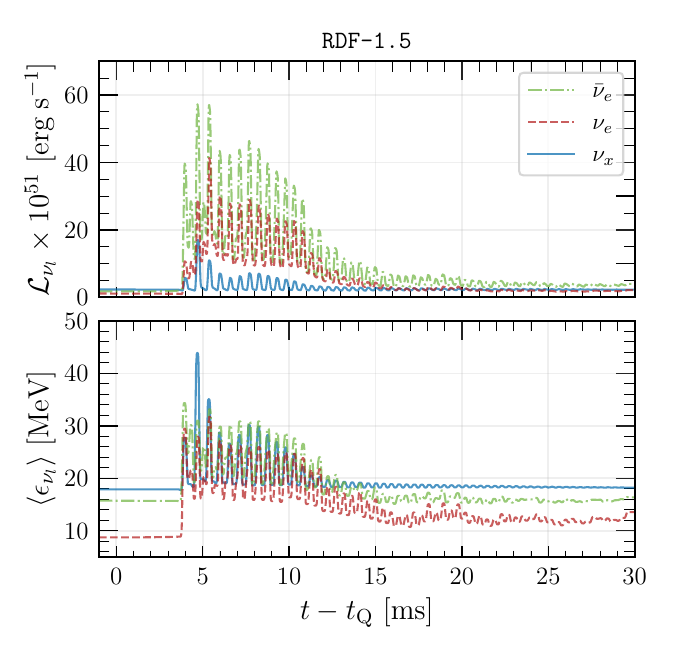}
\caption{\label{lum_eps_rdf15} Time evolution of neutrino luminosity 
		$\mathcal{L}_{\nu_l}$ and average energy $ \langle
		{\epsilon_{\nu_l}} \rangle $ of the second neutrino burst formed
		in the case with \texttt{RDF-1.5} EOS for $\nu_e$ (red dashed), $\bar{\nu}_e$
		(green dash-dotted) and $\nu_x$ (blue solid), respectively.}
\end{figure}

Figure~\ref{lum_eps_rdf15} presents the neutrino luminosity, 
$\mathcal{L}_{\nu_l}$, and average energy, 
$\langle {\epsilon_{\nu_l}} \rangle$, 
for all species during the second millisecond neutrino burst 
in the \texttt{RDF-1.5} model. 
The burst, with luminosities on the order of $10^{52}~\rm{erg~s^{-1}}$, 
is triggered by the second shock as it propagates 
through the PHS neutrinospheres and is observed 
around $4~\rm{ms}$ after $t_{\mathrm{Q}}$.

At the onset of the burst, neutrino luminosities and 
average energies for all species reach a peak, 
as summarized in Table~\ref{table_model} (right columns). 
The burst is dominated by $\bar{\nu}_e$ due to positron ($e^+$) capture 
in the low-proton-fraction cavity during shock propagation. 
Also, the stronger second shock produces neutrinos 
with higher average energy across all species compared to the first shock. 
A comparison of the first and second bursts is detailed in Appendix~\ref{sec:comparison}.

These quantities then oscillate for $\sim 30~\mathrm{ms}$, 
corresponding to the oscillation and damping 
of the PHS following the second bounce 
(Fig.~\ref{central_density_quark_fraction}). 
This oscillatory behavior in the second burst, 
lasting $5$--$30\mathrm{ms}$, 
is consistently observed across all models,
and is also indicated 
in the CCSNe simulations in previous studies~\citep{Zha:2021fbi,Lin:2022lck}.

The second neutrino burst not only confirms 
a strong first-order PT and PHS formation 
but also provides a key observational probe. 
The time interval between the two bursts, 
approximately $t_{\rm Q}$, can potentially constrain the onset density 
and PT properties of the EOS. 
As shown in Table~\ref{table_model}, there are small variations 
in $\rho_{\rm{mixed,c}}$ across different EOSs.
 
\begin{figure}[t!]
\includegraphics[width=1\columnwidth, height=0.48\textwidth]{ 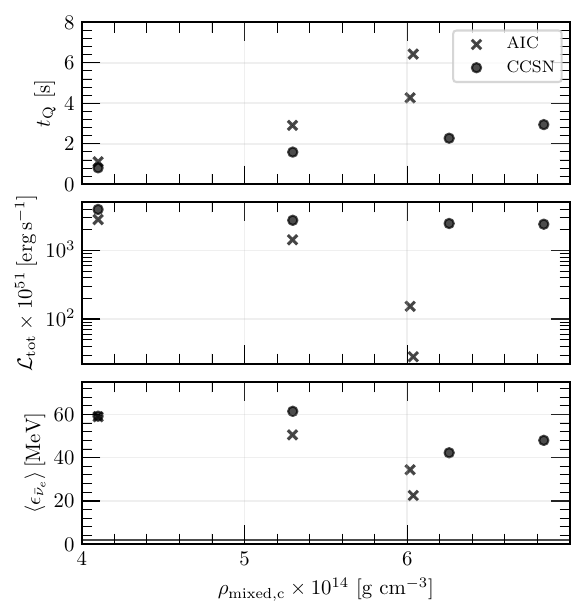}
\caption{\label{nu_observables} Empirical relations of $t_{\rm{Q}}$,
		the total neutrino luminosity $\mathcal{L}_{{\rm{tot}}}$ and $ \langle
		{\epsilon_{\bar\nu_e}} \rangle$ as functions of
		$\rho_{\rm{mixed,c}}$. Relations from PT in CCSN and AIC are marked with circles and 
		crosses respectively.
		The horizontal line in the bottom panel
		indicates $1.8~\rm{MeV}$ inverse beta decay threshold in water Cherenkov detectors
		~\citep{Ricciardi:2022pru,Strumia:2003zx,Super-Kamiokande:2007zsl,Hyper-Kamiokande:2018ofw}.}
\end{figure}

Figure~\ref{nu_observables} presents empirical relations 
between $t_{\rm{Q}}$ (top panel),
total neutrino luminosity $\mathcal{L}_{\rm{tot}}$ (middle panel), 
and $\langle{\epsilon_{\bar\nu_e}} \rangle$ (bottom panel) 
as functions of $\rho_{\rm{mixed,c}}$.
Results from AIC and CCSN simulations in Appendix~\ref{sec:ccsn_pt}  
are shown for comparison to highlight systematic differences
due to the presence of a massive envelope.
The top panel shows that $t_{\rm{Q}}$ increases significantly with 
$\rho_{\rm{mixed,c}}$, unlike QCD PT simulations in CCSNe, 

The discrepancy arises because, in AIC systems, 
the accreting and infalling mass is negligible compared to CCSNe,
model with higher onsity density undergo PT-induced contraction
with less assistance from neutrino cooling, as disucssed in details above,
thus leading to a much longer $t_{\rm{Q}}$.

For these reasons, different relationships between neutrino 
observables and $\rho_{\rm{mixed,c}}$ emerge (see the middle and bottom panels of Fig.~\ref{nu_observables}). 

First, both $t_{\rm Q}$ and $\langle \epsilon_{\bar\nu_e} \rangle$ 
exhibit greater sensitivity to variations in $\rho_{\rm{mixed,c}}$ in AIC systems, 
resulting in a wider dynamic range of response.
In particular, $\langle \epsilon_{\bar\nu_e} \rangle$ in AIC shows 
a clearer and more rapid decrease compared to CCSN.
Despite the substantial drop in $\langle \epsilon_{\bar\nu_e} \rangle$ for AIC, 
it remains above the $1.8~\rm{MeV}$ inverse beta decay threshold in water Cherenkov detectors
~\citep{Ricciardi:2022pru,Strumia:2003zx,Super-Kamiokande:2007zsl,Hyper-Kamiokande:2018ofw},
showing that neutrino bursts remain potentially detectable for all AIC models.

Moreover, the relation between $\mathcal{L}_{\rm{tot}}$ and $\rho_{\rm{mixed,c}}$
unambiguously shows a transition
from linear dependence in CCSN to exponential dependence in the AIC system.
Although this dependence implies a rapid decline
in detectability with increasing $\rho_{\rm{mixed,c}}$,
it establishes a distinct and
potentially powerful relation for constraining the onset density of the first-order PT.
Although a $\sim 10^{51}~\rm{ergs^{-1}}$
neutrino burst from an AIC event occurring within our Galaxy
may remain detectable by supernova-dedicated neutrino observatories,
such as Super-Kamiokande and Hyper-Kamiokande~\cite{Super-Kamiokande:2007zsl,Hyper-Kamiokande:2018ofw},
a comprehensive assessment of detectability across current and
next-generation detectors will be conducted in future work~\cite{Brdar:2018zds}.

These empirical relations directly link potentially detectable neutrino signals to 
$\rho_{\rm{mixed,c}}$ in AIC systems, 
offering distinct advantages for constraining the onset density of the first-order PT. 
Notably, these relations demonstrate that the absence of a
 massive envelope in AIC substantially enhances its sensitivity to 
 the properties of the hybrid EOS compared to CCSN.

Due to the advantageous second-long time separation between the two
neutrino bursts in AIC systems, along with their less massive accretion
envelopes and a narrower total mass range of $1.44$–$2.5~M_\odot$~\cite{Liu:2020qny,Wang:2020pzc} 
, in contrast to the broader $8$–$100~M_\odot$ range in CCSNe~\citep{Woosley:2002zz}, exploring first-order
PT in AIC shows a stronger dependence on the onset properties of hybrid EOS,
particularly in neutrino observables, than PT in CCSNe. 
The strong EOS sensitivity in AIC could enable exceptionally tight constraints on PT and 
QCD matter properties from the detection of a single Galactic AIC event.

\section{Discussions and Conclusions}\label{sec:discussion}
We present the first seconds-long general relativistic 
neutrino-radiation simulations of AIC 
with realistic hadron-quark hybrid EOSs, 
evolving the collapse of WDs. 
A first-order PT triggers a second dynamical collapse and
the formation of a quasistable PHS 
with a deconfined quark core and a distinct second neutrino burst.

Following core bounce, 
the PNS undergoes slow contraction mainly due to neutrino cooling, 
with the appearance of mixed nuclear-quark matter further softening the core 
and slightly accelerating collapse.
We find that the thermally suppressed onset density of mixed phase allows
a low-mass PNS to enter the mixed phase during
a long-term and second-scale evolution for hybrid EOS
characterized by a high onset density for the mixed phase.
We further observe that the absence of a massive envelope allows all AIC models to 
exhibit a tightly constrained onset mass 
in both mixed nuclear-quark phase and deconfined quark phase, 
showing minimal hybrid EOS dependence.
The absence of sustained accretion indicates that, compared to PT in CCSNe, AIC receives substantially 
less support from mass accretion during the mixed phase.
As a consequence, the time spent in the mixed phase 
becomes more strongly dependent on 
the properties of the hadron–quark hybrid EOS.
Therefore, the low-mass nature of AIC progenitors 
enhances its sensitivity to the properties of the hybrid EOS compared to CCSNe.

The transition to a stiffer deconfined quark core halts the collapse, 
generating a second shock breakout and a $\bar\nu_e$-dominated burst 
with higher average neutrino energy across all species compared to the first burst. 
In all cases, a stable PHS forms, consisting of 
a deconfined quark core, a mixed-phase mantle, and an outer hadronic envelope.
Although the formation of hybrid quark stars near the Chandrasekhar mass is possible for hybrid EOSs with low onset densities~\citep{Bastian:2020unt}, 
such scenarios may conflict with observations of pulsars with comparable masses~\citep{Ozel:2016oaf,Miller:2019cac}.
Our results indicate that if the cooled down hybrid star structure persists, certain PT models could be ruled out.

AIC provides a promising avenue for constraining PT thresholds 
and hybrid EOS properties. 
Unlike CCSNe, its narrower mass range 
and a less massive accreting envelope, 
makes them ideal for studying QCD PT 
and PHS properties through observational signatures. 
We establish an empirical relation between PT onset density and 
neutrino observables, particularly the time delay between 
the two bursts, total neutrino luminosity and
the average energy as a function of the PT onset density.
This allows a direct comparison 
with the corresponding relation in CCSN and reveals a new, distinct relation in AIC, 
primarily driven by the low-mass nature of its progenitor.

Although the rate of AIC ($\lesssim 10^{-3}~\rm{yr}^{-1}$) is 
lower than that of CCSN ($\sim 10^{-2}~\rm{yr}^{-1}$), 
our results suggest that a single Galactic AIC neutrino detection 
could impose more precise constraints on PT thresholds, 
hybrid EOS properties, and the existence of HSs. 
These energetic outbursts are also a promising new source of GW emission, 
gamma-ray bursts, 
and $r$-process nucleosynthesis~\citep{Fischer:2020xjl}.

It is observed that PT in CCSN, and the associated neutrino observables, are sensitive to the different physics and numerical setup (see Appendix~\ref{sec:codetest}).
Therefore, future work of PT in AIC should incorporate multidimensional effects, including accretion instabilities~\citep{Batziou:2024ory}.
 Rotation also plays a crucial role in AIC~\citep{LongoMicchi:2023khv,Cheong:2024hrd}, 
where it can potentially regulate PNS cooling and contraction, modify the second collapse dynamics, and thereby shape the detailed evolution of the PT.
In multidimensional simulations convection 
 and anisotropic temperature profiles can also 
 play an essential role in the formation and 
 transport of mixed hadron quark matter as observed 
 in multidimensional simulations of neutron star mergers~\citep{Most2019c,Prakash:2021wpz}.
More advanced
microphysics will affect the Kelvin-Helmholtz cooling strength and, hence,
alter the neutrino observables such as $t_{\mathrm{Q}}$~\citep{Guo:2020tgx, Cheong:2024cnb, Ng:2024zve}.
The strength and geometry of magnetic fields are known to affect AIC dynamics~\citep{Sykes:2024mel,Cheong:2024hrd}, and their potential influence on the second collapse will be crucial.
A comprehensive study of the detectability of the neutrino signal 
in the current and next generation neutrino detector will also be
left for future work~\cite{Brdar:2018zds}.
The stiffness of the hadronic phase also alters the resistance to PNS contraction, creating potential degeneracies with the onset properties of hybrid EOSs. 
A survey of hadronic EOSs with varying stiffness to quantify this degeneracy can assess whether AIC neutrino signals can disentangle EOS stiffness from PT effects.

\begin{acknowledgements}
The authors thank Liang Dai, Christian Ecker, Jose Maria Ezquiaga, 
Tobias Fischer, JJ Hermes, Jin-Liang Jiang, 
Pablo Martínez-Miravé and Luciano Rezzolla for the useful discussion.  
J.C.L.C. acknowledges support from the Villum
Investigator program supported by the VILLUM Foundation (grant no.
VIL37766 and no.~VIL53101) and the DNRF Chair program (grant no. DNRF162)
by the Danish National Research Foundation.  H.H.Y.N. is supported by the
European Research Council Advanced Grant ``JETSET: Launching, propagation
and emission of relativistic jets from binary mergers and across mass
scales'' (grant No. 884631).  P.C.-K.C. gratefully acknowledges support
from NSF Grant PHY-2020275 (Network for Neutrinos, Nuclear Astrophysics,
and Symmetries (N3AS)).  The Tycho supercomputer hosted at the SCIENCE
HPC center at the University of Copenhagen was used for supporting this
work. 
\end{acknowledgements}

\bibliographystyle{apsrev4-1}
\bibliography{references}

\begin{thebibliography}{124}%
\makeatletter
\providecommand \@ifxundefined [1]{%
 \@ifx{#1\undefined}
}%
\providecommand \@ifnum [1]{%
 \ifnum #1\expandafter \@firstoftwo
 \else \expandafter \@secondoftwo
 \fi
}%
\providecommand \@ifx [1]{%
 \ifx #1\expandafter \@firstoftwo
 \else \expandafter \@secondoftwo
 \fi
}%
\providecommand \natexlab [1]{#1}%
\providecommand \enquote  [1]{``#1''}%
\providecommand \bibnamefont  [1]{#1}%
\providecommand \bibfnamefont [1]{#1}%
\providecommand \citenamefont [1]{#1}%
\providecommand \href@noop [0]{\@secondoftwo}%
\providecommand \href [0]{\begingroup \@sanitize@url \@href}%
\providecommand \@href[1]{\@@startlink{#1}\@@href}%
\providecommand \@@href[1]{\endgroup#1\@@endlink}%
\providecommand \@sanitize@url [0]{\catcode `\\12\catcode `\$12\catcode
  `\&12\catcode `\#12\catcode `\^12\catcode `\_12\catcode `\%12\relax}%
\providecommand \@@startlink[1]{}%
\providecommand \@@endlink[0]{}%
\providecommand \url  [0]{\begingroup\@sanitize@url \@url }%
\providecommand \@url [1]{\endgroup\@href {#1}{\urlprefix }}%
\providecommand \urlprefix  [0]{URL }%
\providecommand \Eprint [0]{\href }%
\providecommand \doibase [0]{http://dx.doi.org/}%
\providecommand \selectlanguage [0]{\@gobble}%
\providecommand \bibinfo  [0]{\@secondoftwo}%
\providecommand \bibfield  [0]{\@secondoftwo}%
\providecommand \translation [1]{[#1]}%
\providecommand \BibitemOpen [0]{}%
\providecommand \bibitemStop [0]{}%
\providecommand \bibitemNoStop [0]{.\EOS\space}%
\providecommand \EOS [0]{\spacefactor3000\relax}%
\providecommand \BibitemShut  [1]{\csname bibitem#1\endcsname}%
\let\auto@bib@innerbib\@empty
\bibitem [{\citenamefont {{Yoon}}\ \emph {et~al.}(2007)\citenamefont {{Yoon}},
  \citenamefont {{Podsiadlowski}},\ and\ \citenamefont
  {{Rosswog}}}]{2007MNRAS.380..933Y}%
  \BibitemOpen
  \bibfield  {author} {\bibinfo {author} {\bibfnamefont {S.~C.}\ \bibnamefont
  {{Yoon}}}, \bibinfo {author} {\bibfnamefont {P.}~\bibnamefont
  {{Podsiadlowski}}}, \ and\ \bibinfo {author} {\bibfnamefont {S.}~\bibnamefont
  {{Rosswog}}},\ }\href {\doibase 10.1111/j.1365-2966.2007.12161.x} {\bibfield
  {journal} {\bibinfo  {journal} {\mnras}\ }\textbf {\bibinfo {volume} {380}},\
  \bibinfo {pages} {933} (\bibinfo {year} {2007})},\ \Eprint
  {http://arxiv.org/abs/0704.0297} {arXiv:0704.0297 [astro-ph]} \BibitemShut
  {NoStop}%
\bibitem [{\citenamefont {{Shen}}\ and\ \citenamefont
  {{Bildsten}}(2009)}]{2009ApJ...692..324S}%
  \BibitemOpen
  \bibfield  {author} {\bibinfo {author} {\bibfnamefont {K.~J.}\ \bibnamefont
  {{Shen}}}\ and\ \bibinfo {author} {\bibfnamefont {L.}~\bibnamefont
  {{Bildsten}}},\ }\href {\doibase 10.1088/0004-637X/692/1/324} {\bibfield
  {journal} {\bibinfo  {journal} {\apj}\ }\textbf {\bibinfo {volume} {692}},\
  \bibinfo {pages} {324} (\bibinfo {year} {2009})},\ \Eprint
  {http://arxiv.org/abs/0805.2160} {arXiv:0805.2160 [astro-ph]} \BibitemShut
  {NoStop}%
\bibitem [{\citenamefont {{Shen}}\ \emph {et~al.}(2009)\citenamefont {{Shen}},
  \citenamefont {{Idan}},\ and\ \citenamefont
  {{Bildsten}}}]{2009ApJ...705..693S}%
  \BibitemOpen
  \bibfield  {author} {\bibinfo {author} {\bibfnamefont {K.~J.}\ \bibnamefont
  {{Shen}}}, \bibinfo {author} {\bibfnamefont {I.}~\bibnamefont {{Idan}}}, \
  and\ \bibinfo {author} {\bibfnamefont {L.}~\bibnamefont {{Bildsten}}},\
  }\href {\doibase 10.1088/0004-637X/705/1/693} {\bibfield  {journal} {\bibinfo
   {journal} {\apj}\ }\textbf {\bibinfo {volume} {705}},\ \bibinfo {pages}
  {693} (\bibinfo {year} {2009})},\ \Eprint {http://arxiv.org/abs/0906.3767}
  {arXiv:0906.3767 [astro-ph.HE]} \BibitemShut {NoStop}%
\bibitem [{\citenamefont {{Moore}}\ \emph {et~al.}(2013)\citenamefont
  {{Moore}}, \citenamefont {{Townsley}},\ and\ \citenamefont
  {{Bildsten}}}]{2013ApJ...776...97M}%
  \BibitemOpen
  \bibfield  {author} {\bibinfo {author} {\bibfnamefont {K.}~\bibnamefont
  {{Moore}}}, \bibinfo {author} {\bibfnamefont {D.~M.}\ \bibnamefont
  {{Townsley}}}, \ and\ \bibinfo {author} {\bibfnamefont {L.}~\bibnamefont
  {{Bildsten}}},\ }\href {\doibase 10.1088/0004-637X/776/2/97} {\bibfield
  {journal} {\bibinfo  {journal} {\apj}\ }\textbf {\bibinfo {volume} {776}},\
  \bibinfo {eid} {97} (\bibinfo {year} {2013})},\ \Eprint
  {http://arxiv.org/abs/1308.4193} {arXiv:1308.4193 [astro-ph.SR]} \BibitemShut
  {NoStop}%
\bibitem [{\citenamefont {{Chandrasekhar}}(1931)}]{1931ApJ....74...81C}%
  \BibitemOpen
  \bibfield  {author} {\bibinfo {author} {\bibfnamefont {S.}~\bibnamefont
  {{Chandrasekhar}}},\ }\href {\doibase 10.1086/143324} {\bibfield  {journal}
  {\bibinfo  {journal} {\apj}\ }\textbf {\bibinfo {volume} {74}},\ \bibinfo
  {pages} {81} (\bibinfo {year} {1931})}\BibitemShut {NoStop}%
\bibitem [{\citenamefont {{Nomoto}}\ and\ \citenamefont
  {{Kondo}}(1991)}]{1991ApJ...367L..19N}%
  \BibitemOpen
  \bibfield  {author} {\bibinfo {author} {\bibfnamefont {K.}~\bibnamefont
  {{Nomoto}}}\ and\ \bibinfo {author} {\bibfnamefont {Y.}~\bibnamefont
  {{Kondo}}},\ }\href {\doibase 10.1086/185922} {\bibfield  {journal} {\bibinfo
   {journal} {\apjl}\ }\textbf {\bibinfo {volume} {367}},\ \bibinfo {pages}
  {L19} (\bibinfo {year} {1991})}\BibitemShut {NoStop}%
\bibitem [{\citenamefont {Yoon}\ and\ \citenamefont
  {Langer}(2005)}]{Yoon:2005cz}%
  \BibitemOpen
  \bibfield  {author} {\bibinfo {author} {\bibfnamefont {S.-C.}\ \bibnamefont
  {Yoon}}\ and\ \bibinfo {author} {\bibfnamefont {N.}~\bibnamefont {Langer}},\
  }\href {\doibase 10.1051/0004-6361:20042542} {\bibfield  {journal} {\bibinfo
  {journal} {Astron. Astrophys.}\ }\textbf {\bibinfo {volume} {435}},\ \bibinfo
  {pages} {967} (\bibinfo {year} {2005})},\ \Eprint
  {http://arxiv.org/abs/astro-ph/0502133} {arXiv:astro-ph/0502133} \BibitemShut
  {NoStop}%
\bibitem [{\citenamefont {Nomoto}\ and\ \citenamefont
  {Kondo}(1991)}]{nomoto1991conditions}%
  \BibitemOpen
  \bibfield  {author} {\bibinfo {author} {\bibfnamefont {K.}~\bibnamefont
  {Nomoto}}\ and\ \bibinfo {author} {\bibfnamefont {Y.}~\bibnamefont {Kondo}},\
  }\href@noop {} {\bibfield  {journal} {\bibinfo  {journal} {Astrophysical
  Journal, Part 2-Letters (ISSN 0004-637X), vol. 367, Jan. 20, 1991, p.
  L19-L22.}\ }\textbf {\bibinfo {volume} {367}},\ \bibinfo {pages} {L19}
  (\bibinfo {year} {1991})}\BibitemShut {NoStop}%
\bibitem [{\citenamefont {Tauris}\ \emph {et~al.}(2013)\citenamefont {Tauris},
  \citenamefont {Sanyal}, \citenamefont {Yoon},\ and\ \citenamefont
  {Langer}}]{Tauris:2013zna}%
  \BibitemOpen
  \bibfield  {author} {\bibinfo {author} {\bibfnamefont {T.~M.}\ \bibnamefont
  {Tauris}}, \bibinfo {author} {\bibfnamefont {D.}~\bibnamefont {Sanyal}},
  \bibinfo {author} {\bibfnamefont {S.-C.}\ \bibnamefont {Yoon}}, \ and\
  \bibinfo {author} {\bibfnamefont {N.}~\bibnamefont {Langer}},\ }\href
  {\doibase 10.1051/0004-6361/201321662} {\bibfield  {journal} {\bibinfo
  {journal} {Astron. Astrophys.}\ }\textbf {\bibinfo {volume} {558}},\ \bibinfo
  {pages} {A39} (\bibinfo {year} {2013})},\ \Eprint
  {http://arxiv.org/abs/1308.4887} {arXiv:1308.4887 [astro-ph.SR]} \BibitemShut
  {NoStop}%
\bibitem [{\citenamefont {Wang}(2018)}]{Wang:2018wje}%
  \BibitemOpen
  \bibfield  {author} {\bibinfo {author} {\bibfnamefont {B.}~\bibnamefont
  {Wang}},\ }\href {\doibase 10.1093/mnras/sty2278} {\bibfield  {journal}
  {\bibinfo  {journal} {Mon. Not. Roy. Astron. Soc.}\ }\textbf {\bibinfo
  {volume} {481}},\ \bibinfo {pages} {439} (\bibinfo {year} {2018})},\ \Eprint
  {http://arxiv.org/abs/1808.05992} {arXiv:1808.05992 [astro-ph.SR]}
  \BibitemShut {NoStop}%
\bibitem [{\citenamefont {Schwab}\ \emph {et~al.}(2016)\citenamefont {Schwab},
  \citenamefont {Quataert},\ and\ \citenamefont {Kasen}}]{Schwab:2016lep}%
  \BibitemOpen
  \bibfield  {author} {\bibinfo {author} {\bibfnamefont {J.}~\bibnamefont
  {Schwab}}, \bibinfo {author} {\bibfnamefont {E.}~\bibnamefont {Quataert}}, \
  and\ \bibinfo {author} {\bibfnamefont {D.}~\bibnamefont {Kasen}},\ }\href
  {\doibase 10.1093/mnras/stw2249} {\bibfield  {journal} {\bibinfo  {journal}
  {Mon. Not. Roy. Astron. Soc.}\ }\textbf {\bibinfo {volume} {463}},\ \bibinfo
  {pages} {3461} (\bibinfo {year} {2016})},\ \Eprint
  {http://arxiv.org/abs/1606.02300} {arXiv:1606.02300 [astro-ph.SR]}
  \BibitemShut {NoStop}%
\bibitem [{\citenamefont {Wu}\ \emph {et~al.}(2019)\citenamefont {Wu},
  \citenamefont {Wang},\ and\ \citenamefont {Liu}}]{Wu:2018pbt}%
  \BibitemOpen
  \bibfield  {author} {\bibinfo {author} {\bibfnamefont {C.}~\bibnamefont
  {Wu}}, \bibinfo {author} {\bibfnamefont {B.}~\bibnamefont {Wang}}, \ and\
  \bibinfo {author} {\bibfnamefont {D.}~\bibnamefont {Liu}},\ }\href {\doibase
  10.1093/mnras/sty3176} {\bibfield  {journal} {\bibinfo  {journal} {Mon. Not.
  Roy. Astron. Soc.}\ }\textbf {\bibinfo {volume} {483}},\ \bibinfo {pages}
  {263} (\bibinfo {year} {2019})},\ \Eprint {http://arxiv.org/abs/1811.08638}
  {arXiv:1811.08638 [astro-ph.SR]} \BibitemShut {NoStop}%
\bibitem [{\citenamefont {Ruiter}(2019)}]{Ruiter:2019tam}%
  \BibitemOpen
  \bibfield  {author} {\bibinfo {author} {\bibfnamefont {A.~J.}\ \bibnamefont
  {Ruiter}},\ }\href {\doibase 10.1017/S1743921320000587} {\bibfield  {journal}
  {\bibinfo  {journal} {IAU Symp.}\ }\textbf {\bibinfo {volume} {357}},\
  \bibinfo {pages} {1} (\bibinfo {year} {2019})},\ \Eprint
  {http://arxiv.org/abs/2001.02947} {arXiv:2001.02947 [astro-ph.SR]}
  \BibitemShut {NoStop}%
\bibitem [{\citenamefont {Liu}\ and\ \citenamefont {Wang}(2020)}]{Liu:2020qny}%
  \BibitemOpen
  \bibfield  {author} {\bibinfo {author} {\bibfnamefont {D.}~\bibnamefont
  {Liu}}\ and\ \bibinfo {author} {\bibfnamefont {B.}~\bibnamefont {Wang}},\
  }\href {\doibase 10.1093/mnras/staa963} {\bibfield  {journal} {\bibinfo
  {journal} {Mon. Not. Roy. Astron. Soc.}\ }\textbf {\bibinfo {volume} {494}},\
  \bibinfo {pages} {3422} (\bibinfo {year} {2020})},\ \Eprint
  {http://arxiv.org/abs/2004.03157} {arXiv:2004.03157 [astro-ph.SR]}
  \BibitemShut {NoStop}%
\bibitem [{\citenamefont {Wang}\ and\ \citenamefont
  {Liu}(2020)}]{Wang:2020pzc}%
  \BibitemOpen
  \bibfield  {author} {\bibinfo {author} {\bibfnamefont {B.}~\bibnamefont
  {Wang}}\ and\ \bibinfo {author} {\bibfnamefont {D.}~\bibnamefont {Liu}},\
  }\href {\doibase 10.1088/1674-4527/20/9/135} {\bibfield  {journal} {\bibinfo
  {journal} {Res. Astron. Astrophys.}\ }\textbf {\bibinfo {volume} {20}},\
  \bibinfo {pages} {135} (\bibinfo {year} {2020})},\ \Eprint
  {http://arxiv.org/abs/2005.01880} {arXiv:2005.01880 [astro-ph.SR]}
  \BibitemShut {NoStop}%
\bibitem [{\citenamefont {Schwab}(2021)}]{Schwab:2020law}%
  \BibitemOpen
  \bibfield  {author} {\bibinfo {author} {\bibfnamefont {J.}~\bibnamefont
  {Schwab}},\ }\href {\doibase 10.3847/1538-4357/abc87e} {\bibfield  {journal}
  {\bibinfo  {journal} {Astrophys. J.}\ }\textbf {\bibinfo {volume} {906}},\
  \bibinfo {pages} {53} (\bibinfo {year} {2021})},\ \Eprint
  {http://arxiv.org/abs/2011.03546} {arXiv:2011.03546 [astro-ph.SR]}
  \BibitemShut {NoStop}%
\bibitem [{\citenamefont {Piersanti}\ \emph {et~al.}(2003)\citenamefont
  {Piersanti}, \citenamefont {Gagliardi}, \citenamefont {Iben},\ and\
  \citenamefont {Tornambe}}]{Piersanti:2002sb}%
  \BibitemOpen
  \bibfield  {author} {\bibinfo {author} {\bibfnamefont {L.}~\bibnamefont
  {Piersanti}}, \bibinfo {author} {\bibfnamefont {S.}~\bibnamefont
  {Gagliardi}}, \bibinfo {author} {\bibfnamefont {I.}~\bibnamefont {Iben},
  \bibfnamefont {Jr.}}, \ and\ \bibinfo {author} {\bibfnamefont
  {A.}~\bibnamefont {Tornambe}},\ }\href {\doibase 10.1086/345444} {\bibfield
  {journal} {\bibinfo  {journal} {Astrophys. J.}\ }\textbf {\bibinfo {volume}
  {583}},\ \bibinfo {pages} {885} (\bibinfo {year} {2003})},\ \Eprint
  {http://arxiv.org/abs/astro-ph/0210624} {arXiv:astro-ph/0210624} \BibitemShut
  {NoStop}%
\bibitem [{\citenamefont {Uenishi}\ \emph {et~al.}(2003)\citenamefont
  {Uenishi}, \citenamefont {Nomoto},\ and\ \citenamefont
  {Hachisu}}]{Uenishi:2003sk}%
  \BibitemOpen
  \bibfield  {author} {\bibinfo {author} {\bibfnamefont {T.}~\bibnamefont
  {Uenishi}}, \bibinfo {author} {\bibfnamefont {K.}~\bibnamefont {Nomoto}}, \
  and\ \bibinfo {author} {\bibfnamefont {I.}~\bibnamefont {Hachisu}},\ }\href
  {\doibase 10.1086/377248} {\bibfield  {journal} {\bibinfo  {journal}
  {Astrophys. J.}\ }\textbf {\bibinfo {volume} {595}},\ \bibinfo {pages} {1094}
  (\bibinfo {year} {2003})},\ \Eprint {http://arxiv.org/abs/astro-ph/0309433}
  {arXiv:astro-ph/0309433} \BibitemShut {NoStop}%
\bibitem [{\citenamefont {Saio}\ and\ \citenamefont
  {Nomoto}(2004)}]{Saio:2004gz}%
  \BibitemOpen
  \bibfield  {author} {\bibinfo {author} {\bibfnamefont {H.}~\bibnamefont
  {Saio}}\ and\ \bibinfo {author} {\bibfnamefont {K.}~\bibnamefont {Nomoto}},\
  }\href {\doibase 10.1086/423976} {\bibfield  {journal} {\bibinfo  {journal}
  {Astrophys. J.}\ }\textbf {\bibinfo {volume} {615}},\ \bibinfo {pages} {444}
  (\bibinfo {year} {2004})},\ \Eprint {http://arxiv.org/abs/astro-ph/0401141}
  {arXiv:astro-ph/0401141} \BibitemShut {NoStop}%
\bibitem [{\citenamefont {{Ferrario}}\ \emph {et~al.}(2015)\citenamefont
  {{Ferrario}}, \citenamefont {{de Martino}},\ and\ \citenamefont
  {{G{\"a}nsicke}}}]{2015SSRv..191..111F}%
  \BibitemOpen
  \bibfield  {author} {\bibinfo {author} {\bibfnamefont {L.}~\bibnamefont
  {{Ferrario}}}, \bibinfo {author} {\bibfnamefont {D.}~\bibnamefont {{de
  Martino}}}, \ and\ \bibinfo {author} {\bibfnamefont {B.~T.}\ \bibnamefont
  {{G{\"a}nsicke}}},\ }\href {\doibase 10.1007/s11214-015-0152-0} {\bibfield
  {journal} {\bibinfo  {journal} {\ssr}\ }\textbf {\bibinfo {volume} {191}},\
  \bibinfo {pages} {111} (\bibinfo {year} {2015})},\ \Eprint
  {http://arxiv.org/abs/1504.08072} {arXiv:1504.08072 [astro-ph.SR]}
  \BibitemShut {NoStop}%
\bibitem [{\citenamefont {{Kawka}}(2020)}]{2020IAUS..357...60K}%
  \BibitemOpen
  \bibfield  {author} {\bibinfo {author} {\bibfnamefont {A.}~\bibnamefont
  {{Kawka}}},\ }\href {\doibase 10.1017/S1743921320000745} {\bibfield
  {journal} {\bibinfo  {journal} {IAU Symposium}\ }\textbf {\bibinfo {volume}
  {357}},\ \bibinfo {pages} {60} (\bibinfo {year} {2020})},\ \Eprint
  {http://arxiv.org/abs/2001.10672} {arXiv:2001.10672 [astro-ph.SR]}
  \BibitemShut {NoStop}%
\bibitem [{\citenamefont {{Ferrario}}\ \emph {et~al.}(2020)\citenamefont
  {{Ferrario}}, \citenamefont {{Wickramasinghe}},\ and\ \citenamefont
  {{Kawka}}}]{2020AdSpR..66.1025F}%
  \BibitemOpen
  \bibfield  {author} {\bibinfo {author} {\bibfnamefont {L.}~\bibnamefont
  {{Ferrario}}}, \bibinfo {author} {\bibfnamefont {D.}~\bibnamefont
  {{Wickramasinghe}}}, \ and\ \bibinfo {author} {\bibfnamefont
  {A.}~\bibnamefont {{Kawka}}},\ }\href {\doibase 10.1016/j.asr.2019.11.012}
  {\bibfield  {journal} {\bibinfo  {journal} {Advances in Space Research}\
  }\textbf {\bibinfo {volume} {66}},\ \bibinfo {pages} {1025} (\bibinfo {year}
  {2020})},\ \Eprint {http://arxiv.org/abs/2001.10147} {arXiv:2001.10147
  [astro-ph.SR]} \BibitemShut {NoStop}%
\bibitem [{\citenamefont {{Zhu}}\ \emph {et~al.}(2015)\citenamefont {{Zhu}},
  \citenamefont {{Pakmor}}, \citenamefont {{van Kerkwijk}},\ and\ \citenamefont
  {{Chang}}}]{2015ApJ...806L...1Z}%
  \BibitemOpen
  \bibfield  {author} {\bibinfo {author} {\bibfnamefont {C.}~\bibnamefont
  {{Zhu}}}, \bibinfo {author} {\bibfnamefont {R.}~\bibnamefont {{Pakmor}}},
  \bibinfo {author} {\bibfnamefont {M.~H.}\ \bibnamefont {{van Kerkwijk}}}, \
  and\ \bibinfo {author} {\bibfnamefont {P.}~\bibnamefont {{Chang}}},\ }\href
  {\doibase 10.1088/2041-8205/806/1/L1} {\bibfield  {journal} {\bibinfo
  {journal} {\apjl}\ }\textbf {\bibinfo {volume} {806}},\ \bibinfo {eid} {L1}
  (\bibinfo {year} {2015})},\ \Eprint {http://arxiv.org/abs/1504.01732}
  {arXiv:1504.01732 [astro-ph.SR]} \BibitemShut {NoStop}%
\bibitem [{\citenamefont {Pakmor}\ \emph {et~al.}(2024)\citenamefont {Pakmor}
  \emph {et~al.}}]{Pakmor:2024efc}%
  \BibitemOpen
  \bibfield  {author} {\bibinfo {author} {\bibfnamefont {R.}~\bibnamefont
  {Pakmor}} \emph {et~al.},\ }\href {\doibase 10.1051/0004-6361/202451352}
  {\bibfield  {journal} {\bibinfo  {journal} {Astron. Astrophys.}\ }\textbf
  {\bibinfo {volume} {691}},\ \bibinfo {pages} {A179} (\bibinfo {year}
  {2024})},\ \Eprint {http://arxiv.org/abs/2407.02566} {arXiv:2407.02566
  [astro-ph.SR]} \BibitemShut {NoStop}%
\bibitem [{\citenamefont {Fryer}\ \emph {et~al.}(2002)\citenamefont {Fryer},
  \citenamefont {Holz},\ and\ \citenamefont {Hughes}}]{Fryer:2001zw}%
  \BibitemOpen
  \bibfield  {author} {\bibinfo {author} {\bibfnamefont {C.~L.}\ \bibnamefont
  {Fryer}}, \bibinfo {author} {\bibfnamefont {D.~E.}\ \bibnamefont {Holz}}, \
  and\ \bibinfo {author} {\bibfnamefont {S.~A.}\ \bibnamefont {Hughes}},\
  }\href {\doibase 10.1086/324034} {\bibfield  {journal} {\bibinfo  {journal}
  {Astrophys. J.}\ }\textbf {\bibinfo {volume} {565}},\ \bibinfo {pages} {430}
  (\bibinfo {year} {2002})},\ \Eprint {http://arxiv.org/abs/astro-ph/0106113}
  {arXiv:astro-ph/0106113} \BibitemShut {NoStop}%
\bibitem [{\citenamefont {Dimmelmeier}\ \emph {et~al.}(2008)\citenamefont
  {Dimmelmeier}, \citenamefont {Ott}, \citenamefont {Marek},\ and\
  \citenamefont {Janka}}]{Dimmelmeier:2008iq}%
  \BibitemOpen
  \bibfield  {author} {\bibinfo {author} {\bibfnamefont {H.}~\bibnamefont
  {Dimmelmeier}}, \bibinfo {author} {\bibfnamefont {C.~D.}\ \bibnamefont
  {Ott}}, \bibinfo {author} {\bibfnamefont {A.}~\bibnamefont {Marek}}, \ and\
  \bibinfo {author} {\bibfnamefont {H.~T.}\ \bibnamefont {Janka}},\ }\href
  {\doibase 10.1103/PhysRevD.78.064056} {\bibfield  {journal} {\bibinfo
  {journal} {Phys. Rev. D}\ }\textbf {\bibinfo {volume} {78}},\ \bibinfo
  {pages} {064056} (\bibinfo {year} {2008})},\ \Eprint
  {http://arxiv.org/abs/0806.4953} {arXiv:0806.4953 [astro-ph]} \BibitemShut
  {NoStop}%
\bibitem [{\citenamefont {Abdikamalov}\ \emph {et~al.}(2010)\citenamefont
  {Abdikamalov}, \citenamefont {Ott}, \citenamefont {Rezzolla}, \citenamefont
  {Dessart}, \citenamefont {Dimmelmeier}, \citenamefont {Marek},\ and\
  \citenamefont {Janka}}]{Abdikamalov:2009aq}%
  \BibitemOpen
  \bibfield  {author} {\bibinfo {author} {\bibfnamefont {E.~B.}\ \bibnamefont
  {Abdikamalov}}, \bibinfo {author} {\bibfnamefont {C.~D.}\ \bibnamefont
  {Ott}}, \bibinfo {author} {\bibfnamefont {L.}~\bibnamefont {Rezzolla}},
  \bibinfo {author} {\bibfnamefont {L.}~\bibnamefont {Dessart}}, \bibinfo
  {author} {\bibfnamefont {H.}~\bibnamefont {Dimmelmeier}}, \bibinfo {author}
  {\bibfnamefont {A.}~\bibnamefont {Marek}}, \ and\ \bibinfo {author}
  {\bibfnamefont {H.~T.}\ \bibnamefont {Janka}},\ }\href {\doibase
  10.1103/PhysRevD.81.044012} {\bibfield  {journal} {\bibinfo  {journal} {Phys.
  Rev. D}\ }\textbf {\bibinfo {volume} {81}},\ \bibinfo {pages} {044012}
  (\bibinfo {year} {2010})},\ \Eprint {http://arxiv.org/abs/0910.2703}
  {arXiv:0910.2703 [astro-ph.HE]} \BibitemShut {NoStop}%
\bibitem [{\citenamefont {Longo~Micchi}\ \emph {et~al.}(2023)\citenamefont
  {Longo~Micchi}, \citenamefont {Radice},\ and\ \citenamefont
  {Chirenti}}]{LongoMicchi:2023khv}%
  \BibitemOpen
  \bibfield  {author} {\bibinfo {author} {\bibfnamefont {L.~F.}\ \bibnamefont
  {Longo~Micchi}}, \bibinfo {author} {\bibfnamefont {D.}~\bibnamefont
  {Radice}}, \ and\ \bibinfo {author} {\bibfnamefont {C.}~\bibnamefont
  {Chirenti}},\ }\href {\doibase 10.1093/mnras/stad2420} {\bibfield  {journal}
  {\bibinfo  {journal} {Mon. Not. Roy. Astron. Soc.}\ }\textbf {\bibinfo
  {volume} {525}},\ \bibinfo {pages} {6359} (\bibinfo {year} {2023})},\ \Eprint
  {http://arxiv.org/abs/2306.04711} {arXiv:2306.04711 [astro-ph.HE]}
  \BibitemShut {NoStop}%
\bibitem [{\citenamefont {Fryer}\ \emph {et~al.}(1999)\citenamefont {Fryer},
  \citenamefont {Benz}, \citenamefont {Herant},\ and\ \citenamefont
  {Colgate}}]{Fryer:1998jb}%
  \BibitemOpen
  \bibfield  {author} {\bibinfo {author} {\bibfnamefont {C.~L.}\ \bibnamefont
  {Fryer}}, \bibinfo {author} {\bibfnamefont {W.}~\bibnamefont {Benz}},
  \bibinfo {author} {\bibfnamefont {M.}~\bibnamefont {Herant}}, \ and\ \bibinfo
  {author} {\bibfnamefont {S.~A.}\ \bibnamefont {Colgate}},\ }\href {\doibase
  10.1086/307119} {\bibfield  {journal} {\bibinfo  {journal} {Astrophys. J.}\
  }\textbf {\bibinfo {volume} {516}},\ \bibinfo {pages} {892} (\bibinfo {year}
  {1999})},\ \Eprint {http://arxiv.org/abs/astro-ph/9812058}
  {arXiv:astro-ph/9812058} \BibitemShut {NoStop}%
\bibitem [{\citenamefont {Dessart}\ \emph {et~al.}(2006)\citenamefont
  {Dessart}, \citenamefont {Burrows}, \citenamefont {Ott}, \citenamefont
  {Livne}, \citenamefont {Yoon},\ and\ \citenamefont
  {Langer}}]{Dessart:2006gd}%
  \BibitemOpen
  \bibfield  {author} {\bibinfo {author} {\bibfnamefont {L.}~\bibnamefont
  {Dessart}}, \bibinfo {author} {\bibfnamefont {A.}~\bibnamefont {Burrows}},
  \bibinfo {author} {\bibfnamefont {C.}~\bibnamefont {Ott}}, \bibinfo {author}
  {\bibfnamefont {E.}~\bibnamefont {Livne}}, \bibinfo {author} {\bibfnamefont
  {S.-C.}\ \bibnamefont {Yoon}}, \ and\ \bibinfo {author} {\bibfnamefont
  {N.}~\bibnamefont {Langer}},\ }\href {\doibase 10.1086/503626} {\bibfield
  {journal} {\bibinfo  {journal} {Astrophys. J.}\ }\textbf {\bibinfo {volume}
  {644}},\ \bibinfo {pages} {1063} (\bibinfo {year} {2006})},\ \Eprint
  {http://arxiv.org/abs/astro-ph/0601603} {arXiv:astro-ph/0601603} \BibitemShut
  {NoStop}%
\bibitem [{\citenamefont {Batziou}\ \emph {et~al.}(2024)\citenamefont
  {Batziou}, \citenamefont {Glas}, \citenamefont {Janka}, \citenamefont
  {Ehring}, \citenamefont {Abdikamalov},\ and\ \citenamefont
  {Just}}]{Batziou:2024ory}%
  \BibitemOpen
  \bibfield  {author} {\bibinfo {author} {\bibfnamefont {E.}~\bibnamefont
  {Batziou}}, \bibinfo {author} {\bibfnamefont {R.}~\bibnamefont {Glas}},
  \bibinfo {author} {\bibfnamefont {H.~T.}\ \bibnamefont {Janka}}, \bibinfo
  {author} {\bibfnamefont {J.}~\bibnamefont {Ehring}}, \bibinfo {author}
  {\bibfnamefont {E.}~\bibnamefont {Abdikamalov}}, \ and\ \bibinfo {author}
  {\bibfnamefont {O.}~\bibnamefont {Just}},\ }\href@noop {} {\  (\bibinfo
  {year} {2024})},\ \Eprint {http://arxiv.org/abs/2412.02756} {arXiv:2412.02756
  [astro-ph.HE]} \BibitemShut {NoStop}%
\bibitem [{\citenamefont {Cheong}\ \emph {et~al.}(2025)\citenamefont {Cheong},
  \citenamefont {Pitik}, \citenamefont {Longo~Micchi},\ and\ \citenamefont
  {Radice}}]{Cheong:2024hrd}%
  \BibitemOpen
  \bibfield  {author} {\bibinfo {author} {\bibfnamefont {P.~C.-K.}\
  \bibnamefont {Cheong}}, \bibinfo {author} {\bibfnamefont {T.}~\bibnamefont
  {Pitik}}, \bibinfo {author} {\bibfnamefont {L.~F.}\ \bibnamefont
  {Longo~Micchi}}, \ and\ \bibinfo {author} {\bibfnamefont {D.}~\bibnamefont
  {Radice}},\ }\href {\doibase 10.3847/2041-8213/ada1cc} {\bibfield  {journal}
  {\bibinfo  {journal} {Astrophys. J. Lett.}\ }\textbf {\bibinfo {volume}
  {978}},\ \bibinfo {pages} {L38} (\bibinfo {year} {2025})},\ \Eprint
  {http://arxiv.org/abs/2410.10938} {arXiv:2410.10938 [astro-ph.HE]}
  \BibitemShut {NoStop}%
\bibitem [{\citenamefont {Yip}\ \emph {et~al.}(2024)\citenamefont {Yip},
  \citenamefont {Chu}, \citenamefont {Leung},\ and\ \citenamefont
  {Lin}}]{Yip:2024akb}%
  \BibitemOpen
  \bibfield  {author} {\bibinfo {author} {\bibfnamefont {C.-M.}\ \bibnamefont
  {Yip}}, \bibinfo {author} {\bibfnamefont {M.-C.}\ \bibnamefont {Chu}},
  \bibinfo {author} {\bibfnamefont {S.-C.}\ \bibnamefont {Leung}}, \ and\
  \bibinfo {author} {\bibfnamefont {L.-M.}\ \bibnamefont {Lin}},\ }\href@noop
  {} {\  (\bibinfo {year} {2024})},\ \Eprint {http://arxiv.org/abs/2401.03798}
  {arXiv:2401.03798 [astro-ph.HE]} \BibitemShut {NoStop}%
\bibitem [{\citenamefont {Yi}\ and\ \citenamefont
  {Blackman}(1998)}]{Yi:1997qb}%
  \BibitemOpen
  \bibfield  {author} {\bibinfo {author} {\bibfnamefont {I.}~\bibnamefont
  {Yi}}\ and\ \bibinfo {author} {\bibfnamefont {E.~G.}\ \bibnamefont
  {Blackman}},\ }\href {\doibase 10.1086/311192} {\bibfield  {journal}
  {\bibinfo  {journal} {Astrophys. J. Lett.}\ }\textbf {\bibinfo {volume}
  {494}},\ \bibinfo {pages} {L163} (\bibinfo {year} {1998})},\ \Eprint
  {http://arxiv.org/abs/astro-ph/9710149} {arXiv:astro-ph/9710149} \BibitemShut
  {NoStop}%
\bibitem [{\citenamefont {Metzger}\ \emph {et~al.}(2008)\citenamefont
  {Metzger}, \citenamefont {Quataert},\ and\ \citenamefont
  {Thompson}}]{Metzger:2007cd}%
  \BibitemOpen
  \bibfield  {author} {\bibinfo {author} {\bibfnamefont {B.~D.}\ \bibnamefont
  {Metzger}}, \bibinfo {author} {\bibfnamefont {E.}~\bibnamefont {Quataert}}, \
  and\ \bibinfo {author} {\bibfnamefont {T.~A.}\ \bibnamefont {Thompson}},\
  }\href {\doibase 10.1111/j.1365-2966.2008.12923.x} {\bibfield  {journal}
  {\bibinfo  {journal} {Mon. Not. Roy. Astron. Soc.}\ }\textbf {\bibinfo
  {volume} {385}},\ \bibinfo {pages} {1455} (\bibinfo {year} {2008})},\ \Eprint
  {http://arxiv.org/abs/0712.1233} {arXiv:0712.1233 [astro-ph]} \BibitemShut
  {NoStop}%
\bibitem [{\citenamefont {Perley}\ \emph {et~al.}(2009)\citenamefont {Perley}
  \emph {et~al.}}]{Perley:2008ay}%
  \BibitemOpen
  \bibfield  {author} {\bibinfo {author} {\bibfnamefont {D.~A.}\ \bibnamefont
  {Perley}} \emph {et~al.},\ }\href {\doibase 10.1088/0004-637X/696/2/1871}
  {\bibfield  {journal} {\bibinfo  {journal} {Astrophys. J.}\ }\textbf
  {\bibinfo {volume} {696}},\ \bibinfo {pages} {1871} (\bibinfo {year}
  {2009})},\ \Eprint {http://arxiv.org/abs/0811.1044} {arXiv:0811.1044
  [astro-ph]} \BibitemShut {NoStop}%
\bibitem [{\citenamefont {Waxman}(2017)}]{Waxman:2017zme}%
  \BibitemOpen
  \bibfield  {author} {\bibinfo {author} {\bibfnamefont {E.}~\bibnamefont
  {Waxman}},\ }\href {\doibase 10.3847/1538-4357/aa713e} {\bibfield  {journal}
  {\bibinfo  {journal} {Astrophys. J.}\ }\textbf {\bibinfo {volume} {842}},\
  \bibinfo {pages} {34} (\bibinfo {year} {2017})},\ \Eprint
  {http://arxiv.org/abs/1703.06723} {arXiv:1703.06723 [astro-ph.HE]}
  \BibitemShut {NoStop}%
\bibitem [{\citenamefont {Margalit}\ \emph {et~al.}(2019)\citenamefont
  {Margalit}, \citenamefont {Berger},\ and\ \citenamefont
  {Metzger}}]{Margalit:2019hke}%
  \BibitemOpen
  \bibfield  {author} {\bibinfo {author} {\bibfnamefont {B.}~\bibnamefont
  {Margalit}}, \bibinfo {author} {\bibfnamefont {E.}~\bibnamefont {Berger}}, \
  and\ \bibinfo {author} {\bibfnamefont {B.~D.}\ \bibnamefont {Metzger}},\
  }\href {\doibase 10.3847/1538-4357/ab4c31} {\  (\bibinfo {year} {2019}),\
  10.3847/1538-4357/ab4c31},\ \Eprint {http://arxiv.org/abs/1907.00016}
  {arXiv:1907.00016 [astro-ph.HE]} \BibitemShut {NoStop}%
\bibitem [{\citenamefont {Lyutikov}\ and\ \citenamefont
  {Toonen}(2019)}]{Lyutikov:2018pkv}%
  \BibitemOpen
  \bibfield  {author} {\bibinfo {author} {\bibfnamefont {M.}~\bibnamefont
  {Lyutikov}}\ and\ \bibinfo {author} {\bibfnamefont {S.}~\bibnamefont
  {Toonen}},\ }\href {\doibase 10.1093/mnras/stz1640} {\bibfield  {journal}
  {\bibinfo  {journal} {Mon. Not. Roy. Astron. Soc.}\ }\textbf {\bibinfo
  {volume} {487}},\ \bibinfo {pages} {5618} (\bibinfo {year} {2019})},\ \Eprint
  {http://arxiv.org/abs/1812.07569} {arXiv:1812.07569 [astro-ph.HE]}
  \BibitemShut {NoStop}%
\bibitem [{\citenamefont {Lyutikov}(2022)}]{Lyutikov:2022xri}%
  \BibitemOpen
  \bibfield  {author} {\bibinfo {author} {\bibfnamefont {M.}~\bibnamefont
  {Lyutikov}},\ }\href {\doibase 10.1093/mnras/stac1717} {\bibfield  {journal}
  {\bibinfo  {journal} {Mon. Not. Roy. Astron. Soc.}\ }\textbf {\bibinfo
  {volume} {515}},\ \bibinfo {pages} {2293} (\bibinfo {year} {2022})},\ \Eprint
  {http://arxiv.org/abs/2204.08366} {arXiv:2204.08366 [astro-ph.HE]}
  \BibitemShut {NoStop}%
\bibitem [{\citenamefont {McBrien}\ \emph {et~al.}(2019)\citenamefont {McBrien}
  \emph {et~al.}}]{McBrien:2019wfv}%
  \BibitemOpen
  \bibfield  {author} {\bibinfo {author} {\bibfnamefont {O.~R.}\ \bibnamefont
  {McBrien}} \emph {et~al.},\ }\href {\doibase 10.3847/2041-8213/ab4dae}
  {\bibfield  {journal} {\bibinfo  {journal} {Astrophys. J. Lett.}\ }\textbf
  {\bibinfo {volume} {885}},\ \bibinfo {pages} {L23} (\bibinfo {year}
  {2019})},\ \Eprint {http://arxiv.org/abs/1909.04545} {arXiv:1909.04545
  [astro-ph.HE]} \BibitemShut {NoStop}%
\bibitem [{\citenamefont {Moriya}(2019)}]{Moriya:2019ees}%
  \BibitemOpen
  \bibfield  {author} {\bibinfo {author} {\bibfnamefont {T.~J.}\ \bibnamefont
  {Moriya}},\ }\href {\doibase 10.1093/mnras/stz2627} {\bibfield  {journal}
  {\bibinfo  {journal} {Mon. Not. Roy. Astron. Soc.}\ }\textbf {\bibinfo
  {volume} {490}},\ \bibinfo {pages} {1166} (\bibinfo {year} {2019})},\ \Eprint
  {http://arxiv.org/abs/1909.07183} {arXiv:1909.07183 [astro-ph.HE]}
  \BibitemShut {NoStop}%
\bibitem [{\citenamefont {Gillanders}\ \emph {et~al.}(2020)\citenamefont
  {Gillanders}, \citenamefont {Sim},\ and\ \citenamefont
  {Smartt}}]{Gillanders:2020fhm}%
  \BibitemOpen
  \bibfield  {author} {\bibinfo {author} {\bibfnamefont {J.~H.}\ \bibnamefont
  {Gillanders}}, \bibinfo {author} {\bibfnamefont {S.~A.}\ \bibnamefont {Sim}},
  \ and\ \bibinfo {author} {\bibfnamefont {S.~J.}\ \bibnamefont {Smartt}},\
  }\href {\doibase 10.1093/mnras/staa1822} {\bibfield  {journal} {\bibinfo
  {journal} {Mon. Not. Roy. Astron. Soc.}\ }\textbf {\bibinfo {volume} {497}},\
  \bibinfo {pages} {246} (\bibinfo {year} {2020})},\ \Eprint
  {http://arxiv.org/abs/2007.12110} {arXiv:2007.12110 [astro-ph.HE]}
  \BibitemShut {NoStop}%
\bibitem [{\citenamefont {Kato}\ and\ \citenamefont
  {Hachisu}(2003)}]{Kato:2003qz}%
  \BibitemOpen
  \bibfield  {author} {\bibinfo {author} {\bibfnamefont {M.}~\bibnamefont
  {Kato}}\ and\ \bibinfo {author} {\bibfnamefont {I.}~\bibnamefont {Hachisu}},\
  }\href {\doibase 10.1086/380597} {\bibfield  {journal} {\bibinfo  {journal}
  {Astrophys. J. Lett.}\ }\textbf {\bibinfo {volume} {598}},\ \bibinfo {pages}
  {L107} (\bibinfo {year} {2003})},\ \Eprint
  {http://arxiv.org/abs/astro-ph/0310351} {arXiv:astro-ph/0310351} \BibitemShut
  {NoStop}%
\bibitem [{\citenamefont {{Gvaramadze}}\ \emph {et~al.}(2019)\citenamefont
  {{Gvaramadze}}, \citenamefont {{Gr{\"a}fener}}, \citenamefont {{Langer}},
  \citenamefont {{Maryeva}}, \citenamefont {{Kniazev}}, \citenamefont
  {{Moskvitin}},\ and\ \citenamefont {{Spiridonova}}}]{2019Natur.569..684G}%
  \BibitemOpen
  \bibfield  {author} {\bibinfo {author} {\bibfnamefont {V.~V.}\ \bibnamefont
  {{Gvaramadze}}}, \bibinfo {author} {\bibfnamefont {G.}~\bibnamefont
  {{Gr{\"a}fener}}}, \bibinfo {author} {\bibfnamefont {N.}~\bibnamefont
  {{Langer}}}, \bibinfo {author} {\bibfnamefont {O.~V.}\ \bibnamefont
  {{Maryeva}}}, \bibinfo {author} {\bibfnamefont {A.~Y.}\ \bibnamefont
  {{Kniazev}}}, \bibinfo {author} {\bibfnamefont {A.~S.}\ \bibnamefont
  {{Moskvitin}}}, \ and\ \bibinfo {author} {\bibfnamefont {O.~I.}\ \bibnamefont
  {{Spiridonova}}},\ }\href {\doibase 10.1038/s41586-019-1216-1} {\bibfield
  {journal} {\bibinfo  {journal} {\nat}\ }\textbf {\bibinfo {volume} {569}},\
  \bibinfo {pages} {684} (\bibinfo {year} {2019})},\ \Eprint
  {http://arxiv.org/abs/1904.00012} {arXiv:1904.00012 [astro-ph.SR]}
  \BibitemShut {NoStop}%
\bibitem [{\citenamefont {{Oskinova}}\ \emph {et~al.}(2020)\citenamefont
  {{Oskinova}}, \citenamefont {{Gvaramadze}}, \citenamefont {{Gr{\"a}fener}},
  \citenamefont {{Langer}},\ and\ \citenamefont
  {{Todt}}}]{2020A&A...644L...8O}%
  \BibitemOpen
  \bibfield  {author} {\bibinfo {author} {\bibfnamefont {L.~M.}\ \bibnamefont
  {{Oskinova}}}, \bibinfo {author} {\bibfnamefont {V.~V.}\ \bibnamefont
  {{Gvaramadze}}}, \bibinfo {author} {\bibfnamefont {G.}~\bibnamefont
  {{Gr{\"a}fener}}}, \bibinfo {author} {\bibfnamefont {N.}~\bibnamefont
  {{Langer}}}, \ and\ \bibinfo {author} {\bibfnamefont {H.}~\bibnamefont
  {{Todt}}},\ }\href {\doibase 10.1051/0004-6361/202039232} {\bibfield
  {journal} {\bibinfo  {journal} {\aap}\ }\textbf {\bibinfo {volume} {644}},\
  \bibinfo {eid} {L8} (\bibinfo {year} {2020})},\ \Eprint
  {http://arxiv.org/abs/2008.10612} {arXiv:2008.10612 [astro-ph.SR]}
  \BibitemShut {NoStop}%
\bibitem [{\citenamefont {Caiazzo}\ \emph {et~al.}(2021)\citenamefont {Caiazzo}
  \emph {et~al.}}]{Caiazzo:2021xkk}%
  \BibitemOpen
  \bibfield  {author} {\bibinfo {author} {\bibfnamefont {I.}~\bibnamefont
  {Caiazzo}} \emph {et~al.},\ }\href {\doibase 10.1038/s41586-021-03799-3}
  {\bibfield  {journal} {\bibinfo  {journal} {Nature}\ }\textbf {\bibinfo
  {volume} {595}},\ \bibinfo {pages} {39} (\bibinfo {year} {2021})},\ \bibinfo
  {note} {[Erratum: Nature 596, E15 (2021)]},\ \Eprint
  {http://arxiv.org/abs/2107.08458} {arXiv:2107.08458 [astro-ph.SR]}
  \BibitemShut {NoStop}%
\bibitem [{\citenamefont {Luo}\ \emph {et~al.}(2025)\citenamefont {Luo} \emph
  {et~al.}}]{Luo:2024qwa}%
  \BibitemOpen
  \bibfield  {author} {\bibinfo {author} {\bibfnamefont {C.}~\bibnamefont
  {Luo}} \emph {et~al.},\ }\href {\doibase 10.1007/s11433-024-2630-x}
  {\bibfield  {journal} {\bibinfo  {journal} {Sci. China Phys. Mech. Astron.}\
  }\textbf {\bibinfo {volume} {68}},\ \bibinfo {pages} {269511} (\bibinfo
  {year} {2025})},\ \Eprint {http://arxiv.org/abs/2404.04835} {arXiv:2404.04835
  [astro-ph.SR]} \BibitemShut {NoStop}%
\bibitem [{\citenamefont {Zhao}\ \emph {et~al.}(2024)\citenamefont {Zhao},
  \citenamefont {Wang}, \citenamefont {Wang}, \citenamefont {Zheng},
  \citenamefont {Yuan},\ and\ \citenamefont {Liu}}]{Zhao:2024gmz}%
  \BibitemOpen
  \bibfield  {author} {\bibinfo {author} {\bibfnamefont {X.}~\bibnamefont
  {Zhao}}, \bibinfo {author} {\bibfnamefont {S.}~\bibnamefont {Wang}}, \bibinfo
  {author} {\bibfnamefont {P.}~\bibnamefont {Wang}}, \bibinfo {author}
  {\bibfnamefont {C.}~\bibnamefont {Zheng}}, \bibinfo {author} {\bibfnamefont
  {H.}~\bibnamefont {Yuan}}, \ and\ \bibinfo {author} {\bibfnamefont
  {J.}~\bibnamefont {Liu}},\ }\href {\doibase 10.3847/1538-4357/ad9273}
  {\bibfield  {journal} {\bibinfo  {journal} {Astrophys. J.}\ }\textbf
  {\bibinfo {volume} {977}},\ \bibinfo {pages} {245} (\bibinfo {year}
  {2024})},\ \Eprint {http://arxiv.org/abs/2411.08837} {arXiv:2411.08837
  [astro-ph.SR]} \BibitemShut {NoStop}%
\bibitem [{\citenamefont {Mereghetti}(2024)}]{Mereghetti:2024mxn}%
  \BibitemOpen
  \bibfield  {author} {\bibinfo {author} {\bibfnamefont {S.}~\bibnamefont
  {Mereghetti}}\ }(\bibinfo {year} {2024})\ \Eprint
  {http://arxiv.org/abs/2412.18546} {arXiv:2412.18546 [astro-ph.HE]}
  \BibitemShut {NoStop}%
\bibitem [{\citenamefont {Woosley}\ \emph {et~al.}(2002)\citenamefont
  {Woosley}, \citenamefont {Heger},\ and\ \citenamefont
  {Weaver}}]{Woosley:2002zz}%
  \BibitemOpen
  \bibfield  {author} {\bibinfo {author} {\bibfnamefont {S.~E.}\ \bibnamefont
  {Woosley}}, \bibinfo {author} {\bibfnamefont {A.}~\bibnamefont {Heger}}, \
  and\ \bibinfo {author} {\bibfnamefont {T.~A.}\ \bibnamefont {Weaver}},\
  }\href {\doibase 10.1103/RevModPhys.74.1015} {\bibfield  {journal} {\bibinfo
  {journal} {Rev. Mod. Phys.}\ }\textbf {\bibinfo {volume} {74}},\ \bibinfo
  {pages} {1015} (\bibinfo {year} {2002})}\BibitemShut {NoStop}%
\bibitem [{\citenamefont {Leung}\ \emph {et~al.}(2019)\citenamefont {Leung},
  \citenamefont {Zha}, \citenamefont {Chu}, \citenamefont {Lin},\ and\
  \citenamefont {Nomoto}}]{Leung:2019ctw}%
  \BibitemOpen
  \bibfield  {author} {\bibinfo {author} {\bibfnamefont {S.-C.}\ \bibnamefont
  {Leung}}, \bibinfo {author} {\bibfnamefont {S.}~\bibnamefont {Zha}}, \bibinfo
  {author} {\bibfnamefont {M.-C.}\ \bibnamefont {Chu}}, \bibinfo {author}
  {\bibfnamefont {L.-M.}\ \bibnamefont {Lin}}, \ and\ \bibinfo {author}
  {\bibfnamefont {K.}~\bibnamefont {Nomoto}},\ }\href {\doibase
  10.3847/1538-4357/ab3b5e} {\bibfield  {journal} {\bibinfo  {journal}
  {Astrophys. J.}\ }\textbf {\bibinfo {volume} {884}},\ \bibinfo {pages} {9}
  (\bibinfo {year} {2019})},\ \Eprint {http://arxiv.org/abs/1908.05102}
  {arXiv:1908.05102 [astro-ph.HE]} \BibitemShut {NoStop}%
\bibitem [{\citenamefont {Ablimit}\ and\ \citenamefont
  {Li}(2015)}]{Ablimit:2014vka}%
  \BibitemOpen
  \bibfield  {author} {\bibinfo {author} {\bibfnamefont {I.}~\bibnamefont
  {Ablimit}}\ and\ \bibinfo {author} {\bibfnamefont {X.-D.}\ \bibnamefont
  {Li}},\ }\href {\doibase 10.1088/0004-637X/800/2/98} {\bibfield  {journal}
  {\bibinfo  {journal} {Astrophys. J.}\ }\textbf {\bibinfo {volume} {800}},\
  \bibinfo {pages} {98} (\bibinfo {year} {2015})},\ \Eprint
  {http://arxiv.org/abs/1412.7245} {arXiv:1412.7245 [astro-ph.HE]} \BibitemShut
  {NoStop}%
\bibitem [{\citenamefont {Kuroda}\ \emph {et~al.}(2025)\citenamefont {Kuroda},
  \citenamefont {Kawaguchi},\ and\ \citenamefont {Shibata}}]{Kuroda:2025iyj}%
  \BibitemOpen
  \bibfield  {author} {\bibinfo {author} {\bibfnamefont {T.}~\bibnamefont
  {Kuroda}}, \bibinfo {author} {\bibfnamefont {K.}~\bibnamefont {Kawaguchi}}, \
  and\ \bibinfo {author} {\bibfnamefont {M.}~\bibnamefont {Shibata}},\
  }\href@noop {} {\  (\bibinfo {year} {2025})},\ \Eprint
  {http://arxiv.org/abs/2503.17082} {arXiv:2503.17082 [astro-ph.HE]}
  \BibitemShut {NoStop}%
\bibitem [{\citenamefont {Burrows}\ and\ \citenamefont
  {Lattimer}(1986)}]{Burrows:1986me}%
  \BibitemOpen
  \bibfield  {author} {\bibinfo {author} {\bibfnamefont {A.}~\bibnamefont
  {Burrows}}\ and\ \bibinfo {author} {\bibfnamefont {J.~M.}\ \bibnamefont
  {Lattimer}},\ }\href {\doibase 10.1086/164405} {\bibfield  {journal}
  {\bibinfo  {journal} {Astrophys. J.}\ }\textbf {\bibinfo {volume} {307}},\
  \bibinfo {pages} {178} (\bibinfo {year} {1986})}\BibitemShut {NoStop}%
\bibitem [{\citenamefont {Janka}(2012)}]{Janka:2012wk}%
  \BibitemOpen
  \bibfield  {author} {\bibinfo {author} {\bibfnamefont {H.-T.}\ \bibnamefont
  {Janka}},\ }\href {\doibase 10.1146/annurev-nucl-102711-094901} {\bibfield
  {journal} {\bibinfo  {journal} {Ann. Rev. Nucl. Part. Sci.}\ }\textbf
  {\bibinfo {volume} {62}},\ \bibinfo {pages} {407} (\bibinfo {year} {2012})},\
  \Eprint {http://arxiv.org/abs/1206.2503} {arXiv:1206.2503 [astro-ph.SR]}
  \BibitemShut {NoStop}%
\bibitem [{\citenamefont {{Wilson}}\ and\ \citenamefont
  {{Mayle}}(1989)}]{1989ASIB..216..731W}%
  \BibitemOpen
  \bibfield  {author} {\bibinfo {author} {\bibfnamefont {J.~R.}\ \bibnamefont
  {{Wilson}}}\ and\ \bibinfo {author} {\bibfnamefont {R.~W.}\ \bibnamefont
  {{Mayle}}},\ }in\ \href@noop {} {\emph {\bibinfo {booktitle} {The Nuclear
  Equation of State}}},\ \bibinfo {series} {NATO Advanced Study Institute (ASI)
  Series B}, Vol.\ \bibinfo {volume} {216},\ \bibinfo {editor} {edited by\
  \bibinfo {editor} {\bibfnamefont {W.}~\bibnamefont {{Greiner}}}\ and\
  \bibinfo {editor} {\bibfnamefont {H.}~\bibnamefont {{St{\"o}cker}}}}\
  (\bibinfo {year} {1989})\ p.\ \bibinfo {pages} {731}\BibitemShut {NoStop}%
\bibitem [{\citenamefont {Pons}\ \emph {et~al.}(1999)\citenamefont {Pons},
  \citenamefont {Reddy}, \citenamefont {Prakash}, \citenamefont {Lattimer},\
  and\ \citenamefont {Miralles}}]{Pons:1998mm}%
  \BibitemOpen
  \bibfield  {author} {\bibinfo {author} {\bibfnamefont {J.~A.}\ \bibnamefont
  {Pons}}, \bibinfo {author} {\bibfnamefont {S.}~\bibnamefont {Reddy}},
  \bibinfo {author} {\bibfnamefont {M.}~\bibnamefont {Prakash}}, \bibinfo
  {author} {\bibfnamefont {J.~M.}\ \bibnamefont {Lattimer}}, \ and\ \bibinfo
  {author} {\bibfnamefont {J.~A.}\ \bibnamefont {Miralles}},\ }\href {\doibase
  10.1086/306889} {\bibfield  {journal} {\bibinfo  {journal} {Astrophys. J.}\
  }\textbf {\bibinfo {volume} {513}},\ \bibinfo {pages} {780} (\bibinfo {year}
  {1999})},\ \Eprint {http://arxiv.org/abs/astro-ph/9807040}
  {arXiv:astro-ph/9807040} \BibitemShut {NoStop}%
\bibitem [{\citenamefont {Roberts}(2012)}]{Roberts:2012zza}%
  \BibitemOpen
  \bibfield  {author} {\bibinfo {author} {\bibfnamefont {L.~F.}\ \bibnamefont
  {Roberts}},\ }\href {\doibase 10.1088/0004-637X/755/2/126} {\bibfield
  {journal} {\bibinfo  {journal} {Astrophys. J.}\ }\textbf {\bibinfo {volume}
  {755}},\ \bibinfo {pages} {126} (\bibinfo {year} {2012})},\ \Eprint
  {http://arxiv.org/abs/1205.3228} {arXiv:1205.3228 [astro-ph.HE]} \BibitemShut
  {NoStop}%
\bibitem [{\citenamefont {Fischer}\ \emph {et~al.}(2011)\citenamefont
  {Fischer}, \citenamefont {Sagert}, \citenamefont {Pagliara}, \citenamefont
  {Hempel}, \citenamefont {Schaffner-Bielich}, \citenamefont {Rauscher},
  \citenamefont {Thielemann}, \citenamefont {Kappeli}, \citenamefont
  {Martinez-Pinedo},\ and\ \citenamefont {Liebendorfer}}]{Fischer:2010wp}%
  \BibitemOpen
  \bibfield  {author} {\bibinfo {author} {\bibfnamefont {T.}~\bibnamefont
  {Fischer}}, \bibinfo {author} {\bibfnamefont {I.}~\bibnamefont {Sagert}},
  \bibinfo {author} {\bibfnamefont {G.}~\bibnamefont {Pagliara}}, \bibinfo
  {author} {\bibfnamefont {M.}~\bibnamefont {Hempel}}, \bibinfo {author}
  {\bibfnamefont {J.}~\bibnamefont {Schaffner-Bielich}}, \bibinfo {author}
  {\bibfnamefont {T.}~\bibnamefont {Rauscher}}, \bibinfo {author}
  {\bibfnamefont {F.~K.}\ \bibnamefont {Thielemann}}, \bibinfo {author}
  {\bibfnamefont {R.}~\bibnamefont {Kappeli}}, \bibinfo {author} {\bibfnamefont
  {G.}~\bibnamefont {Martinez-Pinedo}}, \ and\ \bibinfo {author} {\bibfnamefont
  {M.}~\bibnamefont {Liebendorfer}},\ }\href {\doibase
  10.1088/0067-0049/194/2/39} {\bibfield  {journal} {\bibinfo  {journal}
  {Astrophys. J. Suppl.}\ }\textbf {\bibinfo {volume} {194}},\ \bibinfo {pages}
  {39} (\bibinfo {year} {2011})},\ \Eprint {http://arxiv.org/abs/1011.3409}
  {arXiv:1011.3409 [astro-ph.HE]} \BibitemShut {NoStop}%
\bibitem [{\citenamefont {Bastian}(2021)}]{Bastian:2020unt}%
  \BibitemOpen
  \bibfield  {author} {\bibinfo {author} {\bibfnamefont {N.-U.~F.}\
  \bibnamefont {Bastian}},\ }\href {\doibase 10.1103/PhysRevD.103.023001}
  {\bibfield  {journal} {\bibinfo  {journal} {Phys. Rev. D}\ }\textbf {\bibinfo
  {volume} {103}},\ \bibinfo {pages} {023001} (\bibinfo {year} {2021})},\
  \Eprint {http://arxiv.org/abs/2009.10846} {arXiv:2009.10846 [nucl-th]}
  \BibitemShut {NoStop}%
\bibitem [{\citenamefont {Blacker}\ \emph {et~al.}(2023)\citenamefont
  {Blacker}, \citenamefont {Bauswein},\ and\ \citenamefont
  {Typel}}]{Blacker:2023afl}%
  \BibitemOpen
  \bibfield  {author} {\bibinfo {author} {\bibfnamefont {S.}~\bibnamefont
  {Blacker}}, \bibinfo {author} {\bibfnamefont {A.}~\bibnamefont {Bauswein}}, \
  and\ \bibinfo {author} {\bibfnamefont {S.}~\bibnamefont {Typel}},\ }\href
  {\doibase 10.1103/PhysRevD.108.063032} {\bibfield  {journal} {\bibinfo
  {journal} {Phys. Rev. D}\ }\textbf {\bibinfo {volume} {108}},\ \bibinfo
  {pages} {063032} (\bibinfo {year} {2023})},\ \Eprint
  {http://arxiv.org/abs/2304.01971} {arXiv:2304.01971 [astro-ph.HE]}
  \BibitemShut {NoStop}%
\bibitem [{\citenamefont {Bauswein}\ \emph {et~al.}(2019)\citenamefont
  {Bauswein}, \citenamefont {Friedrich~Bastian}, \citenamefont {Blaschke},
  \citenamefont {Chatziioannou}, \citenamefont {Clark}, \citenamefont
  {Fischer}, \citenamefont {Janka}, \citenamefont {Just}, \citenamefont
  {Oertel},\ and\ \citenamefont {Stergioulas}}]{Bauswein:2019skm}%
  \BibitemOpen
  \bibfield  {author} {\bibinfo {author} {\bibfnamefont {A.}~\bibnamefont
  {Bauswein}}, \bibinfo {author} {\bibfnamefont {N.-U.}\ \bibnamefont
  {Friedrich~Bastian}}, \bibinfo {author} {\bibfnamefont {D.}~\bibnamefont
  {Blaschke}}, \bibinfo {author} {\bibfnamefont {K.}~\bibnamefont
  {Chatziioannou}}, \bibinfo {author} {\bibfnamefont {J.~A.}\ \bibnamefont
  {Clark}}, \bibinfo {author} {\bibfnamefont {T.}~\bibnamefont {Fischer}},
  \bibinfo {author} {\bibfnamefont {H.-T.}\ \bibnamefont {Janka}}, \bibinfo
  {author} {\bibfnamefont {O.}~\bibnamefont {Just}}, \bibinfo {author}
  {\bibfnamefont {M.}~\bibnamefont {Oertel}}, \ and\ \bibinfo {author}
  {\bibfnamefont {N.}~\bibnamefont {Stergioulas}},\ }\href {\doibase
  10.1063/1.5117803} {\bibfield  {journal} {\bibinfo  {journal} {AIP Conf.
  Proc.}\ }\textbf {\bibinfo {volume} {2127}},\ \bibinfo {pages} {020013}
  (\bibinfo {year} {2019})},\ \Eprint {http://arxiv.org/abs/1904.01306}
  {arXiv:1904.01306 [astro-ph.HE]} \BibitemShut {NoStop}%
\bibitem [{\citenamefont {Miao}\ \emph {et~al.}(2020)\citenamefont {Miao},
  \citenamefont {Li}, \citenamefont {Zhu},\ and\ \citenamefont
  {Han}}]{Miao:2020yjk}%
  \BibitemOpen
  \bibfield  {author} {\bibinfo {author} {\bibfnamefont {Z.}~\bibnamefont
  {Miao}}, \bibinfo {author} {\bibfnamefont {A.}~\bibnamefont {Li}}, \bibinfo
  {author} {\bibfnamefont {Z.}~\bibnamefont {Zhu}}, \ and\ \bibinfo {author}
  {\bibfnamefont {S.}~\bibnamefont {Han}},\ }\href {\doibase
  10.3847/1538-4357/abbd41} {\bibfield  {journal} {\bibinfo  {journal}
  {Astrophys. J.}\ }\textbf {\bibinfo {volume} {904}},\ \bibinfo {pages} {103}
  (\bibinfo {year} {2020})},\ \Eprint {http://arxiv.org/abs/2006.00839}
  {arXiv:2006.00839 [nucl-th]} \BibitemShut {NoStop}%
\bibitem [{\citenamefont {Blacker}\ \emph {et~al.}(2020)\citenamefont
  {Blacker}, \citenamefont {Bastian}, \citenamefont {Bauswein}, \citenamefont
  {Blaschke}, \citenamefont {Fischer}, \citenamefont {Oertel}, \citenamefont
  {Soultanis},\ and\ \citenamefont {Typel}}]{Blacker:2020nlq}%
  \BibitemOpen
  \bibfield  {author} {\bibinfo {author} {\bibfnamefont {S.}~\bibnamefont
  {Blacker}}, \bibinfo {author} {\bibfnamefont {N.-U.~F.}\ \bibnamefont
  {Bastian}}, \bibinfo {author} {\bibfnamefont {A.}~\bibnamefont {Bauswein}},
  \bibinfo {author} {\bibfnamefont {D.~B.}\ \bibnamefont {Blaschke}}, \bibinfo
  {author} {\bibfnamefont {T.}~\bibnamefont {Fischer}}, \bibinfo {author}
  {\bibfnamefont {M.}~\bibnamefont {Oertel}}, \bibinfo {author} {\bibfnamefont
  {T.}~\bibnamefont {Soultanis}}, \ and\ \bibinfo {author} {\bibfnamefont
  {S.}~\bibnamefont {Typel}},\ }\href {\doibase 10.1103/PhysRevD.102.123023}
  {\bibfield  {journal} {\bibinfo  {journal} {Phys. Rev. D}\ }\textbf {\bibinfo
  {volume} {102}},\ \bibinfo {pages} {123023} (\bibinfo {year} {2020})},\
  \Eprint {http://arxiv.org/abs/2006.03789} {arXiv:2006.03789 [astro-ph.HE]}
  \BibitemShut {NoStop}%
\bibitem [{\citenamefont {Zhu}\ \emph {et~al.}(2025)\citenamefont {Zhu},
  \citenamefont {Zha},\ and\ \citenamefont {Han}}]{Zhu:2025vmz}%
  \BibitemOpen
  \bibfield  {author} {\bibinfo {author} {\bibfnamefont {Z.}~\bibnamefont
  {Zhu}}, \bibinfo {author} {\bibfnamefont {S.}~\bibnamefont {Zha}}, \ and\
  \bibinfo {author} {\bibfnamefont {S.}~\bibnamefont {Han}},\ }\href {\doibase
  10.1103/f8tf-3dn5} {\bibfield  {journal} {\bibinfo  {journal} {Phys. Rev. D}\
  }\textbf {\bibinfo {volume} {112}},\ \bibinfo {pages} {103003} (\bibinfo
  {year} {2025})},\ \Eprint {http://arxiv.org/abs/2506.17569} {arXiv:2506.17569
  [astro-ph.HE]} \BibitemShut {NoStop}%
\bibitem [{\citenamefont {Sagert}\ \emph {et~al.}(2009)\citenamefont {Sagert},
  \citenamefont {Fischer}, \citenamefont {Hempel}, \citenamefont {Pagliara},
  \citenamefont {Schaffner-Bielich}, \citenamefont {Mezzacappa}, \citenamefont
  {Thielemann},\ and\ \citenamefont {Liebendorfer}}]{Sagert:2008ka}%
  \BibitemOpen
  \bibfield  {author} {\bibinfo {author} {\bibfnamefont {I.}~\bibnamefont
  {Sagert}}, \bibinfo {author} {\bibfnamefont {T.}~\bibnamefont {Fischer}},
  \bibinfo {author} {\bibfnamefont {M.}~\bibnamefont {Hempel}}, \bibinfo
  {author} {\bibfnamefont {G.}~\bibnamefont {Pagliara}}, \bibinfo {author}
  {\bibfnamefont {J.}~\bibnamefont {Schaffner-Bielich}}, \bibinfo {author}
  {\bibfnamefont {A.}~\bibnamefont {Mezzacappa}}, \bibinfo {author}
  {\bibfnamefont {F.~K.}\ \bibnamefont {Thielemann}}, \ and\ \bibinfo {author}
  {\bibfnamefont {M.}~\bibnamefont {Liebendorfer}},\ }\href {\doibase
  10.1103/PhysRevLett.102.081101} {\bibfield  {journal} {\bibinfo  {journal}
  {Phys. Rev. Lett.}\ }\textbf {\bibinfo {volume} {102}},\ \bibinfo {pages}
  {081101} (\bibinfo {year} {2009})},\ \Eprint {http://arxiv.org/abs/0809.4225}
  {arXiv:0809.4225 [astro-ph]} \BibitemShut {NoStop}%
\bibitem [{\citenamefont {Fischer}\ \emph {et~al.}(2018)\citenamefont
  {Fischer}, \citenamefont {Bastian}, \citenamefont {Wu}, \citenamefont
  {Baklanov}, \citenamefont {Sorokina}, \citenamefont {Blinnikov},
  \citenamefont {Typel}, \citenamefont {Kl\"ahn},\ and\ \citenamefont
  {Blaschke}}]{Fischer:2017lag}%
  \BibitemOpen
  \bibfield  {author} {\bibinfo {author} {\bibfnamefont {T.}~\bibnamefont
  {Fischer}}, \bibinfo {author} {\bibfnamefont {N.-U.~F.}\ \bibnamefont
  {Bastian}}, \bibinfo {author} {\bibfnamefont {M.-R.}\ \bibnamefont {Wu}},
  \bibinfo {author} {\bibfnamefont {P.}~\bibnamefont {Baklanov}}, \bibinfo
  {author} {\bibfnamefont {E.}~\bibnamefont {Sorokina}}, \bibinfo {author}
  {\bibfnamefont {S.}~\bibnamefont {Blinnikov}}, \bibinfo {author}
  {\bibfnamefont {S.}~\bibnamefont {Typel}}, \bibinfo {author} {\bibfnamefont
  {T.}~\bibnamefont {Kl\"ahn}}, \ and\ \bibinfo {author} {\bibfnamefont
  {D.~B.}\ \bibnamefont {Blaschke}},\ }\href {\doibase
  10.1038/s41550-018-0583-0} {\bibfield  {journal} {\bibinfo  {journal} {Nature
  Astron.}\ }\textbf {\bibinfo {volume} {2}},\ \bibinfo {pages} {980} (\bibinfo
  {year} {2018})},\ \Eprint {http://arxiv.org/abs/1712.08788} {arXiv:1712.08788
  [astro-ph.HE]} \BibitemShut {NoStop}%
\bibitem [{\citenamefont {Zha}\ \emph {et~al.}(2020)\citenamefont {Zha},
  \citenamefont {O'Connor}, \citenamefont {Chu}, \citenamefont {Lin},\ and\
  \citenamefont {Couch}}]{Zha:2020gjw}%
  \BibitemOpen
  \bibfield  {author} {\bibinfo {author} {\bibfnamefont {S.}~\bibnamefont
  {Zha}}, \bibinfo {author} {\bibfnamefont {E.~P.}\ \bibnamefont {O'Connor}},
  \bibinfo {author} {\bibfnamefont {M.-C.}\ \bibnamefont {Chu}}, \bibinfo
  {author} {\bibfnamefont {L.-M.}\ \bibnamefont {Lin}}, \ and\ \bibinfo
  {author} {\bibfnamefont {S.~M.}\ \bibnamefont {Couch}},\ }\href {\doibase
  10.1103/PhysRevLett.127.219901} {\bibfield  {journal} {\bibinfo  {journal}
  {Phys. Rev. Lett.}\ }\textbf {\bibinfo {volume} {125}},\ \bibinfo {pages}
  {051102} (\bibinfo {year} {2020})},\ \bibinfo {note} {[Erratum:
  Phys.Rev.Lett. 127, 219901 (2021)]},\ \Eprint
  {http://arxiv.org/abs/2007.04716} {arXiv:2007.04716 [astro-ph.HE]}
  \BibitemShut {NoStop}%
\bibitem [{\citenamefont {Zha}\ \emph {et~al.}(2021)\citenamefont {Zha},
  \citenamefont {O'Connor},\ and\ \citenamefont
  {da~Silva~Schneider}}]{Zha:2021fbi}%
  \BibitemOpen
  \bibfield  {author} {\bibinfo {author} {\bibfnamefont {S.}~\bibnamefont
  {Zha}}, \bibinfo {author} {\bibfnamefont {E.~P.}\ \bibnamefont {O'Connor}}, \
  and\ \bibinfo {author} {\bibfnamefont {A.}~\bibnamefont
  {da~Silva~Schneider}},\ }\href {\doibase 10.3847/1538-4357/abec4c} {\bibfield
   {journal} {\bibinfo  {journal} {Astrophys. J.}\ }\textbf {\bibinfo {volume}
  {911}},\ \bibinfo {pages} {74} (\bibinfo {year} {2021})},\ \Eprint
  {http://arxiv.org/abs/2103.02268} {arXiv:2103.02268 [astro-ph.HE]}
  \BibitemShut {NoStop}%
\bibitem [{\citenamefont {Fischer}(2021)}]{Fischer:2021tvv}%
  \BibitemOpen
  \bibfield  {author} {\bibinfo {author} {\bibfnamefont {T.}~\bibnamefont
  {Fischer}},\ }\href {\doibase 10.1140/epja/s10050-021-00571-z} {\bibfield
  {journal} {\bibinfo  {journal} {Eur. Phys. J. A}\ }\textbf {\bibinfo {volume}
  {57}},\ \bibinfo {pages} {270} (\bibinfo {year} {2021})},\ \Eprint
  {http://arxiv.org/abs/2108.00196} {arXiv:2108.00196 [astro-ph.HE]}
  \BibitemShut {NoStop}%
\bibitem [{\citenamefont {Kuroda}\ \emph {et~al.}(2022)\citenamefont {Kuroda},
  \citenamefont {Fischer}, \citenamefont {Takiwaki},\ and\ \citenamefont
  {Kotake}}]{Kuroda:2021eiv}%
  \BibitemOpen
  \bibfield  {author} {\bibinfo {author} {\bibfnamefont {T.}~\bibnamefont
  {Kuroda}}, \bibinfo {author} {\bibfnamefont {T.}~\bibnamefont {Fischer}},
  \bibinfo {author} {\bibfnamefont {T.}~\bibnamefont {Takiwaki}}, \ and\
  \bibinfo {author} {\bibfnamefont {K.}~\bibnamefont {Kotake}},\ }\href
  {\doibase 10.3847/1538-4357/ac31a8} {\bibfield  {journal} {\bibinfo
  {journal} {Astrophys. J.}\ }\textbf {\bibinfo {volume} {924}},\ \bibinfo
  {pages} {38} (\bibinfo {year} {2022})},\ \Eprint
  {http://arxiv.org/abs/2109.01508} {arXiv:2109.01508 [astro-ph.HE]}
  \BibitemShut {NoStop}%
\bibitem [{\citenamefont {Jakobus}\ \emph {et~al.}(2022)\citenamefont
  {Jakobus}, \citenamefont {Mueller}, \citenamefont {Heger}, \citenamefont
  {Motornenko}, \citenamefont {Steinheimer},\ and\ \citenamefont
  {Stoecker}}]{Jakobus:2022ucs}%
  \BibitemOpen
  \bibfield  {author} {\bibinfo {author} {\bibfnamefont {P.}~\bibnamefont
  {Jakobus}}, \bibinfo {author} {\bibfnamefont {B.}~\bibnamefont {Mueller}},
  \bibinfo {author} {\bibfnamefont {A.}~\bibnamefont {Heger}}, \bibinfo
  {author} {\bibfnamefont {A.}~\bibnamefont {Motornenko}}, \bibinfo {author}
  {\bibfnamefont {J.}~\bibnamefont {Steinheimer}}, \ and\ \bibinfo {author}
  {\bibfnamefont {H.}~\bibnamefont {Stoecker}},\ }\href {\doibase
  10.1093/mnras/stac2352} {\bibfield  {journal} {\bibinfo  {journal} {Mon. Not.
  Roy. Astron. Soc.}\ }\textbf {\bibinfo {volume} {516}},\ \bibinfo {pages}
  {2554} (\bibinfo {year} {2022})},\ \Eprint {http://arxiv.org/abs/2204.10397}
  {arXiv:2204.10397 [astro-ph.HE]} \BibitemShut {NoStop}%
\bibitem [{\citenamefont {Janka}\ and\ \citenamefont
  {Bauswein}(2023)}]{Janka:2022krt}%
  \BibitemOpen
  \bibfield  {author} {\bibinfo {author} {\bibfnamefont {H.-T.}\ \bibnamefont
  {Janka}}\ and\ \bibinfo {author} {\bibfnamefont {A.}~\bibnamefont
  {Bauswein}},\ }\enquote {\bibinfo {title} {{Dynamics and Equation of State
  Dependencies of Relevance for Nucleosynthesis in Supernovae and Neutron Star
  Mergers}},}\ in\ \href {\doibase 10.1007/978-981-15-8818-1_93-1} {\emph
  {\bibinfo {booktitle} {{Handbook of Nuclear Physics}}}},\ \bibinfo {editor}
  {edited by\ \bibinfo {editor} {\bibfnamefont {I.}~\bibnamefont {Tanihata}},
  \bibinfo {editor} {\bibfnamefont {H.}~\bibnamefont {Toki}}, \ and\ \bibinfo
  {editor} {\bibfnamefont {T.}~\bibnamefont {Kajino}}}\ (\bibinfo {year}
  {2023})\ pp.\ \bibinfo {pages} {1--98},\ \Eprint
  {http://arxiv.org/abs/2212.07498} {arXiv:2212.07498 [astro-ph.HE]}
  \BibitemShut {NoStop}%
\bibitem [{\citenamefont {Bauswein}\ \emph {et~al.}(2022)\citenamefont
  {Bauswein}, \citenamefont {Blaschke},\ and\ \citenamefont
  {Fischer}}]{Bauswein:2022vtq}%
  \BibitemOpen
  \bibfield  {author} {\bibinfo {author} {\bibfnamefont {A.}~\bibnamefont
  {Bauswein}}, \bibinfo {author} {\bibfnamefont {D.}~\bibnamefont {Blaschke}},
  \ and\ \bibinfo {author} {\bibfnamefont {T.}~\bibnamefont {Fischer}},\
  }\enquote {\bibinfo {title} {{Effects of a strong phase transition on
  supernova explosions, compact stars and their mergers}},}\ \ (\bibinfo {year}
  {2022})\ \Eprint {http://arxiv.org/abs/2203.17188} {arXiv:2203.17188
  [nucl-th]} \BibitemShut {NoStop}%
\bibitem [{\citenamefont {Largani}\ \emph {et~al.}(2024)\citenamefont
  {Largani}, \citenamefont {Fischer},\ and\ \citenamefont
  {Bastian}}]{Largani:2023oyk}%
  \BibitemOpen
  \bibfield  {author} {\bibinfo {author} {\bibfnamefont {N.~K.}\ \bibnamefont
  {Largani}}, \bibinfo {author} {\bibfnamefont {T.}~\bibnamefont {Fischer}}, \
  and\ \bibinfo {author} {\bibfnamefont {N.~U.~F.}\ \bibnamefont {Bastian}},\
  }\href {\doibase 10.3847/1538-4357/ad24f2} {\bibfield  {journal} {\bibinfo
  {journal} {Astrophys. J.}\ }\textbf {\bibinfo {volume} {964}},\ \bibinfo
  {pages} {143} (\bibinfo {year} {2024})},\ \Eprint
  {http://arxiv.org/abs/2304.12316} {arXiv:2304.12316 [astro-ph.HE]}
  \BibitemShut {NoStop}%
\bibitem [{\citenamefont {Fischer}\ \emph {et~al.}(2020)\citenamefont
  {Fischer}, \citenamefont {Wu}, \citenamefont {Wehmeyer}, \citenamefont
  {Bastian}, \citenamefont {Mart\'\i{}nez-Pinedo},\ and\ \citenamefont
  {Thielemann}}]{Fischer:2020xjl}%
  \BibitemOpen
  \bibfield  {author} {\bibinfo {author} {\bibfnamefont {T.}~\bibnamefont
  {Fischer}}, \bibinfo {author} {\bibfnamefont {M.-R.}\ \bibnamefont {Wu}},
  \bibinfo {author} {\bibfnamefont {B.}~\bibnamefont {Wehmeyer}}, \bibinfo
  {author} {\bibfnamefont {N.-U.~F.}\ \bibnamefont {Bastian}}, \bibinfo
  {author} {\bibfnamefont {G.}~\bibnamefont {Mart\'\i{}nez-Pinedo}}, \ and\
  \bibinfo {author} {\bibfnamefont {F.-K.}\ \bibnamefont {Thielemann}},\ }\href
  {\doibase 10.3847/1538-4357/ab86b0} {\bibfield  {journal} {\bibinfo
  {journal} {Astrophys. J.}\ }\textbf {\bibinfo {volume} {894}},\ \bibinfo
  {pages} {9} (\bibinfo {year} {2020})},\ \Eprint
  {http://arxiv.org/abs/2003.00972} {arXiv:2003.00972 [astro-ph.HE]}
  \BibitemShut {NoStop}%
\bibitem [{\citenamefont {Huang}\ \emph {et~al.}(2025)\citenamefont {Huang},
  \citenamefont {Zha}, \citenamefont {Chu}, \citenamefont {O'Connor},\ and\
  \citenamefont {Chen}}]{Huang:2024xff}%
  \BibitemOpen
  \bibfield  {author} {\bibinfo {author} {\bibfnamefont {X.-R.}\ \bibnamefont
  {Huang}}, \bibinfo {author} {\bibfnamefont {S.}~\bibnamefont {Zha}}, \bibinfo
  {author} {\bibfnamefont {M.-c.}\ \bibnamefont {Chu}}, \bibinfo {author}
  {\bibfnamefont {E.~P.}\ \bibnamefont {O'Connor}}, \ and\ \bibinfo {author}
  {\bibfnamefont {L.-W.}\ \bibnamefont {Chen}},\ }\href {\doibase
  10.3847/1538-4357/ada146} {\bibfield  {journal} {\bibinfo  {journal}
  {Astrophys. J.}\ }\textbf {\bibinfo {volume} {979}},\ \bibinfo {pages} {151}
  (\bibinfo {year} {2025})},\ \Eprint {http://arxiv.org/abs/2409.16189}
  {arXiv:2409.16189 [astro-ph.HE]} \BibitemShut {NoStop}%
\bibitem [{\citenamefont {Bluhm}\ \emph {et~al.}(2025)\citenamefont {Bluhm},
  \citenamefont {Fujimoto}, \citenamefont {McLerran},\ and\ \citenamefont
  {Nahrgang}}]{Bluhm:2024uhj}%
  \BibitemOpen
  \bibfield  {author} {\bibinfo {author} {\bibfnamefont {M.}~\bibnamefont
  {Bluhm}}, \bibinfo {author} {\bibfnamefont {Y.}~\bibnamefont {Fujimoto}},
  \bibinfo {author} {\bibfnamefont {L.}~\bibnamefont {McLerran}}, \ and\
  \bibinfo {author} {\bibfnamefont {M.}~\bibnamefont {Nahrgang}},\ }\href
  {\doibase 10.1103/PhysRevC.111.044914} {\bibfield  {journal} {\bibinfo
  {journal} {Phys. Rev. C}\ }\textbf {\bibinfo {volume} {111}},\ \bibinfo
  {pages} {044914} (\bibinfo {year} {2025})},\ \Eprint
  {http://arxiv.org/abs/2409.12088} {arXiv:2409.12088 [nucl-th]} \BibitemShut
  {NoStop}%
\bibitem [{\citenamefont {Clarke}\ \emph {et~al.}(2025)\citenamefont {Clarke},
  \citenamefont {Dimopoulos}, \citenamefont {Di~Renzo}, \citenamefont
  {Goswami}, \citenamefont {Schmidt}, \citenamefont {Singh},\ and\
  \citenamefont {Zambello}}]{Clarke:2024ugt}%
  \BibitemOpen
  \bibfield  {author} {\bibinfo {author} {\bibfnamefont {D.~A.}\ \bibnamefont
  {Clarke}}, \bibinfo {author} {\bibfnamefont {P.}~\bibnamefont {Dimopoulos}},
  \bibinfo {author} {\bibfnamefont {F.}~\bibnamefont {Di~Renzo}}, \bibinfo
  {author} {\bibfnamefont {J.}~\bibnamefont {Goswami}}, \bibinfo {author}
  {\bibfnamefont {C.}~\bibnamefont {Schmidt}}, \bibinfo {author} {\bibfnamefont
  {S.}~\bibnamefont {Singh}}, \ and\ \bibinfo {author} {\bibfnamefont
  {K.}~\bibnamefont {Zambello}},\ }\href {\doibase 10.1103/y6kg-ry8x}
  {\bibfield  {journal} {\bibinfo  {journal} {Phys. Rev. D}\ }\textbf {\bibinfo
  {volume} {112}},\ \bibinfo {pages} {L091504} (\bibinfo {year} {2025})},\
  \Eprint {http://arxiv.org/abs/2405.10196} {arXiv:2405.10196 [hep-lat]}
  \BibitemShut {NoStop}%
\bibitem [{\citenamefont {Shah}\ \emph {et~al.}(2024)\citenamefont {Shah},
  \citenamefont {Hippert}, \citenamefont {Noronha}, \citenamefont {Ratti},\
  and\ \citenamefont {Vovchenko}}]{Shah:2024img}%
  \BibitemOpen
  \bibfield  {author} {\bibinfo {author} {\bibfnamefont {H.}~\bibnamefont
  {Shah}}, \bibinfo {author} {\bibfnamefont {M.}~\bibnamefont {Hippert}},
  \bibinfo {author} {\bibfnamefont {J.}~\bibnamefont {Noronha}}, \bibinfo
  {author} {\bibfnamefont {C.}~\bibnamefont {Ratti}}, \ and\ \bibinfo {author}
  {\bibfnamefont {V.}~\bibnamefont {Vovchenko}},\ }\href@noop {} {\  (\bibinfo
  {year} {2024})},\ \Eprint {http://arxiv.org/abs/2410.16206} {arXiv:2410.16206
  [hep-ph]} \BibitemShut {NoStop}%
\bibitem [{\citenamefont {Chatziioannou}\ and\ \citenamefont
  {Han}(2020)}]{Chatziioannou:2019yko}%
  \BibitemOpen
  \bibfield  {author} {\bibinfo {author} {\bibfnamefont {K.}~\bibnamefont
  {Chatziioannou}}\ and\ \bibinfo {author} {\bibfnamefont {S.}~\bibnamefont
  {Han}},\ }\href {\doibase 10.1103/PhysRevD.101.044019} {\bibfield  {journal}
  {\bibinfo  {journal} {Phys. Rev. D}\ }\textbf {\bibinfo {volume} {101}},\
  \bibinfo {pages} {044019} (\bibinfo {year} {2020})},\ \Eprint
  {http://arxiv.org/abs/1911.07091} {arXiv:1911.07091 [gr-qc]} \BibitemShut
  {NoStop}%
\bibitem [{\citenamefont {Pang}\ \emph {et~al.}(2020)\citenamefont {Pang},
  \citenamefont {Dietrich}, \citenamefont {Tews},\ and\ \citenamefont {Van
  Den~Broeck}}]{Pang:2020ilf}%
  \BibitemOpen
  \bibfield  {author} {\bibinfo {author} {\bibfnamefont {P.~T.~H.}\
  \bibnamefont {Pang}}, \bibinfo {author} {\bibfnamefont {T.}~\bibnamefont
  {Dietrich}}, \bibinfo {author} {\bibfnamefont {I.}~\bibnamefont {Tews}}, \
  and\ \bibinfo {author} {\bibfnamefont {C.}~\bibnamefont {Van Den~Broeck}},\
  }\href {\doibase 10.1103/PhysRevResearch.2.033514} {\bibfield  {journal}
  {\bibinfo  {journal} {Phys. Rev. Res.}\ }\textbf {\bibinfo {volume} {2}},\
  \bibinfo {pages} {033514} (\bibinfo {year} {2020})},\ \Eprint
  {http://arxiv.org/abs/2006.14936} {arXiv:2006.14936 [astro-ph.HE]}
  \BibitemShut {NoStop}%
\bibitem [{\citenamefont {Fujimoto}\ \emph {et~al.}(2023)\citenamefont
  {Fujimoto}, \citenamefont {Fukushima}, \citenamefont {Hotokezaka},\ and\
  \citenamefont {Kyutoku}}]{Fujimoto:2022xhv}%
  \BibitemOpen
  \bibfield  {author} {\bibinfo {author} {\bibfnamefont {Y.}~\bibnamefont
  {Fujimoto}}, \bibinfo {author} {\bibfnamefont {K.}~\bibnamefont {Fukushima}},
  \bibinfo {author} {\bibfnamefont {K.}~\bibnamefont {Hotokezaka}}, \ and\
  \bibinfo {author} {\bibfnamefont {K.}~\bibnamefont {Kyutoku}},\ }\href
  {\doibase 10.1103/PhysRevLett.130.091404} {\bibfield  {journal} {\bibinfo
  {journal} {Phys. Rev. Lett.}\ }\textbf {\bibinfo {volume} {130}},\ \bibinfo
  {pages} {091404} (\bibinfo {year} {2023})},\ \Eprint
  {http://arxiv.org/abs/2205.03882} {arXiv:2205.03882 [astro-ph.HE]}
  \BibitemShut {NoStop}%
\bibitem [{\citenamefont {Li}\ \emph {et~al.}(2025)\citenamefont {Li},
  \citenamefont {Sedrakian},\ and\ \citenamefont {Alford}}]{Li:2024sft}%
  \BibitemOpen
  \bibfield  {author} {\bibinfo {author} {\bibfnamefont {J.~J.}\ \bibnamefont
  {Li}}, \bibinfo {author} {\bibfnamefont {A.}~\bibnamefont {Sedrakian}}, \
  and\ \bibinfo {author} {\bibfnamefont {M.}~\bibnamefont {Alford}},\ }\href
  {\doibase 10.1088/1475-7516/2025/02/002} {\bibfield  {journal} {\bibinfo
  {journal} {JCAP}\ }\textbf {\bibinfo {volume} {02}},\ \bibinfo {pages} {002}
  (\bibinfo {year} {2025})},\ \Eprint {http://arxiv.org/abs/2409.05322}
  {arXiv:2409.05322 [astro-ph.HE]} \BibitemShut {NoStop}%
\bibitem [{\citenamefont {Tang}\ \emph {et~al.}(2025)\citenamefont {Tang},
  \citenamefont {Huang},\ and\ \citenamefont {Fan}}]{Tang:2025xib}%
  \BibitemOpen
  \bibfield  {author} {\bibinfo {author} {\bibfnamefont {S.-P.}\ \bibnamefont
  {Tang}}, \bibinfo {author} {\bibfnamefont {Y.-J.}\ \bibnamefont {Huang}}, \
  and\ \bibinfo {author} {\bibfnamefont {Y.-Z.}\ \bibnamefont {Fan}},\ }\href
  {\doibase 10.1103/bmsk-8n85} {\bibfield  {journal} {\bibinfo  {journal}
  {Phys. Rev. D}\ }\textbf {\bibinfo {volume} {112}},\ \bibinfo {pages}
  {083009} (\bibinfo {year} {2025})},\ \Eprint
  {http://arxiv.org/abs/2507.10025} {arXiv:2507.10025 [astro-ph.HE]}
  \BibitemShut {NoStop}%
\bibitem [{\citenamefont {Lin}\ \emph {et~al.}(2006)\citenamefont {Lin},
  \citenamefont {Cheng}, \citenamefont {Chu},\ and\ \citenamefont
  {Suen}}]{Lin:2005zda}%
  \BibitemOpen
  \bibfield  {author} {\bibinfo {author} {\bibfnamefont {L.-M.}\ \bibnamefont
  {Lin}}, \bibinfo {author} {\bibfnamefont {K.~S.}\ \bibnamefont {Cheng}},
  \bibinfo {author} {\bibfnamefont {M.~C.}\ \bibnamefont {Chu}}, \ and\
  \bibinfo {author} {\bibfnamefont {W.~M.}\ \bibnamefont {Suen}},\ }\href
  {\doibase 10.1086/499202} {\bibfield  {journal} {\bibinfo  {journal}
  {Astrophys. J.}\ }\textbf {\bibinfo {volume} {639}},\ \bibinfo {pages} {382}
  (\bibinfo {year} {2006})},\ \Eprint {http://arxiv.org/abs/astro-ph/0509447}
  {arXiv:astro-ph/0509447} \BibitemShut {NoStop}%
\bibitem [{\citenamefont {Mallick}\ \emph {et~al.}(2021)\citenamefont
  {Mallick}, \citenamefont {Singh},\ and\ \citenamefont
  {Prasad}}]{Mallick:2020bdc}%
  \BibitemOpen
  \bibfield  {author} {\bibinfo {author} {\bibfnamefont {R.}~\bibnamefont
  {Mallick}}, \bibinfo {author} {\bibfnamefont {S.}~\bibnamefont {Singh}}, \
  and\ \bibinfo {author} {\bibfnamefont {R.}~\bibnamefont {Prasad}},\ }\href
  {\doibase 10.1093/mnras/stab2217} {\bibfield  {journal} {\bibinfo  {journal}
  {Mon. Not. Roy. Astron. Soc.}\ }\textbf {\bibinfo {volume} {507}},\ \bibinfo
  {pages} {1318} (\bibinfo {year} {2021})},\ \Eprint
  {http://arxiv.org/abs/2003.00693} {arXiv:2003.00693 [astro-ph.HE]}
  \BibitemShut {NoStop}%
\bibitem [{\citenamefont {Typel}\ \emph {et~al.}(2022)\citenamefont {Typel},
  \citenamefont {Oertel}, \citenamefont {Kl\"ahn}, \citenamefont {Chatterjee},
  \citenamefont {Dexheimer}, \citenamefont {Ishizuka}, \citenamefont {Mancini},
  \citenamefont {Novak}, \citenamefont {Pais}, \citenamefont {Providencia},
  \citenamefont {Raduta}, \citenamefont {Servillat}, \citenamefont {Tolos},\
  and\ \citenamefont {the CompOSE~Collaboration}}]{compose}%
  \BibitemOpen
  \bibfield  {author} {\bibinfo {author} {\bibfnamefont {S.}~\bibnamefont
  {Typel}}, \bibinfo {author} {\bibfnamefont {M.}~\bibnamefont {Oertel}},
  \bibinfo {author} {\bibfnamefont {T.}~\bibnamefont {Kl\"ahn}}, \bibinfo
  {author} {\bibfnamefont {D.}~\bibnamefont {Chatterjee}}, \bibinfo {author}
  {\bibfnamefont {V.}~\bibnamefont {Dexheimer}}, \bibinfo {author}
  {\bibfnamefont {C.}~\bibnamefont {Ishizuka}}, \bibinfo {author}
  {\bibfnamefont {M.}~\bibnamefont {Mancini}}, \bibinfo {author} {\bibfnamefont
  {J.}~\bibnamefont {Novak}}, \bibinfo {author} {\bibfnamefont
  {H.}~\bibnamefont {Pais}}, \bibinfo {author} {\bibfnamefont {C.}~\bibnamefont
  {Providencia}}, \bibinfo {author} {\bibfnamefont {A.}~\bibnamefont {Raduta}},
  \bibinfo {author} {\bibfnamefont {M.}~\bibnamefont {Servillat}}, \bibinfo
  {author} {\bibfnamefont {L.}~\bibnamefont {Tolos}}, \ and\ \bibinfo {author}
  {\bibnamefont {the CompOSE~Collaboration}},\ }\href@noop {} {\enquote
  {\bibinfo {title} {{CompOSE Reference Manual: CompStar Online Supernovae
  Equations of State}},}\ }\bibinfo {howpublished}
  {\url{https://compose.obspm.fr}} (\bibinfo {year} {2022}),\ \bibinfo {note}
  {[Online; accessed 2025-06-23]}\BibitemShut {NoStop}%
\bibitem [{\citenamefont {Antoniadis}\ \emph {et~al.}(2013)\citenamefont
  {Antoniadis} \emph {et~al.}}]{Antoniadis:2013pzd}%
  \BibitemOpen
  \bibfield  {author} {\bibinfo {author} {\bibfnamefont {J.}~\bibnamefont
  {Antoniadis}} \emph {et~al.},\ }\href {\doibase 10.1126/science.1233232}
  {\bibfield  {journal} {\bibinfo  {journal} {Science}\ }\textbf {\bibinfo
  {volume} {340}},\ \bibinfo {pages} {6131} (\bibinfo {year} {2013})},\ \Eprint
  {http://arxiv.org/abs/1304.6875} {arXiv:1304.6875 [astro-ph.HE]} \BibitemShut
  {NoStop}%
\bibitem [{\citenamefont {Fonseca}\ \emph {et~al.}(2021)\citenamefont {Fonseca}
  \emph {et~al.}}]{Fonseca:2021wxt}%
  \BibitemOpen
  \bibfield  {author} {\bibinfo {author} {\bibfnamefont {E.}~\bibnamefont
  {Fonseca}} \emph {et~al.},\ }\href {\doibase 10.3847/2041-8213/ac03b8}
  {\bibfield  {journal} {\bibinfo  {journal} {Astrophys. J. Lett.}\ }\textbf
  {\bibinfo {volume} {915}},\ \bibinfo {pages} {L12} (\bibinfo {year}
  {2021})},\ \Eprint {http://arxiv.org/abs/2104.00880} {arXiv:2104.00880
  [astro-ph.HE]} \BibitemShut {NoStop}%
\bibitem [{\citenamefont {Stergioulas}\ and\ \citenamefont
  {Friedman}(1995)}]{Stergioulas:1994ea}%
  \BibitemOpen
  \bibfield  {author} {\bibinfo {author} {\bibfnamefont {N.}~\bibnamefont
  {Stergioulas}}\ and\ \bibinfo {author} {\bibfnamefont {J.~L.}\ \bibnamefont
  {Friedman}},\ }\href {\doibase 10.1086/175605} {\bibfield  {journal}
  {\bibinfo  {journal} {Astrophys. J.}\ }\textbf {\bibinfo {volume} {444}},\
  \bibinfo {pages} {306} (\bibinfo {year} {1995})},\ \Eprint
  {http://arxiv.org/abs/astro-ph/9411032} {arXiv:astro-ph/9411032} \BibitemShut
  {NoStop}%
\bibitem [{\citenamefont {Dessart}\ \emph {et~al.}(2007)\citenamefont
  {Dessart}, \citenamefont {Burrows}, \citenamefont {Livne},\ and\
  \citenamefont {Ott}}]{Dessart:2007kh}%
  \BibitemOpen
  \bibfield  {author} {\bibinfo {author} {\bibfnamefont {L.}~\bibnamefont
  {Dessart}}, \bibinfo {author} {\bibfnamefont {A.}~\bibnamefont {Burrows}},
  \bibinfo {author} {\bibfnamefont {E.}~\bibnamefont {Livne}}, \ and\ \bibinfo
  {author} {\bibfnamefont {C.}~\bibnamefont {Ott}},\ }\href {\doibase
  10.1086/521701} {\bibfield  {journal} {\bibinfo  {journal} {Astrophys. J.}\
  }\textbf {\bibinfo {volume} {669}},\ \bibinfo {pages} {585} (\bibinfo {year}
  {2007})},\ \Eprint {http://arxiv.org/abs/0705.3678} {arXiv:0705.3678
  [astro-ph]} \BibitemShut {NoStop}%
\bibitem [{\citenamefont {Cheong}\ \emph {et~al.}(2020)\citenamefont {Cheong},
  \citenamefont {Lin},\ and\ \citenamefont {Li}}]{Cheong:2020dxe}%
  \BibitemOpen
  \bibfield  {author} {\bibinfo {author} {\bibfnamefont {P.~C.-K.}\
  \bibnamefont {Cheong}}, \bibinfo {author} {\bibfnamefont {L.-M.}\
  \bibnamefont {Lin}}, \ and\ \bibinfo {author} {\bibfnamefont {T.~G.-F.}\
  \bibnamefont {Li}},\ }\href {\doibase 10.1088/1361-6382/ab8e9c} {\  (\bibinfo
  {year} {2020}),\ 10.1088/1361-6382/ab8e9c},\ \Eprint
  {http://arxiv.org/abs/2001.05723} {arXiv:2001.05723 [gr-qc]} \BibitemShut
  {NoStop}%
\bibitem [{\citenamefont {Cheong}\ \emph {et~al.}(2021)\citenamefont {Cheong},
  \citenamefont {Lam}, \citenamefont {Ng},\ and\ \citenamefont
  {Li}}]{Cheong:2020kpv}%
  \BibitemOpen
  \bibfield  {author} {\bibinfo {author} {\bibfnamefont {P.~C.-K.}\
  \bibnamefont {Cheong}}, \bibinfo {author} {\bibfnamefont {A.~T.-L.}\
  \bibnamefont {Lam}}, \bibinfo {author} {\bibfnamefont {H.~H.-Y.}\
  \bibnamefont {Ng}}, \ and\ \bibinfo {author} {\bibfnamefont {T.~G.~F.}\
  \bibnamefont {Li}},\ }\href {\doibase 10.1093/mnras/stab2606} {\bibfield
  {journal} {\bibinfo  {journal} {Mon. Not. Roy. Astron. Soc.}\ }\textbf
  {\bibinfo {volume} {508}},\ \bibinfo {pages} {2279} (\bibinfo {year}
  {2021})},\ \Eprint {http://arxiv.org/abs/2012.07322} {arXiv:2012.07322
  [astro-ph.IM]} \BibitemShut {NoStop}%
\bibitem [{\citenamefont {Cheong}\ \emph {et~al.}(2022)\citenamefont {Cheong},
  \citenamefont {Pong}, \citenamefont {Yip},\ and\ \citenamefont
  {Li}}]{Cheong:2022vev}%
  \BibitemOpen
  \bibfield  {author} {\bibinfo {author} {\bibfnamefont {P.~C.-K.}\
  \bibnamefont {Cheong}}, \bibinfo {author} {\bibfnamefont {D.~Y.~T.}\
  \bibnamefont {Pong}}, \bibinfo {author} {\bibfnamefont {A.~K.~L.}\
  \bibnamefont {Yip}}, \ and\ \bibinfo {author} {\bibfnamefont {T.~G.~F.}\
  \bibnamefont {Li}},\ }\href {\doibase 10.3847/1538-4365/ac6cec} {\bibfield
  {journal} {\bibinfo  {journal} {Astrophys. J. Suppl.}\ }\textbf {\bibinfo
  {volume} {261}} (\bibinfo {year} {2022}),\
  10.3847/1538-4365/ac6cec}\BibitemShut {NoStop}%
\bibitem [{\citenamefont {Cheong}\ \emph {et~al.}(2023)\citenamefont {Cheong},
  \citenamefont {Ng}, \citenamefont {Lam},\ and\ \citenamefont
  {Li}}]{Cheong:2023fgh}%
  \BibitemOpen
  \bibfield  {author} {\bibinfo {author} {\bibfnamefont {P.~C.-K.}\
  \bibnamefont {Cheong}}, \bibinfo {author} {\bibfnamefont {H.~H.-Y.}\
  \bibnamefont {Ng}}, \bibinfo {author} {\bibfnamefont {A.~T.-L.}\ \bibnamefont
  {Lam}}, \ and\ \bibinfo {author} {\bibfnamefont {T.~G.~F.}\ \bibnamefont
  {Li}},\ }\href {\doibase 10.3847/1538-4365/acd931} {\bibfield  {journal}
  {\bibinfo  {journal} {Astrophys. J. Suppl.}\ }\textbf {\bibinfo {volume}
  {267}},\ \bibinfo {pages} {38} (\bibinfo {year} {2023})},\ \Eprint
  {http://arxiv.org/abs/2303.03261} {arXiv:2303.03261 [astro-ph.IM]}
  \BibitemShut {NoStop}%
\bibitem [{\citenamefont {Ng}\ \emph {et~al.}(2024{\natexlab{a}})\citenamefont
  {Ng}, \citenamefont {Cheong}, \citenamefont {Lam},\ and\ \citenamefont
  {Li}}]{Ng:2023syk}%
  \BibitemOpen
  \bibfield  {author} {\bibinfo {author} {\bibfnamefont {H.~H.-Y.}\
  \bibnamefont {Ng}}, \bibinfo {author} {\bibfnamefont {P.~C.-K.}\ \bibnamefont
  {Cheong}}, \bibinfo {author} {\bibfnamefont {A.~T.-L.}\ \bibnamefont {Lam}},
  \ and\ \bibinfo {author} {\bibfnamefont {T.~G.~F.}\ \bibnamefont {Li}},\
  }\href {\doibase 10.3847/1538-4365/ad2fbd} {\bibfield  {journal} {\bibinfo
  {journal} {Astrophys. J. Suppl.}\ }\textbf {\bibinfo {volume} {272}},\
  \bibinfo {pages} {9} (\bibinfo {year} {2024}{\natexlab{a}})},\ \Eprint
  {http://arxiv.org/abs/2309.03526} {arXiv:2309.03526 [astro-ph.HE]}
  \BibitemShut {NoStop}%
\bibitem [{\citenamefont {Cordero-Carrion}\ \emph {et~al.}(2009)\citenamefont
  {Cordero-Carrion}, \citenamefont {Cerda-Duran}, \citenamefont {Dimmelmeier},
  \citenamefont {Jaramillo}, \citenamefont {Novak},\ and\ \citenamefont
  {Gourgoulhon}}]{Cordero-Carrion:2008grk}%
  \BibitemOpen
  \bibfield  {author} {\bibinfo {author} {\bibfnamefont {I.}~\bibnamefont
  {Cordero-Carrion}}, \bibinfo {author} {\bibfnamefont {P.}~\bibnamefont
  {Cerda-Duran}}, \bibinfo {author} {\bibfnamefont {H.}~\bibnamefont
  {Dimmelmeier}}, \bibinfo {author} {\bibfnamefont {J.~L.}\ \bibnamefont
  {Jaramillo}}, \bibinfo {author} {\bibfnamefont {J.}~\bibnamefont {Novak}}, \
  and\ \bibinfo {author} {\bibfnamefont {E.}~\bibnamefont {Gourgoulhon}},\
  }\href {\doibase 10.1103/PhysRevD.79.024017} {\bibfield  {journal} {\bibinfo
  {journal} {Phys. Rev. D}\ }\textbf {\bibinfo {volume} {79}},\ \bibinfo
  {pages} {024017} (\bibinfo {year} {2009})},\ \Eprint
  {http://arxiv.org/abs/0809.2325} {arXiv:0809.2325 [gr-qc]} \BibitemShut
  {NoStop}%
\bibitem [{\citenamefont {Ng}\ \emph {et~al.}(2024{\natexlab{b}})\citenamefont
  {Ng}, \citenamefont {Jiang}, \citenamefont {Musolino}, \citenamefont {Ecker},
  \citenamefont {Tootle},\ and\ \citenamefont {Rezzolla}}]{Ng:2023yyg}%
  \BibitemOpen
  \bibfield  {author} {\bibinfo {author} {\bibfnamefont {H.~H.-Y.}\
  \bibnamefont {Ng}}, \bibinfo {author} {\bibfnamefont {J.-L.}\ \bibnamefont
  {Jiang}}, \bibinfo {author} {\bibfnamefont {C.}~\bibnamefont {Musolino}},
  \bibinfo {author} {\bibfnamefont {C.}~\bibnamefont {Ecker}}, \bibinfo
  {author} {\bibfnamefont {S.~D.}\ \bibnamefont {Tootle}}, \ and\ \bibinfo
  {author} {\bibfnamefont {L.}~\bibnamefont {Rezzolla}},\ }\href {\doibase
  10.1103/PhysRevD.109.064061} {\bibfield  {journal} {\bibinfo  {journal}
  {Phys. Rev. D}\ }\textbf {\bibinfo {volume} {109}},\ \bibinfo {pages}
  {064061} (\bibinfo {year} {2024}{\natexlab{b}})},\ \Eprint
  {http://arxiv.org/abs/2312.11358} {arXiv:2312.11358 [gr-qc]} \BibitemShut
  {NoStop}%
\bibitem [{\citenamefont {{Radice}}\ \emph {et~al.}(2022)\citenamefont
  {{Radice}}, \citenamefont {{Bernuzzi}}, \citenamefont {{Perego}},\ and\
  \citenamefont {{Haas}}}]{2022MNRAS.512.1499R}%
  \BibitemOpen
  \bibfield  {author} {\bibinfo {author} {\bibfnamefont {D.}~\bibnamefont
  {{Radice}}}, \bibinfo {author} {\bibfnamefont {S.}~\bibnamefont
  {{Bernuzzi}}}, \bibinfo {author} {\bibfnamefont {A.}~\bibnamefont
  {{Perego}}}, \ and\ \bibinfo {author} {\bibfnamefont {R.}~\bibnamefont
  {{Haas}}},\ }\href {\doibase 10.1093/mnras/stac589} {\bibfield  {journal}
  {\bibinfo  {journal} {\mnras}\ }\textbf {\bibinfo {volume} {512}},\ \bibinfo
  {pages} {1499} (\bibinfo {year} {2022})},\ \Eprint
  {http://arxiv.org/abs/2111.14858} {arXiv:2111.14858 [astro-ph.HE]}
  \BibitemShut {NoStop}%
\bibitem [{\citenamefont {Cheong}\ \emph
  {et~al.}(2024{\natexlab{a}})\citenamefont {Cheong}, \citenamefont {Foucart},
  \citenamefont {Duez}, \citenamefont {Offermans}, \citenamefont {Muhammed},\
  and\ \citenamefont {Chawhan}}]{Cheong:2024buu}%
  \BibitemOpen
  \bibfield  {author} {\bibinfo {author} {\bibfnamefont {P.~C.-K.}\
  \bibnamefont {Cheong}}, \bibinfo {author} {\bibfnamefont {F.}~\bibnamefont
  {Foucart}}, \bibinfo {author} {\bibfnamefont {M.~D.}\ \bibnamefont {Duez}},
  \bibinfo {author} {\bibfnamefont {A.}~\bibnamefont {Offermans}}, \bibinfo
  {author} {\bibfnamefont {N.}~\bibnamefont {Muhammed}}, \ and\ \bibinfo
  {author} {\bibfnamefont {P.}~\bibnamefont {Chawhan}},\ }\href {\doibase
  10.3847/1538-4357/ad7825} {\bibfield  {journal} {\bibinfo  {journal}
  {Astrophys. J.}\ }\textbf {\bibinfo {volume} {975}},\ \bibinfo {pages} {116}
  (\bibinfo {year} {2024}{\natexlab{a}})},\ \Eprint
  {http://arxiv.org/abs/2407.16017} {arXiv:2407.16017 [astro-ph.HE]}
  \BibitemShut {NoStop}%
\bibitem [{\citenamefont {Pareschi}\ and\ \citenamefont
  {Russo}(2005)}]{pareschi2005implicit}%
  \BibitemOpen
  \bibfield  {author} {\bibinfo {author} {\bibfnamefont {L.}~\bibnamefont
  {Pareschi}}\ and\ \bibinfo {author} {\bibfnamefont {G.}~\bibnamefont
  {Russo}},\ }\href@noop {} {\bibfield  {journal} {\bibinfo  {journal} {Journal
  of Scientific computing}\ }\textbf {\bibinfo {volume} {25}},\ \bibinfo
  {pages} {129} (\bibinfo {year} {2005})}\BibitemShut {NoStop}%
\bibitem [{\citenamefont {Foucart}\ \emph {et~al.}(2016)\citenamefont
  {Foucart}, \citenamefont {O'Connor}, \citenamefont {Roberts}, \citenamefont
  {Kidder}, \citenamefont {Pfeiffer},\ and\ \citenamefont
  {Scheel}}]{Foucart:2016rxm}%
  \BibitemOpen
  \bibfield  {author} {\bibinfo {author} {\bibfnamefont {F.}~\bibnamefont
  {Foucart}}, \bibinfo {author} {\bibfnamefont {E.}~\bibnamefont {O'Connor}},
  \bibinfo {author} {\bibfnamefont {L.}~\bibnamefont {Roberts}}, \bibinfo
  {author} {\bibfnamefont {L.~E.}\ \bibnamefont {Kidder}}, \bibinfo {author}
  {\bibfnamefont {H.~P.}\ \bibnamefont {Pfeiffer}}, \ and\ \bibinfo {author}
  {\bibfnamefont {M.~A.}\ \bibnamefont {Scheel}},\ }\href {\doibase
  10.1103/PhysRevD.94.123016} {\bibfield  {journal} {\bibinfo  {journal} {Phys.
  Rev. D}\ }\textbf {\bibinfo {volume} {94}},\ \bibinfo {pages} {123016}
  (\bibinfo {year} {2016})},\ \Eprint {http://arxiv.org/abs/1607.07450}
  {arXiv:1607.07450 [astro-ph.HE]} \BibitemShut {NoStop}%
\bibitem [{\citenamefont {Andresen}\ \emph {et~al.}(2024)\citenamefont
  {Andresen}, \citenamefont {O'Connor}, \citenamefont {Andersen},\ and\
  \citenamefont {Couch}}]{Andresen:2024mtt}%
  \BibitemOpen
  \bibfield  {author} {\bibinfo {author} {\bibfnamefont {H.}~\bibnamefont
  {Andresen}}, \bibinfo {author} {\bibfnamefont {E.~P.}\ \bibnamefont
  {O'Connor}}, \bibinfo {author} {\bibfnamefont {O.~E.}\ \bibnamefont
  {Andersen}}, \ and\ \bibinfo {author} {\bibfnamefont {S.~M.}\ \bibnamefont
  {Couch}},\ }\href {\doibase 10.1051/0004-6361/202449776} {\bibfield
  {journal} {\bibinfo  {journal} {Astron. Astrophys.}\ }\textbf {\bibinfo
  {volume} {687}},\ \bibinfo {pages} {A55} (\bibinfo {year} {2024})},\ \Eprint
  {http://arxiv.org/abs/2402.18303} {arXiv:2402.18303 [astro-ph.HE]}
  \BibitemShut {NoStop}%
\bibitem [{\citenamefont {Nagakura}\ \emph {et~al.}(2019)\citenamefont
  {Nagakura}, \citenamefont {Furusawa}, \citenamefont {Togashi}, \citenamefont
  {Richers}, \citenamefont {Sumiyoshi},\ and\ \citenamefont
  {Yamada}}]{Nagakura:2018qpg}%
  \BibitemOpen
  \bibfield  {author} {\bibinfo {author} {\bibfnamefont {H.}~\bibnamefont
  {Nagakura}}, \bibinfo {author} {\bibfnamefont {S.}~\bibnamefont {Furusawa}},
  \bibinfo {author} {\bibfnamefont {H.}~\bibnamefont {Togashi}}, \bibinfo
  {author} {\bibfnamefont {S.}~\bibnamefont {Richers}}, \bibinfo {author}
  {\bibfnamefont {K.}~\bibnamefont {Sumiyoshi}}, \ and\ \bibinfo {author}
  {\bibfnamefont {S.}~\bibnamefont {Yamada}},\ }\href {\doibase
  10.3847/1538-4365/aafac9} {\bibfield  {journal} {\bibinfo  {journal}
  {Astrophys. J. Suppl.}\ }\textbf {\bibinfo {volume} {240}},\ \bibinfo {pages}
  {38} (\bibinfo {year} {2019})},\ \Eprint {http://arxiv.org/abs/1812.09811}
  {arXiv:1812.09811 [astro-ph.HE]} \BibitemShut {NoStop}%
\bibitem [{\citenamefont {Liebendoerfer}(2005)}]{Liebendoerfer:2005gm}%
  \BibitemOpen
  \bibfield  {author} {\bibinfo {author} {\bibfnamefont {M.}~\bibnamefont
  {Liebendoerfer}},\ }\href {\doibase 10.1086/466517} {\bibfield  {journal}
  {\bibinfo  {journal} {Astrophys. J.}\ }\textbf {\bibinfo {volume} {633}},\
  \bibinfo {pages} {1042} (\bibinfo {year} {2005})},\ \Eprint
  {http://arxiv.org/abs/astro-ph/0504072} {arXiv:astro-ph/0504072} \BibitemShut
  {NoStop}%
\bibitem [{\citenamefont {Lin}\ \emph {et~al.}(2024)\citenamefont {Lin},
  \citenamefont {Zha}, \citenamefont {O'Connor},\ and\ \citenamefont
  {Steiner}}]{Lin:2022lck}%
  \BibitemOpen
  \bibfield  {author} {\bibinfo {author} {\bibfnamefont {Z.}~\bibnamefont
  {Lin}}, \bibinfo {author} {\bibfnamefont {S.}~\bibnamefont {Zha}}, \bibinfo
  {author} {\bibfnamefont {E.~P.}\ \bibnamefont {O'Connor}}, \ and\ \bibinfo
  {author} {\bibfnamefont {A.~W.}\ \bibnamefont {Steiner}},\ }\href {\doibase
  10.1103/PhysRevD.109.023005} {\bibfield  {journal} {\bibinfo  {journal}
  {Phys. Rev. D}\ }\textbf {\bibinfo {volume} {109}},\ \bibinfo {pages}
  {023005} (\bibinfo {year} {2024})},\ \Eprint
  {http://arxiv.org/abs/2203.05141} {arXiv:2203.05141 [astro-ph.HE]}
  \BibitemShut {NoStop}%
\bibitem [{\citenamefont {Ricciardi}\ \emph {et~al.}(2022)\citenamefont
  {Ricciardi}, \citenamefont {Vignaroli},\ and\ \citenamefont
  {Vissani}}]{Ricciardi:2022pru}%
  \BibitemOpen
  \bibfield  {author} {\bibinfo {author} {\bibfnamefont {G.}~\bibnamefont
  {Ricciardi}}, \bibinfo {author} {\bibfnamefont {N.}~\bibnamefont
  {Vignaroli}}, \ and\ \bibinfo {author} {\bibfnamefont {F.}~\bibnamefont
  {Vissani}},\ }\href {\doibase 10.1007/JHEP08(2022)212} {\bibfield  {journal}
  {\bibinfo  {journal} {JHEP}\ }\textbf {\bibinfo {volume} {08}},\ \bibinfo
  {pages} {212} (\bibinfo {year} {2022})},\ \Eprint
  {http://arxiv.org/abs/2206.05567} {arXiv:2206.05567 [hep-ph]} \BibitemShut
  {NoStop}%
\bibitem [{\citenamefont {Strumia}\ and\ \citenamefont
  {Vissani}(2003)}]{Strumia:2003zx}%
  \BibitemOpen
  \bibfield  {author} {\bibinfo {author} {\bibfnamefont {A.}~\bibnamefont
  {Strumia}}\ and\ \bibinfo {author} {\bibfnamefont {F.}~\bibnamefont
  {Vissani}},\ }\href {\doibase 10.1016/S0370-2693(03)00616-6} {\bibfield
  {journal} {\bibinfo  {journal} {Phys. Lett. B}\ }\textbf {\bibinfo {volume}
  {564}},\ \bibinfo {pages} {42} (\bibinfo {year} {2003})},\ \Eprint
  {http://arxiv.org/abs/astro-ph/0302055} {arXiv:astro-ph/0302055} \BibitemShut
  {NoStop}%
\bibitem [{\citenamefont {Ikeda}\ \emph {et~al.}(2007)\citenamefont {Ikeda}
  \emph {et~al.}}]{Super-Kamiokande:2007zsl}%
  \BibitemOpen
  \bibfield  {author} {\bibinfo {author} {\bibfnamefont {M.}~\bibnamefont
  {Ikeda}} \emph {et~al.} (\bibinfo {collaboration} {Super-Kamiokande}),\
  }\href {\doibase 10.1086/521547} {\bibfield  {journal} {\bibinfo  {journal}
  {Astrophys. J.}\ }\textbf {\bibinfo {volume} {669}},\ \bibinfo {pages} {519}
  (\bibinfo {year} {2007})},\ \Eprint {http://arxiv.org/abs/0706.2283}
  {arXiv:0706.2283 [astro-ph]} \BibitemShut {NoStop}%
\bibitem [{\citenamefont {Abe}\ \emph {et~al.}(2018)\citenamefont {Abe} \emph
  {et~al.}}]{Hyper-Kamiokande:2018ofw}%
  \BibitemOpen
  \bibfield  {author} {\bibinfo {author} {\bibfnamefont {K.}~\bibnamefont
  {Abe}} \emph {et~al.} (\bibinfo {collaboration} {Hyper-Kamiokande}),\
  }\href@noop {} {\  (\bibinfo {year} {2018})},\ \Eprint
  {http://arxiv.org/abs/1805.04163} {arXiv:1805.04163 [physics.ins-det]}
  \BibitemShut {NoStop}%
\bibitem [{\citenamefont {Brdar}\ \emph {et~al.}(2018)\citenamefont {Brdar},
  \citenamefont {Lindner},\ and\ \citenamefont {Xu}}]{Brdar:2018zds}%
  \BibitemOpen
  \bibfield  {author} {\bibinfo {author} {\bibfnamefont {V.}~\bibnamefont
  {Brdar}}, \bibinfo {author} {\bibfnamefont {M.}~\bibnamefont {Lindner}}, \
  and\ \bibinfo {author} {\bibfnamefont {X.-J.}\ \bibnamefont {Xu}},\ }\href
  {\doibase 10.1088/1475-7516/2018/04/025} {\bibfield  {journal} {\bibinfo
  {journal} {JCAP}\ }\textbf {\bibinfo {volume} {04}},\ \bibinfo {pages} {025}
  (\bibinfo {year} {2018})},\ \Eprint {http://arxiv.org/abs/1802.02577}
  {arXiv:1802.02577 [hep-ph]} \BibitemShut {NoStop}%
\bibitem [{\citenamefont {{\"O}zel}\ and\ \citenamefont
  {Freire}(2016)}]{Ozel:2016oaf}%
  \BibitemOpen
  \bibfield  {author} {\bibinfo {author} {\bibfnamefont {F.}~\bibnamefont
  {{\"O}zel}}\ and\ \bibinfo {author} {\bibfnamefont {P.}~\bibnamefont
  {Freire}},\ }\href {\doibase 10.1146/annurev-astro-081915-023322} {\bibfield
  {journal} {\bibinfo  {journal} {Ann. Rev. Astron. Astrophys.}\ }\textbf
  {\bibinfo {volume} {54}},\ \bibinfo {pages} {401} (\bibinfo {year} {2016})},\
  \Eprint {http://arxiv.org/abs/1603.02698} {arXiv:1603.02698 [astro-ph.HE]}
  \BibitemShut {NoStop}%
\bibitem [{\citenamefont {Miller}\ \emph {et~al.}(2019)\citenamefont {Miller}
  \emph {et~al.}}]{Miller:2019cac}%
  \BibitemOpen
  \bibfield  {author} {\bibinfo {author} {\bibfnamefont {M.~C.}\ \bibnamefont
  {Miller}} \emph {et~al.},\ }\href {\doibase 10.3847/2041-8213/ab50c5}
  {\bibfield  {journal} {\bibinfo  {journal} {Astrophys. J. Lett.}\ }\textbf
  {\bibinfo {volume} {887}},\ \bibinfo {pages} {L24} (\bibinfo {year}
  {2019})},\ \Eprint {http://arxiv.org/abs/1912.05705} {arXiv:1912.05705
  [astro-ph.HE]} \BibitemShut {NoStop}%
\bibitem [{\citenamefont {{Most}}\ \emph {et~al.}(2020)\citenamefont {{Most}},
  \citenamefont {{Jens Papenfort}}, \citenamefont {{Dexheimer}}, \citenamefont
  {{Hanauske}}, \citenamefont {{Stoecker}},\ and\ \citenamefont
  {{Rezzolla}}}]{Most2019c}%
  \BibitemOpen
  \bibfield  {author} {\bibinfo {author} {\bibfnamefont {E.~R.}\ \bibnamefont
  {{Most}}}, \bibinfo {author} {\bibfnamefont {L.}~\bibnamefont {{Jens
  Papenfort}}}, \bibinfo {author} {\bibfnamefont {V.}~\bibnamefont
  {{Dexheimer}}}, \bibinfo {author} {\bibfnamefont {M.}~\bibnamefont
  {{Hanauske}}}, \bibinfo {author} {\bibfnamefont {H.}~\bibnamefont
  {{Stoecker}}}, \ and\ \bibinfo {author} {\bibfnamefont {L.}~\bibnamefont
  {{Rezzolla}}},\ }\href {\doibase 10.1140/epja/s10050-020-00073-4} {\bibfield
  {journal} {\bibinfo  {journal} {European Physical Journal A}\ }\textbf
  {\bibinfo {volume} {56}},\ \bibinfo {eid} {59} (\bibinfo {year} {2020})},\
  \Eprint {http://arxiv.org/abs/1910.13893} {arXiv:1910.13893 [astro-ph.HE]}
  \BibitemShut {NoStop}%
\bibitem [{\citenamefont {Prakash}\ \emph {et~al.}(2021)\citenamefont
  {Prakash}, \citenamefont {Radice}, \citenamefont {Logoteta}, \citenamefont
  {Perego}, \citenamefont {Nedora}, \citenamefont {Bombaci}, \citenamefont
  {Kashyap}, \citenamefont {Bernuzzi},\ and\ \citenamefont
  {Endrizzi}}]{Prakash:2021wpz}%
  \BibitemOpen
  \bibfield  {author} {\bibinfo {author} {\bibfnamefont {A.}~\bibnamefont
  {Prakash}}, \bibinfo {author} {\bibfnamefont {D.}~\bibnamefont {Radice}},
  \bibinfo {author} {\bibfnamefont {D.}~\bibnamefont {Logoteta}}, \bibinfo
  {author} {\bibfnamefont {A.}~\bibnamefont {Perego}}, \bibinfo {author}
  {\bibfnamefont {V.}~\bibnamefont {Nedora}}, \bibinfo {author} {\bibfnamefont
  {I.}~\bibnamefont {Bombaci}}, \bibinfo {author} {\bibfnamefont
  {R.}~\bibnamefont {Kashyap}}, \bibinfo {author} {\bibfnamefont
  {S.}~\bibnamefont {Bernuzzi}}, \ and\ \bibinfo {author} {\bibfnamefont
  {A.}~\bibnamefont {Endrizzi}},\ }\href {\doibase 10.1103/PhysRevD.104.083029}
  {\bibfield  {journal} {\bibinfo  {journal} {Phys. Rev. D}\ }\textbf {\bibinfo
  {volume} {104}},\ \bibinfo {pages} {083029} (\bibinfo {year} {2021})},\
  \Eprint {http://arxiv.org/abs/2106.07885} {arXiv:2106.07885 [astro-ph.HE]}
  \BibitemShut {NoStop}%
\bibitem [{\citenamefont {Guo}\ \emph {et~al.}(2020)\citenamefont {Guo},
  \citenamefont {Mart\'\i{}nez-Pinedo}, \citenamefont {Lohs},\ and\
  \citenamefont {Fischer}}]{Guo:2020tgx}%
  \BibitemOpen
  \bibfield  {author} {\bibinfo {author} {\bibfnamefont {G.}~\bibnamefont
  {Guo}}, \bibinfo {author} {\bibfnamefont {G.}~\bibnamefont
  {Mart\'\i{}nez-Pinedo}}, \bibinfo {author} {\bibfnamefont {A.}~\bibnamefont
  {Lohs}}, \ and\ \bibinfo {author} {\bibfnamefont {T.}~\bibnamefont
  {Fischer}},\ }\href {\doibase 10.1103/PhysRevD.102.023037} {\bibfield
  {journal} {\bibinfo  {journal} {Phys. Rev. D}\ }\textbf {\bibinfo {volume}
  {102}},\ \bibinfo {pages} {023037} (\bibinfo {year} {2020})},\ \Eprint
  {http://arxiv.org/abs/2006.12051} {arXiv:2006.12051 [hep-ph]} \BibitemShut
  {NoStop}%
\bibitem [{\citenamefont {Cheong}\ \emph
  {et~al.}(2024{\natexlab{b}})\citenamefont {Cheong}, \citenamefont {Foucart},
  \citenamefont {Ng}, \citenamefont {Offermans}, \citenamefont {Duez},
  \citenamefont {Muhammed},\ and\ \citenamefont {Chawhan}}]{Cheong:2024cnb}%
  \BibitemOpen
  \bibfield  {author} {\bibinfo {author} {\bibfnamefont {P.~C.-K.}\
  \bibnamefont {Cheong}}, \bibinfo {author} {\bibfnamefont {F.}~\bibnamefont
  {Foucart}}, \bibinfo {author} {\bibfnamefont {H.~H.-Y.}\ \bibnamefont {Ng}},
  \bibinfo {author} {\bibfnamefont {A.}~\bibnamefont {Offermans}}, \bibinfo
  {author} {\bibfnamefont {M.~D.}\ \bibnamefont {Duez}}, \bibinfo {author}
  {\bibfnamefont {N.}~\bibnamefont {Muhammed}}, \ and\ \bibinfo {author}
  {\bibfnamefont {P.}~\bibnamefont {Chawhan}},\ }\href@noop {} {\  (\bibinfo
  {year} {2024}{\natexlab{b}})},\ \Eprint {http://arxiv.org/abs/2410.20681}
  {arXiv:2410.20681 [astro-ph.HE]} \BibitemShut {NoStop}%
\bibitem [{\citenamefont {Ng}\ \emph {et~al.}(2024{\natexlab{c}})\citenamefont
  {Ng}, \citenamefont {Musolino}, \citenamefont {Tootle},\ and\ \citenamefont
  {Rezzolla}}]{Ng:2024zve}%
  \BibitemOpen
  \bibfield  {author} {\bibinfo {author} {\bibfnamefont {H.~H.-Y.}\
  \bibnamefont {Ng}}, \bibinfo {author} {\bibfnamefont {C.}~\bibnamefont
  {Musolino}}, \bibinfo {author} {\bibfnamefont {S.~D.}\ \bibnamefont
  {Tootle}}, \ and\ \bibinfo {author} {\bibfnamefont {L.}~\bibnamefont
  {Rezzolla}},\ }\href@noop {} {\  (\bibinfo {year} {2024}{\natexlab{c}})},\
  \Eprint {http://arxiv.org/abs/2411.19178} {arXiv:2411.19178 [astro-ph.HE]}
  \BibitemShut {NoStop}%
\bibitem [{\citenamefont {Sykes}\ and\ \citenamefont
  {M{\"u}ller}(2025)}]{Sykes:2024mel}%
  \BibitemOpen
  \bibfield  {author} {\bibinfo {author} {\bibfnamefont {B.}~\bibnamefont
  {Sykes}}\ and\ \bibinfo {author} {\bibfnamefont {B.}~\bibnamefont
  {M{\"u}ller}},\ }\href {\doibase 10.1103/PhysRevD.111.063042} {\bibfield
  {journal} {\bibinfo  {journal} {Phys. Rev. D}\ }\textbf {\bibinfo {volume}
  {111}},\ \bibinfo {pages} {063042} (\bibinfo {year} {2025})},\ \Eprint
  {http://arxiv.org/abs/2412.01155} {arXiv:2412.01155 [astro-ph.HE]}
  \BibitemShut {NoStop}%
\bibitem [{\citenamefont {Shen}\ \emph {et~al.}(1998)\citenamefont {Shen},
  \citenamefont {Toki}, \citenamefont {Oyamatsu},\ and\ \citenamefont
  {Sumiyoshi}}]{Shen:1998by}%
  \BibitemOpen
  \bibfield  {author} {\bibinfo {author} {\bibfnamefont {H.}~\bibnamefont
  {Shen}}, \bibinfo {author} {\bibfnamefont {H.}~\bibnamefont {Toki}}, \bibinfo
  {author} {\bibfnamefont {K.}~\bibnamefont {Oyamatsu}}, \ and\ \bibinfo
  {author} {\bibfnamefont {K.}~\bibnamefont {Sumiyoshi}},\ }\href {\doibase
  10.1143/PTP.100.1013} {\bibfield  {journal} {\bibinfo  {journal} {Prog.
  Theor. Phys.}\ }\textbf {\bibinfo {volume} {100}},\ \bibinfo {pages} {1013}
  (\bibinfo {year} {1998})},\ \Eprint {http://arxiv.org/abs/nucl-th/9806095}
  {arXiv:nucl-th/9806095} \BibitemShut {NoStop}%
\bibitem [{\citenamefont {Farhi}\ and\ \citenamefont
  {Jaffe}(1984)}]{Farhi:1984qu}%
  \BibitemOpen
  \bibfield  {author} {\bibinfo {author} {\bibfnamefont {E.}~\bibnamefont
  {Farhi}}\ and\ \bibinfo {author} {\bibfnamefont {R.~L.}\ \bibnamefont
  {Jaffe}},\ }\href {\doibase 10.1103/PhysRevD.30.2379} {\bibfield  {journal}
  {\bibinfo  {journal} {Phys. Rev. D}\ }\textbf {\bibinfo {volume} {30}},\
  \bibinfo {pages} {2379} (\bibinfo {year} {1984})}\BibitemShut {NoStop}%
\bibitem [{\citenamefont {Sagert}\ \emph {et~al.}(2010)\citenamefont {Sagert},
  \citenamefont {Fischer}, \citenamefont {Hempel}, \citenamefont {Pagliara},
  \citenamefont {Schaffner-Bielich}, \citenamefont {Thielemann},\ and\
  \citenamefont {Liebendorfer}}]{Sagert:2010yg}%
  \BibitemOpen
  \bibfield  {author} {\bibinfo {author} {\bibfnamefont {I.}~\bibnamefont
  {Sagert}}, \bibinfo {author} {\bibfnamefont {T.}~\bibnamefont {Fischer}},
  \bibinfo {author} {\bibfnamefont {M.}~\bibnamefont {Hempel}}, \bibinfo
  {author} {\bibfnamefont {G.}~\bibnamefont {Pagliara}}, \bibinfo {author}
  {\bibfnamefont {J.}~\bibnamefont {Schaffner-Bielich}}, \bibinfo {author}
  {\bibfnamefont {F.~K.}\ \bibnamefont {Thielemann}}, \ and\ \bibinfo {author}
  {\bibfnamefont {M.}~\bibnamefont {Liebendorfer}},\ }\href {\doibase
  10.1088/0954-3899/37/9/094064} {\bibfield  {journal} {\bibinfo  {journal} {J.
  Phys. G}\ }\textbf {\bibinfo {volume} {37}},\ \bibinfo {pages} {094064}
  (\bibinfo {year} {2010})},\ \Eprint {http://arxiv.org/abs/1003.2320}
  {arXiv:1003.2320 [astro-ph.HE]} \BibitemShut {NoStop}%
\end{thebibliography}%

\clearpage 

\begin{figure}[h]
\includegraphics[width=\columnwidth]{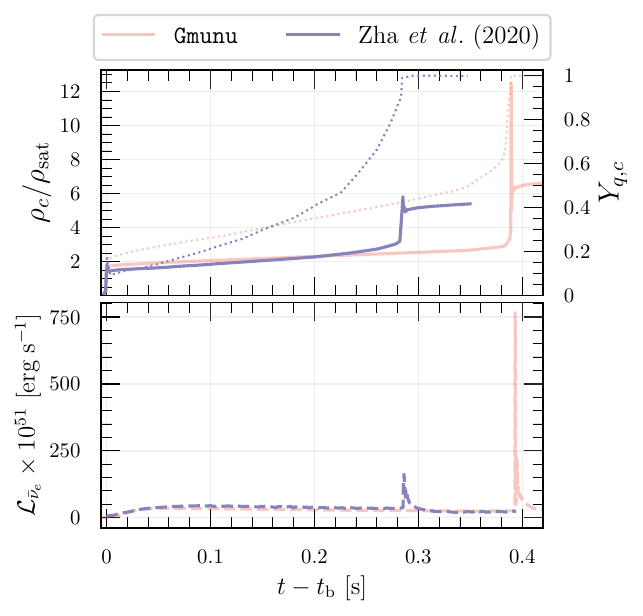}
\caption{\label{ccsn_neutrino}
Time evolution of central density $\rho_c$ (solid), central quark fraction $Y_{q,c}$ (dotted)
and luminosity of electron antineutrino $\mathcal{L}_{\bar\nu_e}$ (dashed). Results from \texttt{Gmunu}
and Ref.~\cite{Zha:2020gjw} are colored in red and purple respectively.
}
\end{figure}

\appendix
\section{Core collapse of a 12 $M_{\odot}$ star with hybrid EOS}
\label{sec:codetest}

In this section, we present general relativistic neutrino-radiation hydrodynamics simulations on 
the first-order QCD PT in CCSNe using hybrid EOS as a reference result of QCD PT in \texttt{Gmunu}.

Our simulation is identical to that of the main text except for employing a different hybrid EOS, progenitor and
computational domain. 
This hybrid EOS employs the STOS for hadronic phase~\citep{Shen:1998by} and MIT bag model with bag constant $(B=165~\rm{MeV})^4$
for deconfined quark phase~\citep{Farhi:1984qu}. This EOS adopts Gibbs construction for the mixed nuclear-quark phase. 
This EOS is adopted from several studies of PT in CCSNe~\citep{Sagert:2008ka,Sagert:2010yg,Zha:2020gjw} and is 
publicly available~\citep{compose}.
We take the same $12M_{\odot}$, solar metallicity,
presupernova progenitor $s12$ from~\cite{Woosley:2002zz}, 
as in~\cite{Zha:2020gjw}.
The computational domain covers from 
$0$ to $2\times 10^4~\rm{km}$, with the resolution $N_{r}=128$ 
and allowing 11 AMR level. The finest grid size $\Delta r $
at the star’s center is $153~\rm{m}$.

We compare our results with those from the axisymmetric simulation in~\cite{Zha:2020gjw},
which employs an energy-independent neutrino transport scheme. 
Additionally, it solves the Newtonian Euler equations with general relativistic corrections. 
The computational domain extends from $0$ to $2 \times 10^4$ km in the radial direction 
and spans $\pm 2 \times 10^4~\rm{km}$ along the cylindrical axis, 
with a finest spatial resolution of $150~\rm{m}$ as reported in~\cite{Zha:2020gjw}.

The time evolution of the central density $\rho_c$,
central quark fraction $Y_{q,c}$, 
and electron antineutrino luminosity $\mathcal{L}_{\bar\nu_e}$
is shown in Fig.~\ref{ccsn_neutrino}.
The iron core collapses until the central density $\rho_c$
exceeds nuclear saturation density ($\gtrsim 10^{14}~\rm{g~cm^{-3}}$),
at which point the stiffening of the EOS halts the collapse and triggers a core bounce,
followed by continued mass accretion.
In this model, the PNS enters the mixed phase even before the core bounce,
near $\rho_{\rm{sat}}$, due to the low onset density in the hybrid EOS.
With ongoing mass accretion, Kelvin–Helmholtz cooling,
and EOS softening in the mixed phase,
the PNS contracts rapidly, accompanied by a substantial rise in the central quark fraction $Y_{q,c}$.
This is reflected in the significant increase in both $\rho_c$ and $Y_{q,c}$
within approximately $0.2~\rm{s}$ after the initial core bounce.

\begin{figure}[]
\includegraphics[width=\columnwidth]{ 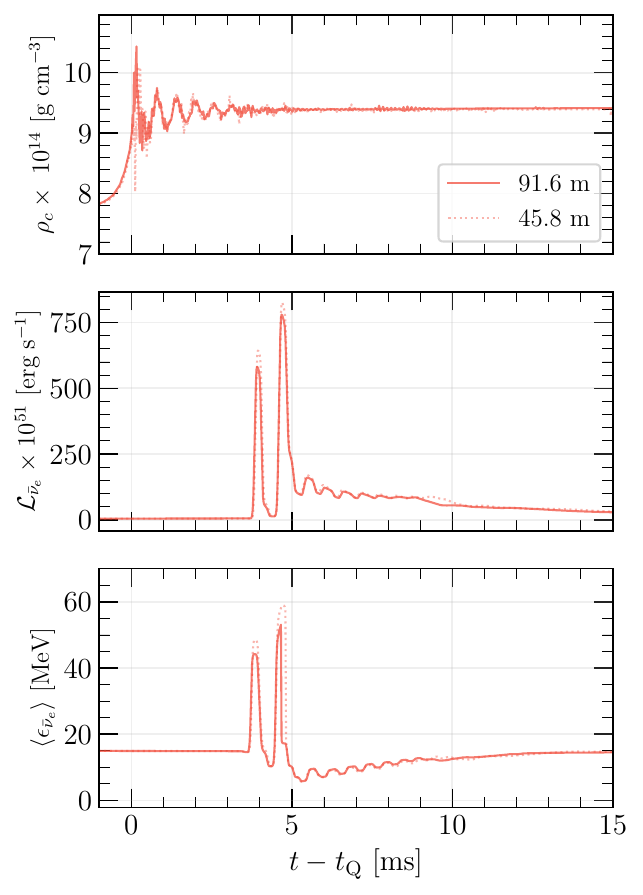}
\caption{\label{code_test_resolution}
		Time evolution of the central density $\rho_{c}$,
		luminosity of electron antineutrino $\mathcal{L}_{\bar\nu_e}$ and
		corresponding average energy $\langle{\epsilon_{\bar\nu_e}}\rangle$ after
 		the second core bounce for \texttt{RDF-1.9} models with
		different resolution.
		The models with finest resolutions of $91.6~\rm{m}$ 
		and $45.8~\rm{m}$ are shown in solid and dotted lines, respectively.
}
\end{figure}

\begin{figure*}[]
\includegraphics[width=\textwidth]{ 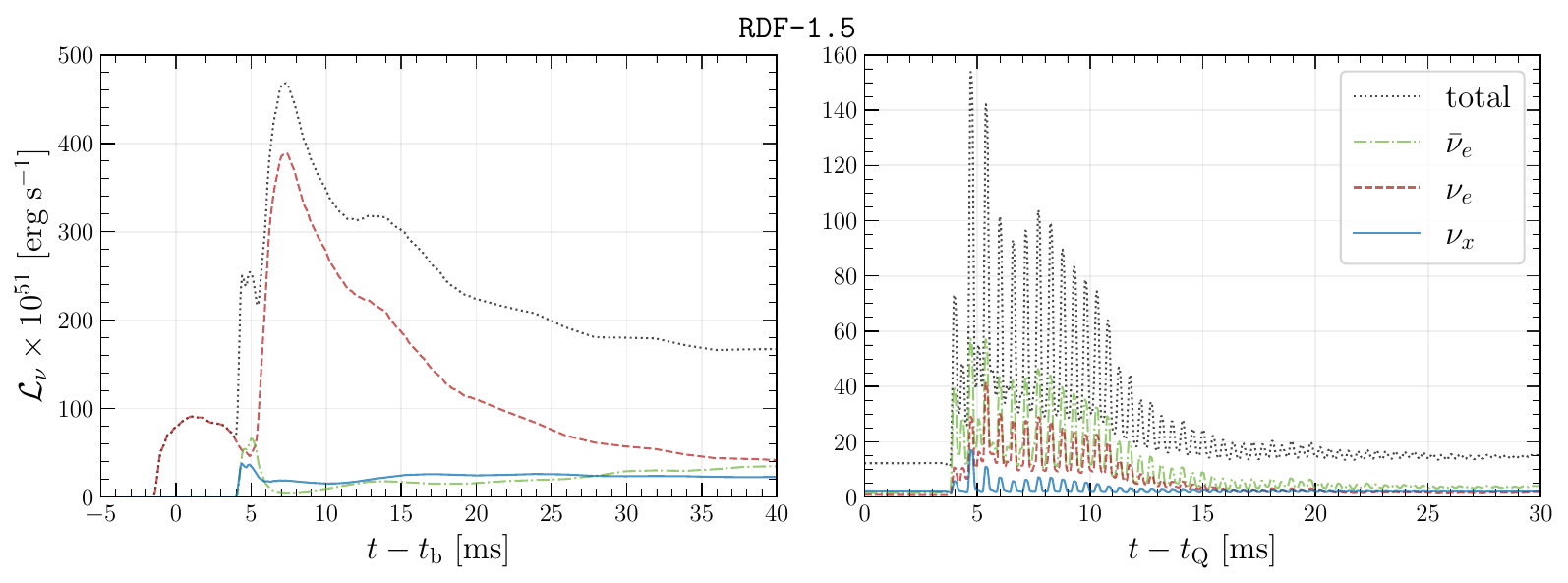}
\caption{\label{EM_nu} Neutrino luminosity $\mathcal{L}_{\nu}$ 
		of the first and the second burst in model with \texttt{RDF-1.5} 
		for total luminosity (black dotted), $\mathcal{L}_{\nu_e}$ (red dashed), $\mathcal{L}_{\bar\nu_e}$
		(green dash-dotted) and $\mathcal{L}_{\nu_x}$ (blue solid), respectively.
}
\end{figure*}

Here, we observe a difference in the rate of increase of $\rho_c$ and $Y_{q,c}$,
primarily due to differences in the neutrino transport schemes.
Our simulation adopts an energy-averaged transport scheme,
while Ref.~\citep{Zha:2020gjw}
employs an energy-dependent transport approach.

We observe a second dynamical collapse and the formation of a PHS in both our simulation and Ref.~\cite{Zha:2020gjw},
occurring when the central quark fraction reaches $Y_{q,c} \sim 0.3$.
This transition leads to a strong collapse,
with the central density reaching up to $\sim 12~\rho_{\rm{sat}}$ in our simulation and
$\sim 6~\rho_{\rm{sat}}$ in~\cite{Zha:2020gjw}.
The timing of the second collapse also differs:
$t - t_b = 0.389~\rm{s}$ in our simulation compared to $t - t_b = 0.286~\rm{s}$
in~\cite{Zha:2020gjw}, primarily due to differences in the neutrino transport scheme, gravitational treatment,
and simulation dimensionality.
Additionally, the luminosity of the second electron antineutrino burst 
differs by a factor of approximately 3, driven by the same factors.
Despite these quantitative differences arising from the numerical setups, 
our results remain qualitatively consistent with those of Ref.~\cite{Zha:2020gjw}.

With major differences in the setup, this set of simulations does not constitute a systematic comparison of specific neutrino microphysics or numerical choices. 
However, the results highlight how numerical schemes and dimensionality can alter PT dynamics and the associated neutrino observables in CCSN. 
It is therefore reasonable to expect 
that PT in AIC systems will be subject to similar sensitivities.  
A key observable difference between AIC systems and CCSNe is that AIC systems exhibit
a significantly longer time delay between the first and second neutrino bursts.

\section{Resolution dependence}
\label{sec:resolution_dependence}
We perform a set of simulations with different resolution
with the \texttt{RDF-1.9} model to compare
the main result discussed in the main text.
The numerical setup is identical to that in the main text,
except that $9$ and $10$ AMR levels are allowed,
with a finest solution of $91.6~\rm{m}$ and $45.8~\rm{m}$.

The PNS cores of both models enter the mixed phase at 
$t_{\rm{mixed}} = 0.58~\rm{s}$ and 
the deconfined quark phase at $t_{\rm{Q}} = 1.1~\rm{s}$.
Figure~\ref{code_test_resolution} shows the evolution 
of $\rho_c$, $\mathcal{L}_{\bar\nu_e}$, 
and $\langle{\epsilon_{\bar\nu_e}}\rangle$ around $t_{\rm{Q}}$.
The core oscillations agree qualitatively well in both frequency and amplitude 
across the two resolutions.
Similarly, the amplitude and oscillatory features of the second neutrino burst 
are in good agreement between the two resolutions.
\\

\section{Comparison of two neutrino bursts in AIC}
\label{sec:comparison}

Electron captures prior to core bounce steadily 
increase $\mathcal{L}_{\nu_e}$. 
The collapse halts due to hadronic matter stiffening, 
triggering an outward shock that dissociates heavy nuclei,
enhancing $\nu_e$ production via $e^-$ captures. 
This results in a luminous $\nu_e$ burst from 
proton captures in shock-heated matter, 
peaking at $t = t_{\rm{b}}$ in Fig.~\ref{EM_nu}. 
Compression and Doppler redshifting 
cause a dip in $\mathcal{L}_{\nu_e}$ around $4~\rm{ms}$ later.

As the shock breaks out, 
$\mathcal{L}_{\rm{tot}}$ peaks at 
$4.6\times10^{53}~\rm{erg~s}^{-1}$,
with $\mathcal{L}_{\nu_e}$ 
peaks at $3.8\times10^{53}~\rm{erg~s}^{-1}$, 
while shock-heated matter enhances pair processes, 
leading to $\bar{\nu}_e$ and $\nu_x$ peaks 
at $4\times10^{52}~\rm{erg~s}^{-1}$ 
and $6\times10^{52}~\rm{erg~s}^{-1}$, 
respectively. Given the similarity of hadronic 
matter modeling near nuclear saturation density 
across RDF EOS models, this first neutrino burst behavior is generic.

In contrast, the second neutrino burst 
differs significantly due to a different emission mechanism. 
The stiffening of deconfined quark matter halts 
the PT-induced collapse, generating a second shock. 
The low-proton-fraction cavity facilitates $e^+$ capture, 
leading to a $\bar{\nu}_e$-dominated burst. 
Unlike the first burst, all species reach a local maximum 
in Fig.~\ref{EM_nu}.
$\mathcal{L}_{\rm{tot}}$ peaks at $1.6\times10^{53}~\rm{erg~s}^{-1}$ with a
$\mathcal{L}_{\nu_e}$, 
$\mathcal{L}_{\bar\nu_e}$ and 
$\mathcal{L}_{\nu_x}$
peak at $6\times10^{52}~\rm{ergs}^{-1}$, $4\times10^{52}~\rm{erg~s}^{-1}$,
$2\times10^{52}~\rm{erg~s}^{-1}$ respectively. 
Multiple follow-up shocks 
from PHS oscillations introduce an oscillatory pattern.

\begin{table*}
	\centering
	\makebox[\columnwidth]{
	\scalebox{0.85}{
	\begin{tabular}{cccccc|cccccccccc}
	\hline
	\hline
          EOS & $t_{\rm{mixed}}$ & $\rho_{\rm{mixed,c}}$ & $T_{\rm{mixed,c}}$ & $M^{\rm b}_{\rm{mixed}}$ & $\mathcal{L}_{\rm{mixed,tot}}$ & $t_{\rm{Q}}$ &  $M^{\rm b}_{t\rm{_Q}}$ & $\mathcal{L}_{\nu_e}$ & $\mathcal{L}_{\bar\nu_e}$ &  $\mathcal{L}_{\nu_x}$& $\langle{\epsilon_{\nu_e} \rangle }$ &$ \langle {\epsilon_{\bar\nu_e}\rangle }$ & $ \langle {\epsilon_{\nu_x}}\rangle$ \\ \hline 
          & $[\rm{s}]$ & [$10^{14}~\rm{g~cm}^{-3}$] & [$\rm{MeV}$] & [$M_{\odot} $] & [$10^{51}\rm{erg~s^{-1}}$] &  $[\rm{s}]$ & [$M_{\odot} $] & & [$10^{51}\rm{erg~s^{-1}}$] &  & & [$\rm{MeV}$] & \\ 
	 \texttt{RDF-1.9} & 0.4 & 4.1 & 14.3 & 1.58 & 79 & 0.81 & 1.66 & 363 & 914 & 669 & 52.7 & 59.2 & 98.4 \\ \hline 
	 \texttt{RDF-1.2} & 1.0 & 5.3 & 23.9 & 1.69 & 55.9 & 1.59  & 1.76 & 178 & 1234 & 333 & 40 & 61.4 & 90.4 \\ \hline 
	 \texttt{RDF-1.5} & 1.9 & 6.26 & 22.9 & 1.78 & 39.8 & 2.27 & 1.8 & 188 & 752 & 372 & 29 & 42.2 & 112.4 \\ \hline 
         \texttt{RDF-1.1} &  2.7 &  6.84 & 24.7 & 1.81 & 23.3 & 2.94 & 1.82 & 183 & 875 & 338 & 42.2 & 48 & 147.9 \\  \hline   
        \end{tabular}
	}}
\caption{Various quantities extracted at the moment of first
		reaching the mixed nuclear-quark phase (columns in the left part) 
		in our CCSN simulations, using different RDF EOS.
		$t_{\rm mixed}$, $\rho_{\rm{mixed,c}}$, $T_{\rm{mixed,c}}$ 
		, $M^{\rm b}_{\rm{mixed}}$ and $\mathcal{L}_{\rm{mixed,tot}}$ denote the 
		the time after the first core bounce at $t=t_{\rm b}$, 
		central density, central temperature, the enclosed baryon mass of the PNS
		and the total neutrino luminosity for the system first reaching the mixed 
		nuclear-quark phase.
		$t_{Q}$ and $M^{\rm b}_{t\rm{_Q}}$ is the time and the enclosed baryon mass of the PHS
		first reaching the deconfined quark phase 
		with respect to $t_{\rm b}$.
		The values of neutrino luminosity and averaged energy 
		are evaluated at $r=500~\rm{km}$
		The values of the second neutrino burst are represented as $\mathcal{L}_{\nu_l}$ 
		and $ \langle{\epsilon_{\nu_l}}\rangle$, respectively for electron
		neutrino $\nu_e$, electron antineutrino $\bar{\nu}_e$, and
		a collective species describing heavy
		lepton neutrinos and their antineutrinos as $\nu_x$.
		Note that the PNS or PHS are defined as the regions of 
		$\rho > 10^{11}~\rm{g~cm^{-3}}$.}
\label{table_ccsn}
\end{table*}

The relative luminosities of the first and 
second bursts depend on EOS properties, 
particularly the onset density, which affects the shock strength. 
(See discussion of Fig.~\ref{nu_observables}). 
Consequently, the total luminosity of the model with 
\texttt{RDF-1.9} and \texttt{RDF-1.2} are larger than the first neutrino burst, with
 $\mathcal{L}_{\rm{tot}} = 2.99\times10^{54}~\rm{ergs}^{-1}$ and
 $1.35\times10^{54}~\rm{ergs}^{-1}$ respectively.
On the other hand, $\mathcal{L}_{\rm{tot}}$ of 
the second neutrino burst in the \texttt{RDF-1.5} 
and \texttt{RDF-1.1} models are weaker than the first, with
$\mathcal{L}_{\rm{tot}} = 2\times10^{53}~\rm{erg~s}^{-1}$ and
$5.47\times10^{52}~\rm{erg~s}^{-1}$, respectively.
Thus, no clear correlations exist between the total
luminosities of the first and second bursts, as these are highly
sensitive to EOS.

\section{Comparison of phase transition in AIC and CCSN}
\label{sec:ccsn_pt}
To investigate how a massive envelope in the progenitor 
influences the onset conditions and neutrino observables,
we perform four CCSN simulations using the same progenitor and setup as in Appendix~\ref{sec:codetest},
except that the computational domain extends from $[0, 1\times10^{4}]~\rm{km}$,
with a resolution of $N_r = 128$ and allowing 11 AMR levels.
The finest grid size $\Delta r$ at the center of the star is $76.3~\rm{m}$.
The four RDF EOSs used are identical to those in the main text.

Similar to Appendix~\ref{sec:codetest}, the collapse of the iron core is halted 
by the stiffening of the EOS when the central density $\rho_c$ exceeds 
nuclear saturation density ($\gtrsim 10^{14}~\rm{g~cm^{-3}}$), leading to a core bounce.
In contrast to the AIC case, the PNS enters the mixed phase only after 
core bounce due to the higher onset density characteristic of the RDF EOS family.
The left columns of Table~\ref{table_ccsn} present the time of mixed-phase 
entry $t_{\rm{mixed}}$, central density $\rho_{\rm{mixed,c}}$, 
central temperature $T_{\rm{mixed,c}}$, 
the enclosed baryonic mass of the PNS $M^{\rm b}_{\rm{mixed}}$, 
and the total neutrino luminosity $\mathcal{L}_{\rm{mixed,tot}}$ 
at the moment the system first enters the mixed phase, defined by $Y_q > 0$.

The baryonic mass at the onset of the mixed phase, $M^{\rm b}_{\rm{mixed}}$,
varies by up to $15.8\%$ across the CCSN models,
indicating a prolonged accretion phase due to the presence of a massive envelope.
In contrast, the AIC models exhibit a variation of less than $0.07\%$ in $M^{\rm b}_{\rm{mixed}}$,
reflecting the absence of extended mass accretion.

Due to the elevated central temperature $T_c$
and low proton fraction $Y_p$ in the PNS core,
the modified onset density $\rho_{\rm{mixed,c}}$ is lower than the cold,
beta-equilibrated onset density $\rho_{\rm{onset}}$ listed in Table~\ref{model}.
For instance, in model \texttt{RDF-1.1}, $\rho_{\rm{mixed,c}}$ is 
$22\%$ lower than $\rho_{\rm{onset}}$, 
corresponding to $T_{\rm{mixed,c}} = 27.5~\rm{MeV}$ 
(or $17.8~\rm{MeV}$) at $t = t_{\rm{mixed}}$.
This thermal reduction in the onset density is less pronounced in the CCSN system
compared to the $30\%$ reduction observed in the AIC system, as reported in the main text.

The presence of a massive envelope in CCSN leads to
ongoing mass accretion and a faster contraction rate compared to AIC.
As a result, PNSs in CCSN enter the mixed phase earlier than their
AIC counterparts when evolved with the same RDF EOS,
and generally exhibit similar or higher values of $\rho_{\rm{mixed,c}}$.
For example, the CCSN model with \texttt{RDF-1.5} enters the mixed phase approximately
$1~\rm{s}$ earlier than in the corresponding AIC case, with a higher central density at the onset.
Therefore, continuous accretion in CCSN can significantly influence the onset conditions
for the QCD PT in the PNS.

Upon the onset of the first-order PT,
the contraction of the PNS accelerates
due to EOS softening in the mixed phase and receives
reduced support from Kelvin–Helmholtz cooling.
This support is less pronounced at higher onset densities,
as indicated by the decreasing trend in $\mathcal{L}_{\rm{mixed,tot}}$.
Although this trend is qualitatively similar to that observed in AIC,
$\mathcal{L}_{\rm{mixed,tot}}$ in CCSN is approximately a factor of 2 larger,
primarily due to the larger $M^{\rm b}_{\rm{mixed}}$,
and thus a greater gravitational binding energy.
Consequently, the presence of a massive envelope in CCSN leads to
a stronger neutrino cooling effect compared to AIC systems,
resulting in a more rapid increase in central density.

With continued contraction and ongoing softening of the EOS,
the PNS undergoes a supersonic collapse that is halted by
the formation of a stiffer deconfined quark core.
This results in the formation of a hydrodynamical shock,
whose propagation generates a second neutrino burst primarily due to positron capture.
This behavior is qualitatively consistent with the PT dynamics observed in the AIC models presented in the main text
and is well documented in various studies of QCD PT in CCSNe
~\citep{Sagert:2008ka,Zha:2020gjw,Kuroda:2021eiv,Fischer:2021tvv,Zha:2021fbi,Jakobus:2022ucs,Largani:2023oyk}.

The right columns in Table~\ref{table_ccsn} present the onset time 
of the deconfined quark phase $t_{\rm{Q}}$, the enclosed baryon mass of the PHS
at the onset of the deconfined quark phase $M^{\rm b}_{t\rm{_Q}}$ ,
the neutrino luminosity  $\mathcal{L}_{\nu_l}$ 
and average energy $ \langle{\epsilon_{\nu_l}}\rangle$
for all neutrino species associated with the second neutrino burst.

The secular mass accretion in CCSN continues throughout the PT,
as indicated by the increase in the enclosed baryonic mass of the PNS ($M^{\rm b}_{t_{\rm Q}} > M^{\rm b}_{\rm{mixed}}$).
In contrast, the change in baryonic mass during the PT in AIC is negligible.
This supports a more rapid progression of the PT in CCSN,
as reflected by the shorter transition duration $t_{\rm{Q}} - t_{\rm{mixed}} \lesssim 0.6~\rm{s}$ across all models,
compared to the longer timescale of up to $\sim 3~\rm{s}$ in the AIC case, such as for model \texttt{RDF-1.1}.
These results highlight the significant role of a massive envelope
in facilitating the PT in CCSN, making the evolution of the PNS less sensitive
to the specific properties of the hybrid EOS described in the main text.
 
The onset condition of the deconfined quark phase determines
the characteristics of the second hydrodynamical shock,
and consequently, the associated neutrino observables.
Since $M^{\rm b}_{t_{\rm Q}}$ varies among different CCSN models,
an EOS with a higher onset density can lead to collapse in a more massive PNS,
complicating the collapse dynamics.
It is evident that, in CCSN, none of the neutrino observables
exhibits a strictly consistent decreasing trend across all neutrino species.
In contrast, both $\mathcal{L}_{\bar\nu_e}$ and $\langle{\epsilon_{\nu_l}}\rangle$
show a clear decreasing trend with increasing onset density in AIC models.
Although a more detailed and systematic
investigation of the PHS structure and collapse dynamics is required to
fully understand how the onset condition influences the neutrino signal,
it is clear that the presence of a massive envelope in CCSN
can modify the onset conditions and alter the resulting neutrino observables.
The presence of a massive envelope introduces a potential systematic uncertainty
when using these observables to constrain the properties of the hybrid EOS.
In contrast, such systematics are negligible in AIC due to the absence of a massive envelope,
making AIC a cleaner environment for probing the EOS-dependent features of the PT.

\end{document}